# Fator de modificação nuclear e anisotropia azimutal evento-a-evento de quarks pesados em colisões de íons pesados

*Caio Alves Garcia Prado*

Orientador: Prof. Dr. Alexandre Alarcon do Passo Suaide

Tese de doutorado apresentada ao Instituto de Física para a obtenção do título de Doutor em Ciências

Banca Examinadora:
Prof. Dr. Alexandre Alarcon do Passo Suaide (IFUSP)
Prof. Dr. Airton Deppman (IFUSP)
Prof.ª Dr.ª Frederique Marie-Brigitte Sylvie Grassi (IFUSP)
Prof. Dr. Mauro Rogerio Cosentino (UFABC)
Prof. Dr. Eduardo de Moraes Gregores (UFABC)

São Paulo

2017

Universidade de São Paulo, Instituto de Física, São Paulo, SP.

---

---

17 18 19 20 21     5 4 3 2 1

agosto de 2017

# Heavy-flavor nuclear modification factor and event-by-event azimuthal anisotropy correlations in heavy ion collisions

*Caio Alves Garcia Prado*

Advisor: Prof. Dr. Alexandre Alarcon do Passo Suaide

A dissertation submitted to Instituto de Física in partial fulfillment of the requirements for the degree of Doctor of Phylosophy in Physics

Examiner's Committee:
Prof. Dr. Alexandre Alarcon do Passo Suaide (IFUSP)
Prof. Dr. Airton Deppman (IFUSP)
Prof. Dr. Frederique Marie-Brigitte Sylvie Grassi (IFUSP)
Prof. Dr. Mauro Rogerio Cosentino (UFABC)
Prof. Dr. Eduardo de Moraes Gregores (UFABC)

São Paulo

2017

Universidade de São Paulo, Instituto de Física, São Paulo, SP.

# Heavy-flavor nuclear modification factor and event-by-event azimuthal anisotropy correlations in heavy ion collisions

**Abstract**

Relativistic heavy ion collisions, which are performed at large experimental programs such as Relativistic Heavy Ion Collider's (RHIC) STAR experiment and the Large Hadron Collider's (LHC) experiments, can create an extremely hot and dense state of the matter known as the quark gluon plasma (QGP). A huge amount of sub-nucleonic particles are created in the collision processes and their interaction and subsequent evolution after the collision takes place is at the core of the understanding of the matter that builds up the Universe. It has recently been shown that event-by-event fluctuations in the spatial distribution between different collision events have great impact on the particle distributions that are measured after the evolution of the created system. Specifically, these distributions are greatly responsible for generating the observed azimuthal anisotropy in measurements. Furthermore, the eventual cooling and expansion of the fluctuating system can become very complex due to lumps of energy density and temperature, which affects the interaction of the particles that traverse the medium.

In this configuration, heavy flavor particles play a special role, as they are generally created at the initial stages of the process and have properties that allow them to retain "memory" from the interactions within the whole evolution of the system. However, the comparison between experimental data and theoretical or phenomenological predictions on the heavy flavor sector cannot fully explain the heavy quarks coupling with the medium and their subsequent hadronization process.

This work presents a phenomenological study of the evolution of heavy quarks, namely bottom and charm, within the QGP. In order to accomplish that, a computer simulation framework has been developed in which heavy quarks are sampled from fluctuating events and travel the underlying hydrodynamical expanding medium background. The interaction of the heavy quarks with the medium is explored via parametrization of energy loss models and the final spectra of particles are obtained after the hadronization of these quarks and futher decays of the heavy mesons. The observables are tested within the phenomenological framework under different parameters setup and analyzed.

The simulation was able to generate results that are consistent with currently available experimental data within error bars for the high-$p_\mathrm{T}$ regime. It was also possible to predict, for the first time, the third Fourier harmonic cumulant $v_3\{2\}$ for heavy mesons. Furthermore, a different observable that encodes the event-by-event correlations between the heavy flavor particles and charged soft particles in the collision is proposed, using event-shape engineering techniques.

# Resumo

Colisões de íons pesados relativísticos, que são realizadas em grandes experimentos tais como o experimento STAR no *Relativistic Heavy Ion Collider* (RHIC) e os experimentos do *Large Hadron Collider* (LHC), são capazes de gerar um estado da matéria nuclear extremamente quente e denso, conhecido como plasma de quarks e glúons (QGP). Uma enorme quantidade de partículas sub-nucleônicas são geradas no processo de colisão e suas interações e a subsequente evolução do sistema após a colisão estão no cerne do compreendimento da estrutura da matéria que forma o Universo. Recentemente foi mostrado que as flutuações event-a-evento na distribuição espacial de partículas entre eventos diferentes são de extrema importância na distribuição final de momento das partículas que são detectadas após a evolução do sistema. Especialmente, essas flutuações são responsáveis por gerar a anisotropia azimutal que é observada nos experimentos. Além disso, a expansão e resfriamento do meio ao longo de sua evolução podem resultar em uma dinâmica complexa devido aos focos de diferentes densidades de energia e temperatura e isso deve influenciar a interação das partículas que o atravessam.

Partículas pesadas são essenciais nessa configuração pois são geradas nos instantes iniciais da colisão e suas propriedades permitem que elas retenham "memória" das interações ao longo de toda a evolução do sistema. Entretanto, a comparação entre dados experimentais e predições teóricas ou fenomenológicas no setor de partículas pesadas não é capaz de explicar completamente o acoplamento de quarks pesados com o meio e os subsequentes processos de hadronização.

Este trabalho apresenta um estudo fenomenológico da evolução de quarks pesados, estritamente *bottom* e *charm*, dentro do QGP. Para isso, foi desenvolvida uma simulação computacional em que quarks pesados são sorteados de eventos com diferentes flutuações e em seguida atravessam o meio hidrodinâmico evoluído desses eventos. A interação dos quarks pesados com o meio é explorada através de parametrizações de modelos de perda de energia e os espectros finais de partículas são obtidos após a hadronização destes quarks e subsequente decaimento dos mésons pesados. Os observáveis são testados dentro dos moldes da fenomenologia sob diferentes configurações de parâmetros e analisados.

Com a simulação, foi possível obter resultados consistentes com os dados experimentais disponíveis atualmente dentro das margens de erros para o regime de alto $p_{\mathrm{T}}$. Também foi possível prever, pela primeira vez, o terceiro cumulante dos harmônicos de Fourier $v_3\{2\}$ para quarks pesados. Além disso, um novo observável que agrega as correlações evento-a-evento entre quarks pesados e partículas carregadas leves na colisão é proposto, usando como base técnicas de engenharia de forma de eventos.

Unitermos: 1. Física de alta energia; 2. Quark; 3. Hádrons.

**Acknowledgments**

To my parents and my brother, for all the love in the world that has always given me strength.

To my advisor, Alexandre Suaide, for his amazing character which gives me hope, for believing in my work, and for the great discussions and lessons that kept me growing.

To the HEPIC group at the Institute of Physics, for the meetings and discussions that have always helped improve this work.

To Roland Katz, Jacquelyn Noronha-Hostler, Jorge Noronha, and Marcelo Munhoz for all the work we did together, for pushing me forward and for sometimes believing in me more than myself.

To my friends, so many and so dear, for always trusting in me, believing in my work, and giving me enough happiness to keep moving on.

To the Physics Institute of the University of São Paulo, for giving me the tools to my academic formation.

To CNPq, for the financial support to this research project.

# Contents

# List of Figures

# List of Tables

# Introduction

It's been long since the discovery of the quark gluon plasma (QGP) by the Relativistic Heavy Ion Collider (RHIC) in 2005.[1–4] Since then, the investigation of strongly interacting matter under extreme conditions of temperature and energy density has been the bleeding-edge field in the nuclear physics research. The universe as we know, under low temperature and densities, remains in a confined state known as hadronic matter, however, when exposed to extremely high temperatures or density, such hadronic matter undergoes a phase transition to a deconfined state, the QGP. These same conditions have prevailed during the formation of the early universe, and is still existing inside of neutron stars, where extremely condensed matter is maintained at small temperatures. Therefore, at the core of the nucleons, there lies the probes and, hopefully, some answers for a variety of fields in physics such as particle physics, astrophysics, cosmology and, of course, nuclear physics.

Quantum chromodynamics (QCD)[5,6] is the fundamental theory that describe the strong interactions. Its formulation predicted the existence of the quark gluon plasma[7–9] as a system where the dominant degrees of freedom are the constituents of hadrons, the quarks and gluons, or in other words, quoted from the early STAR collaboration: *"a (locally) thermally equilibrated state of matter in which quarks and gluons are deconfined from hadrons, so that color degrees of freedom become manifest over nuclear, rather than merely nucleonic, volumes."*[2] The understanding of such a state and its transition from the hadronic matter is an important tool in order to fully comprehend the QCD phase diagram and the properties of the fundamental particles that build up the universe.

In order to study the properties of this strongly-interactive matter, large experimental programs have been built and remain under heavy development. The Large Hadron Collider (LHC) in Geneva, from the European Research Center (CERN— *Conseil Européen pour la Recherche Nucléaire*) and the already mentioned RHIC in Upton, from the Brookhaven National Laboratory (BNL), are dedicated to the study of these properties by colliding particles at extremely high energies. These collisions are able to replicate the extreme thermodynamical conditions that generates the QGP. Furthermore, with the beam energy scan project at RHIC, one can systematically map the phase diagram by the variation of the beam energy. Specifically, the search for the critical point is one of the main goals of the project. The critical point is a second order transition point expected to mark the end of the first order transitions line and the beginning of analytical crossover transitions.[10] The existence and exact location of the critical point is still one of the biggest questions in the field.

Heavy quarks, namely bottom (b) and charm (c), have their masses much greater than the QCD scale parameter $\Lambda_{QCD}$. Due to this fact, they are important assets in the study of the medium created in the collisions.[11] Furthermore, they are mostly produced at the very beginning of the system's evolution. As they traverse the medium throughout the whole lifetime of the expansion, since the thermal relaxation time of heavy flavor particles is much longer than the thermalization time of the bulk medium, they retain this information which is transfered to their measurable products.

Specifically, the heavy quarks interaction with the medium gives information on the opacity of the QGP. This information is usually studied by means of energy loss and jet quenching, comparing results from a small medium such as the one formed in proton-proton collisions, for instance, with nuclei collisions that can create the plasma. The nuclear modification factor is one of the main observables in this regard. On the other hand, the azimuthal distribution of particles is one of the most important probes of the QGP and is related to studies on the collectivity motion of the particles in the plasma. By expanding this distribution into a Fourier series, one can investigate the harmonic coefficients in order to describe this distribution. However, due to fluctuations on the spatial distribution of the medium formed in the collisions, a proper study of these harmonics can only be achieved in event-by-event studies, in which these fluctuations are taken into account. Although average evaluations of the observables related to the azimuthal distributions require much less statistics from the measurements and computations, they tend to wash out important effects of the fluctuations even for the lowest orders of the azimuthal expansion.

Currently, multiple phenomenological models have been proposed in order to describe available data for heavy flavor particles. However, the correct description of both the nuclear modification factor and the elliptic flow, which is the lowest order Fourier harmonic, at the same time, seems to be a non-trivial task. Moreover, in order to further constrain the various available models, one need to further investigate other observables such as higher order Fourier harmonics. These harmonics, such as the triangular coefficient, have not yet been fully explored, mainly due to the amount of statistics needed in order to obtain reasonable measurements, however, the increasing particle multiplicity of the experiments should allow for these measurements to improve in the near future. One of the main concerns on the heavy flavor studies in ultrarelativistic heavy ion collisions is to understand how and to what extent the heavy quarks couple with the medium created during the collision. Furthermore, heavy flavor azimuthal distribution is generated from completely dif-ferent mechanisms than light particles. Therefore, in order to search for observables that can be sensitive to this coupling, one could investigate the correlation of particles between the heavy sector and the soft sector.

This work presents a study on the effects of an event-by-event approach to evaluate the azimuthal anisotropy of heavy flavor mesons from lead-lead collisions as the ones realized at LHC. In this study, a computer simulation have been developed in order to compute the heavy flavor sector Fourier harmonics, including predictions for the triangular coefficient. Furthermore, the correlation of particles between the heavy and soft sectors have been explored.

This document is organized as follows: in this first chapter, the basic tools that support the phenomenological and theoretical development on high energy physics is presented. Chapters 2 and 3 describe the evolution of the QGP and the interaction of the heavy quarks with it. Chapter 4 presents the development of the simulation program. Finally, the results obtained from the simulation are presented and discussed in chapter 5.

## 1.1 Quantum chromodynamics

Quantum chromodynamics (QCD) is the subfield in Quantum Field Theory (QFT) that describes the strong interactions. These interactions occur between elementary particles called *quarks*,[12,13] which were proposed back in 1964 as constituents of baryons and mesons in order to explain the interactions between these particles. The existence of the quarks was first observed at the Stanford Linear Accelerator Center (SLAC)[14,15] in inelastic scattering of electrons into protons and neutrons targets. The Deep Inelastic Scattering (DIS) led to different theoretical interpretations that constitutes the basis for the parton model. The *scaling* property, predicted by Bjorken[16] to hold in the DIS, expected that the cross section of the scattering would scale with the ratio between the momentum transfer to the energy transfer from the electrons to the hadrons:

$$x = \frac{-q^2}{2M\nu}, \qquad (1.1)$$

where $q$ is the momentum transfer from the electron to the hadrons, $\nu$ is the energy transfer in the rest frame of the hadron and $M$ is the mass of the target hadron. Feynman interpreted the Bjorken $x$ as the fraction of the hadron momentum carried by a given parton, a point-like constituent inside the hadrons.[17]

According to Pauli exclusion principle, the quarks inside the nucleons could not coexist as they would have identical quantum states. The idea of the color charge was then proposed in 1964[18] as a hidden quantum number that would explain how the Pauli principle was not violated. Evidence for the existence of the color charge came from baryon spectrometry[18] and from $\pi^0$ decay rate.[19,20] The gauge theory of color was first proposed in 1965 and introduced the *gluons*[21,22] as an octet in an SU(3) symmetry group theory and as mediators of the strong force between the quarks. The color as the strong interaction charge and the gluon as mediator of these interaction were finally described as known today in 1973.[23,24] The group of theories that describe the elementary particles is known as *Standard Model* which describes these particles as composed of leptons, quarks, and mediators. The quarks carry fractional electric charge and color charges, conventionally called *red*, *green*, and *blue*. In addition, quarks have 6 flavours: up, down, strange, charm, bottom, and top, listed in increasing mass. The interactions between them is mediated by gluons, which are massless particles that carry color charge. This implies that the gluons also interact with each other.

The quantum chromodynamics was developed from the color gauge theory and the Yang-Mills theory[25] in 1973.[24] The non-abelian characteristics of this theory leads to the interactions among gluons and the further development of the theory leads to very important properties such as color confinement and asymptotic freedom.[5,6] The QCD Lagrangian is given by:

$$\mathcal{L}_{\text{QCD}} = \bar{\Psi}(i\slashed{D} - m)\Psi - \frac{1}{2}\text{Tr}(G_{\mu\nu}G^{\mu\nu}),\tag{1.2}$$

in which $\mathcal{D}_\mu$ is the covariant derivative, $G_{\mu\nu}$ the gluon fields, and $\Psi$ the color triplet spinor defined as:

$$\Psi = \begin{pmatrix} \text{r} \\ \text{g} \\ \text{b} \end{pmatrix},\tag{1.3}$$

for red (r), green (g), and blue (b) colors. In this configuration the singlet formation is always preferable.

In order to develop a perturbative theory of QCD (pQCD), the physical observables are defined using the renormalized strong coupling $\alpha_\text{s}(\mu^2)$, in which $\mu$ is the renormalization scale. As consequence of the non-abelian property of the theory, the running coupling $\alpha_\text{s}(q^2)$ for the scale energy $q$ can be obtained as:

$$\frac{1}{\alpha_\text{s}(q^2)} = \frac{1}{\alpha_\text{s}(\mu^2)} + \frac{33 - 2n_\text{f}}{12\pi}\ln\frac{-q^2}{\mu^2},\tag{1.4}$$

in which $n_\text{f}$ is the number of active flavours at $q$. The running coupling can still be expressed in terms of the QCD scale $\Lambda_{\text{QCD}}$ leading to the following expression:

$$\alpha_\text{s}(q^2) = \frac{12\pi}{33 - 2n_\text{f}\ln\frac{q^2}{\Lambda_{\text{QCD}}^2}}.\tag{1.5}$$

From equation 1.5, if $n_\text{f} \leq 16$, the running coupling decreases with the increasing of $q$ or with the decreasing of quarks distance, this implies that for $q \gg \Lambda_{\text{QCD}}$, the strong interaction becomes very weak so that the partons can be treated as free particles. This is the regime in which perturbative QCD is valid. On the other hand, the opposite is expected for large distances between a quark and an anti-quark. The amount of energy needed to separate the quarks is so that it becomes increasingly favorable to produce a new pair $q\bar{q}$ from the vacuum which in turn forms a new meson. The plot in figure 1.1 shows measurements of the running coupling $\alpha_s$ as function of the energy scale $q$ in comparison with QCD calculations.

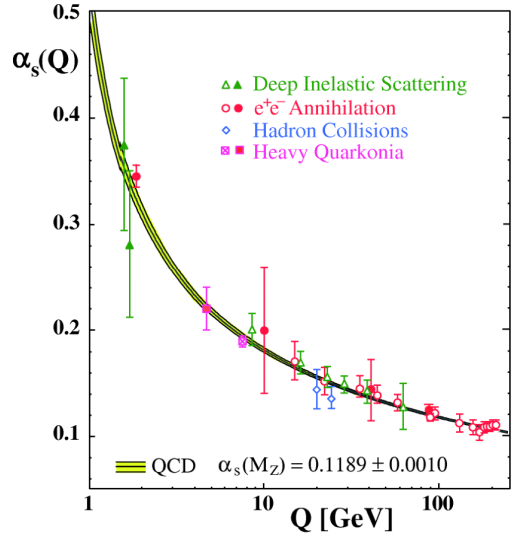

FIGURE 1.1 – Measurements of the running coupling $\alpha_s(q)$ as function of the energy scale $q$. Open symbols indicate NLO calculations while filled symbols indicate NNLO used in the analysis.[26]

## 1.2 Phase transitions and the quark gluon plasma

The analysis of phase transitions and the properties of the nuclear matter close to such transitions requires the application of non-perturbative calculations.[27] As such, lattice QCD[28] calculations at zero chemical potential $\mu_0$ is an important formulation for these studies in which the space-time is discretized in order to obtain finite path integrals evaluations. These calculations predict that the critical temperature $T_c \approx 155$ MeV for the crossover transition.[29] The crossover transition differs from first-order and second-order transitions as the transition is not discontinuous, whereas first-order transitions occurs when the temperature derivative of the free energy of the matter $\partial F/\partial T$ is not continuous, analogously, second-order transitions have the second derivative $\partial^2 F/\partial^2 T$ discontinuous. A schematic depiction of the expected phase transitions is illustrated in figure 1.2 as a phase diagram. In the figure, the crossover transition from hadronic matter to the quark gluon plasma is represented as a dashed line while the first order transition is represented by a continuous line. The critical point, as exposed in the previous sections, is a second-order transition in the diagram and is still one of the many puzzles yet to be solved.

The thermodynamical properties of the quark gluon plasma can be evaluated from the equation of state, current conservation laws and the energy-momentum.

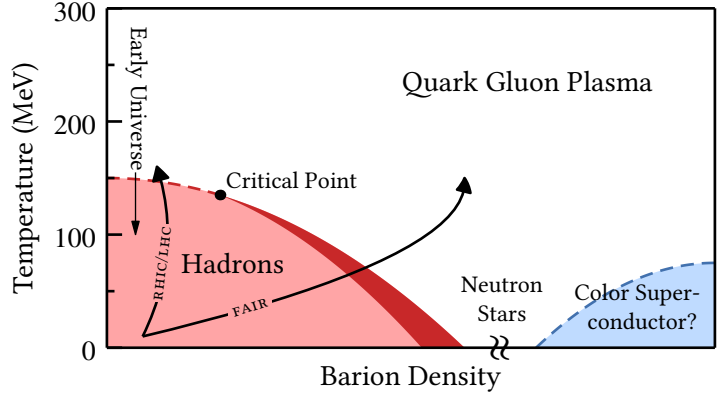

Figure 1.2 – Phase diagram of QCD. The arrows indicate the transitions that are executed at current experiments or planned for future experimental programs.[30]

The equation of state of an ideal gas of massless quarks and gluons is predicted by theoretical calculations to be determined only the number of degrees of freedom:[2]

$$\frac{\varepsilon_{\text{SB}}}{3T^4} = \frac{p_{\text{SB}}}{T^4} = \left[ 2(n_{\text{c}} - 1) + \frac{7}{2} n_{\text{c}} n_{\text{f}} \right] \frac{\pi^2}{90}, \qquad (1.6)$$

in which $\varepsilon$ is energy density, $p$ is the pressure, $T$ is the temperature, $n_{\text{c}}$ the number of colors, and $n_{\text{f}}$ the number of flavors. The SB subscript stands for Stefan-Boltzmann distributions.

Once created, the plasma is not static, instead it evolves with time. The scheme in figure 1.3 illustrates the stages of the system's evolution after the collision takes place. After the Lorentz contracted nuclei collide with each other a transient phase, where particles interact inelastically, is created and must be modeled, as no direct experimental measurements are possible. One of the main concerns about the system's evolution is whether it establishes, maintains, and to what extent, a local thermal equilibrium. One of the experimental indications of thermal equilibrium is observed in the relative abundances of produced particles.[31,32] When thermalization is assumed, it is reasonable to treat the resulting matter as a relativistic fluid undergoing collective, hydrodynamic flow.[2,33] This macroscopic approach has been used for some time[34,35] and have been used to successfully describe some of the experimental data from RHIC[33] in an ideal framework. The QGP was then regarded as a strongly-coupled plasma that flows like a liquid.[1,36] The ideal hydrodynamics though, was not the proper way to describe the evolution of the QGP[37–39] and specially, hydrodynamics could not be used to describe the later hadronic stage, after the hadronization, and should be limited to the QGP phase.[40–43] Two main ingredients should still be considered: the viscosity and the fact that the initial conditions fluctuate event-by-event.[44–48] The implementation of relativistic dissipative fluid dynamics requires second order terms to be included,[49] and one of the most widely used theoretical frameworks is the Israel-Stewart theory.[50,51]

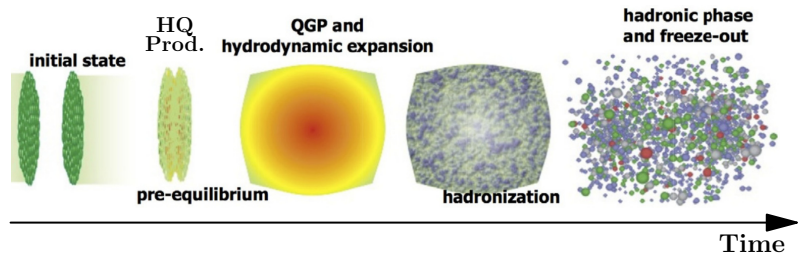

Figure 1.3 – Illustration of the stages of evolution of the matter created in relativistic heavy ion collisions.

When the system reaches the critical temperature $T_c$, the partons start to group into hadrons, this stage cannot be described by hydrodynamics and needs to be modeled differently. One of the most used models do describe the hadronization of the QGP is the Cooper-Frye model.[52]

This work focuses on the study of the heavy flavor observables affected by event-by-event fluctuating initial conditions on top of an evolving quark gluon plasma. The hydrodynamics is implemented using the Smoothed-Particle Hydrodynamics algorithm with viscous corrections on the system's evolution and the Cooper-Frye freeze-out.

# Evolution of the quark gluon plasma

One of the main aspects to be considered in the study of the quark gluon plasma is its evolution over time. As explained in section 1.2, if thermalization of the medium is assumed to occur shortly after the collision and be sustained until the hadronization phase, the QGP can be described by means of a hydrodynamical model. One has to consider the fluctuations on the initial conditions generated by the collisions, so one of the main ingredients for hydrodynamical calculations consist of the initial conditions modeling. Those are, in fact, the input of any hydrodynamics algorithm. It has also been shown that the large degrees of collectivity evidenced by the Fourier harmonics are consistent with a strongly-coupled medium with small shear viscosity, with $\eta/s$ close to $1/4\pi$. Ideal hydrodynamics predict a continually increasing elliptic flow while experiment data shows that the elliptic flow tents to saturate at large transverse momentum. It also fails to predict results for more peripheral collisions and only central collisions happens to be well described by ideal hydrodynamics. On the other hand, dissipative hydrodynamics has many difficulties. First order calculations such as Bjorken model[53] lead to instabilities[54] and violates causality,[55] with perturbations that can propagate at infinite speeds.

The solution to some of these problems is to use second order theories such as the Israel-Stewart.[50,51] In this case, the expansion of entropy 4-current contains terms of second order in dissipative fluxes. This leads to a problem which these fluxes cannot be described as function of the state variables only and the space of thermodynamic variables has to be extended to include them.

This chapter will describe the hydrodynamic modeling of the medium and one of the most commonly used algorithms for numerical calculations, the smoothed-particle hydrodynamics. Furthermore, a description of the initial conditions modeling and the Cooper-Frye prescription is also presented.

## 2.1 Equations of motion

In the following sections, Greek indexes are employed in the usual way for 4-vectors while the bold letters are used for 3-vectors with Latin indexes for their components.

The dissipative hydrodynamics theories following non-equilibrium thermodynamics were derived by Eckart and Landau-Lifshitz.[56,57] It starts with the conservation laws for the particle current $N^\mu$, the energy-momentum tensor $T^{\mu\nu}$ and the entropy current $S^\mu$:[50,51,58]

$$\partial_\mu N^\mu = 0 \tag{2.1}$$

$$\partial_\mu T^{\mu\nu} = 0\,, \tag{2.2}$$

and the second law of thermodynamics:

$$\partial_\mu S^\mu \geq 0\,. \tag{2.3}$$

The hydrodynamic 4-velocity is defined as $u^\mu$, with $u^2 = 1$ and the projector operator $\Delta^{\mu\nu} = g^{\mu\nu} - u^\mu u^\nu$ is defined as orthogonal to the 4-velocity in such a way that $\Delta^{\mu\nu}u_\nu = 0$. The densities variables can be obtained from the 4-velocity when in equilibrium as:[50,51,58]

$$N^\mu_{\text{eq}} = nu^\mu\,, \tag{2.4}$$

$$T^{\mu\nu} = \varepsilon u^\mu u^\nu - p\Delta^{\mu\nu}\,, \tag{2.5}$$

$$S^\mu_{\text{eq}} = su^\mu\,. \tag{2.6}$$

It is then equivalent to fully specify the equilibrium state either with the parameters $(n, \varepsilon, \boldsymbol{u})$ or with the thermal potential $\alpha = \mu/T$ and the inverse 4-temperature $\beta^\mu = u^\mu/T$, in which the quantity $\mu$ is the chemical potential. The pressure $p$ can be obtained from:[50,51,58]

$$S^\mu_{\text{eq}} = p\beta^\mu - \alpha N^\mu_{\text{eq}} + \beta_\lambda T^{\lambda\mu}_{\text{eq}}\,. \tag{2.7}$$

It follows from the above by using the Gibbs-Duhem equation $\mathrm{d}(p\beta^\mu) = N^\mu_{\text{eq}}\,\mathrm{d}\alpha - T^{\mu\nu}_{\text{eq}}\,\mathrm{d}\beta_\lambda$ that the entropy current rate can be expressed as:[50,51,58]

$$\mathrm{d}S^\mu_{\text{eq}} = -\alpha\,\mathrm{d}N^\mu_{\text{eq}} + \beta_\lambda\,\mathrm{d}T^{\mu\nu}_{\text{eq}}\,, \tag{2.8}$$

valid for the equilibrium hyper-surface $\Sigma_{\text{eq}}(\alpha, \beta^\mu)$.

Finally, if the system is not in equilibrium, the particle current, entropy current, and energy-momentum tensor have all an additional term:

$$N^\mu = N^\mu_{\text{eq}} + \delta N^\mu = nu^\mu + V^\mu\,, \tag{2.9}$$

$$S^\mu = S^\mu_{\text{eq}} + \delta S^\mu = su^\mu + \Phi^\mu\,, \tag{2.10}$$

$$T^{\mu\nu} = T^{\mu\nu}_{\text{eq}} + \delta T^{\mu\nu} = \left[\varepsilon u^\mu u^\nu - p\Delta^{\mu\nu}\right] + \Pi\Delta^{\mu\nu} + \pi^{\mu\nu} + \left(W^\mu u^\nu + W^\nu u^\mu\right)\,, \tag{2.11}$$

in which $V^\mu$ describe the net flow of charge, $\Phi^\mu$ is the entropy flow, and $W^\mu$ the energy flow. In the last equation the term $\Pi = -(1/3)\Delta_{\mu\nu}T^{\mu\nu} - p$ is the pressure associated with the bulk viscosity, and $\pi^{\mu\nu} = \left[(1/2)\left(\Delta^{\mu\sigma}\Delta^{\nu\tau} + \Delta^{\nu\sigma}\Delta^{\mu\tau}\right) - (1/3)\Delta^{\mu\nu}\Delta^{\sigma\tau}\right]T_{\sigma\tau}$ is the shear stress tensor.

The Landau definition for the local rest frame is used:

$$u_\nu T^{\mu\nu} = \varepsilon u^\mu.$$ (2.12)

Also, it is made the assumption that the equilibrium relation from equation 2.8 is valid in a "near equilibrium state" so that it can be generalized as:[50,51,58]

$$\mathrm{d}S^\mu = -\delta N^\mu \,\partial_\mu \alpha + \delta T^{\mu\nu}\,\partial_\mu\beta_\nu + \partial_\mu Q^\mu,$$ (2.13)

in which the term $Q^\mu$ describes the deviations of the particle current, and the energy-momentum tensor. This quantity must be chosen in order to obtain first order theories ($Q^\mu = 0$) or second order theories. In the Israel-Stewart theory, the most general form of $Q^\mu$ can be written as:

$$Q^\mu = -\left(\beta_0 \Pi^2 - \beta_1 q^\nu q_\nu + \beta_2 \pi_{\nu\lambda}\pi^{\nu\lambda}\right)\frac{u^\mu}{2T} - \frac{\alpha_0 \Pi q^\mu}{T} + \frac{\alpha_1 \pi^{\mu\nu}q_\nu}{T},$$ (2.14)

in which $\beta_i$ and $\alpha_i$ are thermodynamic coefficients, and $q^\mu$ is the heat flow.

The entropy production rate can be expressed, after generalization to the near equilibrium state, as:[50,51,58]

$$T\partial_\mu S^\mu = \Pi X - q^\mu X_\mu + \pi^{\mu\nu}X_{\mu\nu},$$ (2.15)

in which the thermodynamic forces are defined as: $X = -\nabla\cdot u$, $X^\mu = \nabla^\mu \mu/T - u^\nu\,\partial_\nu u^\mu$, and $X^{\mu\nu} = \nabla^{\langle\mu}u^{\nu\rangle}$. The angle brackets around the indexes indicate the symmetric and trace-free projection.[*]

One can postulate a linear relation between the dissipative flows and the thermodynamics forces in order to satisfy the equation 2.3 and, considering the Curie principle, the currents can be expressed by:

$$\begin{aligned}
\Pi &= -\zeta\theta,\\
q^\mu &= -\lambda\frac{nT^2}{\varepsilon + p}\,\nabla^\mu\left(\frac{\mu}{T}\right),\\
\pi^{\mu\nu} &= 2\eta\,\nabla^{\langle\mu}u^{\nu\rangle},
\end{aligned}$$ (2.16)

which defines the transport coefficients for bulk viscosity $\zeta$, heat conductivity $\lambda$, and shear viscosity $\eta$.

---

[*]The symmetric trace-free projection is the traceless part of the spatial projection defined, for instance, like:

$$A_{\langle\lambda\mu\rangle} = \left(\Delta^\alpha_{(\lambda}\Delta^\beta_{\mu)} - \frac{1}{3}\Delta_{\lambda\mu}\Delta^{\alpha\beta}\right)A_{\alpha\beta},$$

in which the parenthesis around the indexes are the usual symmetric projection.[51]

After those considerations, the differential equations for the dissipative flow can finally be obtained as:[58]

$$\tau_\Pi \dot{\Pi} + \Pi = -\zeta\theta - \left[\frac{1}{2}\zeta T \partial_\mu \left(\frac{\tau_0}{\zeta T} u^\mu\right)\Pi\right] + I_{\Pi q} \nabla_\mu q^\mu , \tag{2.17}$$

$$\tau_\pi \Delta_\mu^\alpha \Delta_\nu^\beta \dot{\pi}_{\alpha\beta} + \pi_{\mu\nu} = 2\eta\sigma_{\mu\nu} - \left[\eta T \partial_\lambda \left(\frac{\tau_2}{2\eta T} u^\lambda\right)\pi_{\mu\nu}\right] + I_{\pi q} \nabla_{\langle\mu} q_{\nu\rangle} , \tag{2.18}$$

$$\tau_q \Delta_\mu^\nu \dot{q}_\nu + q_\mu = \lambda\left(\nabla_\mu T - T\dot{u}_\mu\right) + \left[\frac{1}{2}\lambda T^2 \partial_\lambda \left(\frac{\tau_1}{\lambda T^2} u^\nu\right) q_\mu\right]$$
$$- I_{q\Pi} \nabla_\mu - I_{q\pi} \nabla_\mu \pi_\mu^\nu , \tag{2.19}$$

in which the dot over the letter denotes the temporal derivative. The above equations define the relaxation times:

$$\tau_\Pi = \zeta\beta_0 , \tag{2.20}$$

$$\tau_\pi = 2\eta\beta_2 , \tag{2.21}$$

$$\tau_q = \lambda T\beta_1 , \tag{2.22}$$

and the coupling coefficients:

$$I_{\Pi q} = \zeta\alpha_0 , \qquad\qquad I_{q\Pi} = \lambda T\alpha_0 , \tag{2.23}$$

$$I_{\pi q} = 2\eta\alpha_1 , \qquad\qquad I_{q\pi} = \lambda T\alpha_1 . \tag{2.24}$$

The addition of the new parameters, the relaxation times, in second order theories solves the problem with causality that is seen in first order theories by suppressing the superluminal propagation modes.[59] Also, the equilibrium assumption that is crucial to first order theories has been relaxed to a near equilibrium state.

Other phenomenological approaches such as the Memory Function Method[60–62] take a different approach and tries to solve some of the complications that arise from second order theories such as new degrees of freedom that could not be known for the QCD dynamics and the highly coupled differential equations that need to be solved. This approach aims at removing the generalities of second order theories while at the same time maintaining causality and a correct treatment of the viscosity in the quark gluon plasma. It provides a minimal requirement for causality by introducing a single additional parameter, the relaxation time $\tau_R$, related to a memory effect introduced in the usual Navier-Stokes equations.

The argument starts from the continuity equation:[60]

$$\frac{\partial n}{\partial t} + \nabla \cdot \boldsymbol{j} = 0 , \tag{2.25}$$

in which $n$ is a number density and $\boldsymbol{j}$ is its associated irreversible current. In non-equilibrium thermodynamics, this current is assumed to be proportional to the gradient of $n$ leading to:

$$\frac{\partial n}{\partial t} - \nabla \cdot (L \nabla n) = 0 . \tag{2.26}$$

The coefficient $L$ is assumed to be constant. The above equation leads to the known diffusion equation:

$$\frac{\partial n}{\partial t} = \zeta \nabla^2 n \,, \tag{2.27}$$

in which $\zeta$ is the diffusion coefficient. This parabolic equation leads to the acausality observed in second order theories. In order to "fix" this problem, one could convert this equation to a hyperbolic form, as in:[60]

$$\tau_{\mathrm{R}} \frac{\partial^2 n}{\partial t^2} + \frac{\partial n}{\partial t} = \zeta \nabla^2 n \,, \tag{2.28}$$

in which the parameter $\tau_{\mathrm{R}}$ has been introduced. This could easily be done if one considers a memory function that accounts for the time delay that takes for the space inhomogeneity to generate the irreversible currents due to Fick's law. This memory function could have the following form:

$$G(t, t') = \begin{cases} 0 & \text{for } t < t' \,, \\ \frac{1}{\tau_{\mathrm{R}}} \exp\left(-\frac{t-t'}{\tau_{\mathrm{R}}}\right) & \text{for } t \geq t' \,. \end{cases} \tag{2.29}$$

The irreversible current can then be written as:

$$\boldsymbol{j} = -\int_{-\infty}^{t} G(t, t') L \, \nabla n(t') \, \mathrm{d}t' \,, \tag{2.30}$$

and one can obtain the Maxwell-Cattaneo equation from its time derivative:[60]

$$\frac{\partial \boldsymbol{j}}{\partial t} = -\frac{1}{\tau_{\mathrm{R}}} L \, \nabla n(t) - \frac{1}{\tau_{\mathrm{R}}} \boldsymbol{j} \,. \tag{2.31}$$

Imposing the continuity in the $\boldsymbol{j}$ current obtained above one gets the resulting hyperbolical equation 2.28.

In order to satisfy the special relativity, the relaxation time $\tau_{\mathrm{R}}$ must have a lower bound, given by:

$$\tau_{\mathrm{R}} \geq \frac{\zeta}{c^2} \,. \tag{2.32}$$

One can then apply this phenomenology to the currents. By recalling the Landau-Lifshitz theory, a slightly different linearization than equations 2.16 leads to:[60]

$$\Pi = -\zeta_{\mathrm{MF}} \, \partial_\alpha u^\alpha \,, \tag{2.33}$$

$$\pi_{\mu\nu} = \eta_{\mathrm{MF}} \, \partial^{\langle\alpha} u^{\beta\rangle} \,, \tag{2.34}$$

$$q_\mu = -\lambda_{\mathrm{MF}} \Delta_{\mu\nu} \partial^\nu \alpha \,, \tag{2.35}$$

in which the coefficients for bulk viscosity, shear viscosity and heat conductivity with indexes MF for the Memory Function method are defined differently from the Israel-Stewart formulation. The projector operators are needed here in order to ensure that the shear tensor and the heat flow are orthogonal to $u^\mu$. The unprojected versions of these quantities, $\tilde{\pi}_{\mu\nu}$ and $\tilde{q}_\mu$, can be defined by:[60]

$$\pi^{\mu\nu} = \left[ \frac{1}{2} \left( \Delta^{\mu\alpha} \Delta^{\nu\beta} + \Delta^{\mu\beta} \Delta^{\nu\alpha} \right) - \frac{1}{3} \Delta^{\mu\nu} \Delta^{\alpha\beta} \right] \tilde{\pi}_{\alpha\beta} , \tag{2.36}$$

$$q^\mu = \Delta^{\mu\nu} \tilde{q}_\nu . \tag{2.37}$$

Now, the introduction of the memory function as defined by equation 2.29 leads to:[60]

$$\Pi(\tau) = - \int_{-\infty}^{\tau} G(\tau, \tau') \zeta \, \partial_\alpha u^\alpha(\tau') \, \mathrm{d}\tau' , \tag{2.38}$$

$$\tilde{\pi}^{\mu\nu}(\tau) = - \int_{-\infty}^{\tau} G(\tau, \tau') \eta \, \partial^\mu u^\nu(\tau') \, \mathrm{d}\tau' , \tag{2.39}$$

$$\tilde{q}^\mu(\tau) = - \int_{-\infty}^{\tau} G(\tau, \tau') \kappa \, \partial^\mu \alpha(\tau') \, \mathrm{d}\tau' , \tag{2.40}$$

in which $\tau = \tau(\boldsymbol{r}, t)$ is the local proper time. The above expressions can be expressed in terms of differential equations as:

$$\Pi = -\zeta \, \partial_\alpha u^\alpha - \tau_\mathrm{R} \frac{\mathrm{d}\Pi}{\mathrm{d}\tau} , \tag{2.41}$$

$$\tilde{\pi}^{\mu\nu} = \eta \, \partial^\mu u^\nu - \tau_\mathrm{R} \frac{\mathrm{d}\tilde{\pi}^{\mu\nu}}{\mathrm{d}\tau} , \tag{2.42}$$

$$\tilde{q}^\mu = -\kappa \, \partial^\mu \alpha - \tau_\mathrm{R} \frac{\mathrm{d}\tilde{q}^\mu}{\mathrm{d}\tau} , \tag{2.43}$$

for $\mathrm{d}/\mathrm{d}\tau = u^\mu \, \partial_\mu$.

Both phenomenological approaches described in this section can be implemented using either the Eulerian or the Lagrangian approach.[63] In the Eulerian methods, the hydrodynamical fields are described within a fixed grid which can lead to much heavier computations and numerical instabilities.[64,65] The Lagrangian approach, on the other hand, defines unreal particles that encodes the local properties of the fluid. This approach is called the Smoothed-Particle Hydrodynamics,[66–72] and next section will review its formulation.

## 2.2 Formulation of the smoothed particle hydrodynamics

Initially, let us define the hyperbolic coordinates $x^\mu = (\tau, \boldsymbol{r}, \eta)$, in which:

$$\tau = \sqrt{t^2 - z^2} , \tag{2.44}$$

$$\eta = \frac{1}{2} \ln \left( \frac{t + z}{t - z} \right) . \tag{2.45}$$

In this coordinate system, that is going to be used from now on, the metric is defined as $g_{\mu\nu} = (1, -1, -1, -\tau^2)$. The longitudinal motion is assumed to be uniform, so that the velocity of all particles in the medium is given by $u_\mu = \left( \sqrt{1 + u_x^2 + u_y^2}, u_x, u_y, 0 \right)$, therefore, the computations follow a boost-invariant longitudinal expansion prescription.[53,73] This assumption is valid for the mid-rapidity regime, in which $|\eta| \lesssim 1$.

In the SPH method a conserved reference density current $J^\mu = \sigma u^\mu$ is defined using $\sigma$ as the local density of a fluid element in its rest frame. The density will then obey the relation $\partial_\mu(\tau \sigma u^\mu) = 0$, in hyperbolic coordinates. The general idea behind the method is then to parametrize the extensive densities, associated with some conserved charge, using:[63]

$$\tau \gamma \sigma \rightarrow \sigma^*(\boldsymbol{r}, \tau) = \sum_{\alpha=1}^{N_{\text{SPH}}} \nu_\alpha \, W[\boldsymbol{r} - \boldsymbol{r}_\alpha(\tau); h] \,, \tag{2.46}$$

in which $\nu_\alpha$ is a constant, $\sigma^*(\boldsymbol{r}, \tau)$ is the parametrized charge density in the laboratory frame, and $\boldsymbol{r}_\alpha$ is the Lagrangian coordinate associated with an "SPH particle", which encodes the local properties of the medium. The positive $W[\boldsymbol{r} - \boldsymbol{r}_\alpha(\tau); h]$ is called the *kernel* function is normalized:

$$\int W[\boldsymbol{r}; h] \, \mathrm{d}^2\boldsymbol{r} = 1 \,. \tag{2.47}$$

The parameter $h$ must vanish for $|\boldsymbol{r}| \gg h$ and is defined as the kernel width, which roughly describes the extension of the particle $\alpha$, however, one must note that the total density at the position $\boldsymbol{r}$ is given by the summation of all the SPH particles. The kernel function must also obey the following limit:

$$\lim_{h \to 0} W(\boldsymbol{r}; h) = \delta(\boldsymbol{r}) \,, \tag{2.48}$$

for the Dirac's delta $\delta(\boldsymbol{r})$, which results in a point-like SPH particle for $h \to 0$.

The current associated with the density $\sigma^*(\boldsymbol{r}, \tau)$ is given by:[63]

$$\boldsymbol{j}^*(\boldsymbol{r}, \tau) = \sum_{\alpha=1}^{N_{\text{SPH}}} \nu_\alpha \frac{\mathrm{d}\boldsymbol{r}_\alpha(\tau)}{\mathrm{d}\tau} W[\boldsymbol{r} - \boldsymbol{r}_\alpha(\tau); h] \,, \tag{2.49}$$

and the continuity equation is automatically satisfied.

Let us consider the relation between the fixed frame density $\sigma^*(\boldsymbol{r}_\alpha(\tau))$ and the proper frame density of the SPH particle $\sigma(\boldsymbol{r}_\alpha(\tau))$:

$$\sigma^*(\boldsymbol{r}_\alpha(\tau)) = \gamma_\alpha \sigma(\boldsymbol{r}_\alpha(\tau)) \,, \tag{2.50}$$

then the "specific volume" associated with the particle $\alpha$ can be defined as:[72]

$$V_\alpha = \frac{\nu_\alpha}{\sigma(\boldsymbol{r}_\alpha(\tau))} = \frac{\gamma_\alpha \nu_\alpha}{\sigma^*(\boldsymbol{r}_\alpha(\tau))} \,, \tag{2.51}$$

for the Lorentz factor of the particle $\gamma_\alpha = 1/\sqrt{1 - v_\alpha^2}$. Then it becomes easy to obtain any other extensive quantity $a^*(\boldsymbol{r}, \tau)$ carried by the particle $\alpha$. It suffices to obtain the ratio between the new density $a(\boldsymbol{r}(\tau))$ and the reference density $\sigma(\boldsymbol{r}_\alpha(\tau))$:[63,72]

$$a^*(\boldsymbol{r}, \tau) = \sum_{\alpha=1}^{N_{\text{SPH}}} a(\boldsymbol{r}_\alpha(\tau)) \frac{\nu_\alpha}{\sigma(\boldsymbol{r}_\alpha(\tau))} W[\boldsymbol{r} - \boldsymbol{r}_\alpha(\tau); h] \,. \tag{2.52}$$

As an example, let us consider the case for ideal hydrodynamics.[72] The Lagrangian formulation in the case of the Minkowski metric is given by the action:

$$I = \int \mathcal{L}\, dt = -\int \epsilon\, d^4 x\,,\tag{2.53}$$

in which $\epsilon$ is the proper energy density of the fluid, and can be regarded as the Lagrangian density in the fixed frame, which can then be written in the local frame using equation 2.50. The SPH parametrization of this density follows from equation 2.52 and can be written as:[72]

$$I_{\text{SPH}} = -\int \sum_\alpha \frac{\epsilon_\alpha}{\gamma_\alpha} \frac{\nu_\alpha}{\sigma_\alpha}\, dt = -\int \sum_\alpha \left(\frac{E}{\gamma}\right)_\alpha dt\,,\tag{2.54}$$

in which $E_\alpha = (\nu\epsilon/\sigma)_\alpha$ is the rest energy of the particle $\alpha$.

The variational procedure $\delta I_{\text{SPH}} = 0$ can then be applied, which leads to the equations of motion for each of the SPH particles to be numerically implemented.

As another example, the simplest second-order formulation of the fluid dynamical equations that can be causal and stable, obtained from the Memory Function method, can be written as:[63]

$$\tau_\Pi (D\Pi + \Pi\theta) + \Pi + \zeta\theta = 0\,,\tag{2.55}$$

in which $D = u^\mu \partial_\mu$ is the comoving covariant derivative, $\theta = \tau^{-1} \partial_\mu(\tau u^\mu)$ is the fluid expansion rate, and $\tau_\Pi$ is the relaxation time. The equations of motion for this case can be expressed as:[63,72]

$$\gamma \frac{d}{d\tau}\left(\frac{\epsilon + p + \Pi}{\sigma} u^\mu\right) = \frac{1}{\sigma} \partial^\mu(p + \Pi)\,,\tag{2.56}$$

$$\gamma \frac{d}{d\tau}\left(\frac{s}{\sigma}\right) + \left(\frac{\Pi}{\sigma}\right)\frac{\theta}{T} = 0\,,\tag{2.57}$$

$$\tau_\Pi \gamma \frac{d}{d\tau}\left(\frac{\Pi}{\sigma}\right) + \frac{\Pi}{\sigma} + \left(\frac{\xi}{\sigma}\right)\theta = 0\,.\tag{2.58}$$

Within the SPH scheme, the dynamical variables of the above equations can then be defined:[63,72]

$$\left\{\boldsymbol{r}_\alpha, \boldsymbol{u}_\alpha, \left(\frac{s}{\sigma}\right)_\alpha, \left(\frac{\Pi}{\sigma}\right)_\alpha; \alpha = 1, \ldots, N_{\text{SPH}}\right\}\,,\tag{2.59}$$

in which each variable is associated with an SPH variable $\alpha$. Finally, by employing the density definitions one can find the equations of motion for each SPH particle:

$$\sigma^* \frac{d}{d\tau}\left[\frac{(\epsilon + p + \Pi)_\alpha}{\sigma_\alpha} u_{i\alpha}\right] = \tau \sum_{\beta=1}^{N_{\text{SPH}}} \nu_\beta \sigma_\alpha^* \left[\frac{p_\beta + \Pi_\beta}{(\sigma_\beta^*)^2} + \frac{p_\alpha + \Pi_\alpha}{(\sigma_\alpha^*)^2}\right]$$
$$\times\, \partial_i W[\boldsymbol{r}_\alpha - \boldsymbol{r}_\beta(\tau); h]\,,\tag{2.60}$$

$$\gamma_\alpha \frac{d}{d\tau}\left(\frac{s}{\sigma}\right)_\alpha + \left(\frac{\Pi}{\sigma}\right)_\alpha \left(\frac{\theta}{T}\right)_\alpha = 0\,,\tag{2.61}$$

$$\tau_{\Pi_\alpha} \gamma_\alpha \frac{d}{d\tau}\left(\frac{\Pi}{\sigma}\right)_\alpha + \left(\frac{\Pi}{\sigma}\right)_\alpha + \left(\frac{\xi}{\sigma}\right)_\alpha \theta_\alpha = 0\,,\tag{2.62}$$

in which the fluid expansion rate $\theta_\alpha$ is defined as:

$$\theta_\alpha = (D_\mu u^\mu)_\alpha = \frac{\mathrm{d}\gamma_\alpha}{\mathrm{d}\tau} + \frac{\gamma_\alpha}{\tau} - \frac{\gamma_\alpha}{\sigma_\alpha^*} \frac{\mathrm{d}\sigma_\alpha^*}{\mathrm{d}\tau} \,. \tag{2.63}$$

The SPH method described above can be used to numerically implement the equations of motion by discretizing the medium into local volumes, leading to coupled differential equations for generalized Lagrangian variables. Once the evolution of the system is computed, one must address the decoupling of the medium and the applicability of hydrodynamics end. In order to do that, a common solution is to implement the Cooper-Frye prescription for the medium freeze-out.

## 2.3 Cooper-Frye freeze-out

During the hydrodynamical evolution of the system, inelastic collisions will eventually become too rare and the hadronic abundances will remain fixed. This is called the chemical freeze-out. However, the medium will continue to expand and cool due to elastic collisions among the hadrons. Due to the expansion, the mean free-path of the particles that constitute the medium will increase until, eventually, the medium cannot be regarded as a fluid any longer. The local thermal equilibrium will not be maintained and the hydrodynamic description will not be applicable. This is called the kinetic freeze-out. The Cooper-Frye prescription[74] is commonly used to describe this process.

In the Cooper-Frye prescription, a freeze-out hypersurface $\Sigma$ is defined by some thermodynamical constraint. Hydrodynamic implementations usually adopt the temperature $T_{\mathrm{FO}}$ as a parameter that is obtained from the equation of state at a given instant. The choice of this temperature is constrained to some experimental data fitting. In the simulations, once a particle has crossed the hypersurface $\Sigma$, it is considered to have decoupled from the medium and becomes a free particle.

In order to obtain the final invariant particle distributions, one have to integrate over the hypersurface and evaluate the flux of particles that crosses it. Let us then define the particle density rate that crosses the hypersurface at a given point $x$:

$$\mathrm{d}n = f(x, p)\,\mathrm{d}^3\boldsymbol{p}\,, \tag{2.64}$$

in which $p$ is the particle four-momentum and $f(x, p)$ is the one-body distribution function. In ideal hydrodynamics it can be written as:

$$f(x, p) = \frac{g}{(2\pi)^3} \frac{1}{\exp\left[\beta(u_\mu p^\mu - \mu)\right] \pm 1}\,, \tag{2.65}$$

in which $g$ is the degeneracy factor of the particle, $\mu$ is the chemical potential and $\beta = 1/T$. The positive and negative signs are associated with fermions and bosons respectively. One can now define the current $j^\mu$ associated with the density $n$ as:

$$j^\mu = \int \frac{p^\mu}{E}\,\mathrm{d}n = \int f(x, p)\frac{p^\mu}{E}\,\mathrm{d}^3\boldsymbol{p}\,. \tag{2.66}$$

By integrating the above equation over the hypersurface the total number of particles is obtained:

$$N = \int_{\Sigma} j^{\mu} \, \mathrm{d}\sigma_{\mu} \,, \qquad (2.67)$$

and the invariant distribution is obtained as:

$$E \frac{\mathrm{d}N}{\mathrm{d}^3 \boldsymbol{p}} = \frac{g}{(2\pi)^3} \int_{\Sigma} \frac{p^{\mu} \, \mathrm{d}\sigma_{\mu}}{\exp\left[\beta(u_{\mu}p^{\mu} - \mu)\right] \pm 1} \,, \qquad (2.68)$$

which is the Cooper-Frye equation.

In viscous hydrodynamics though, the equation 2.65 is not valid and must be generalized. The slightly off-equilibrium distribution function can be approximated by:[58]

$$f'(x, p) = f(x, p)\left[1 + rf(x, p)\right]\left[1 + \phi(x, p)\right] \,, \qquad (2.69)$$

for $r = 1$ for Fermi gas, $-1$ for Bose gas, and 0 for Boltzmann gas. The distribution $\phi(x, p)$ is the deviation from equilibrium distribution which can be approximated for shear viscosity as:

$$\phi(x, p) = \frac{1}{2(\epsilon + p)T^2} \pi_{\mu\nu} p^{\mu} p^{\nu} \,, \qquad (2.70)$$

and for bulk viscosity as:

$$\phi(x, p) = \left[D_0 p_{\mu} u^{\mu} + B_0 p_{\mu} p_{\nu} \Delta^{\mu\nu} + \bar{B}_0 p_{\mu} p_{\nu} u^{\mu} u^{\nu}\right] \Pi \,, \qquad (2.71)$$

for $D_0$, $B_0$, and $\bar{B}_0$ space-time dependent parameters.

In the case of shear viscosity one can note the quadratically dependence on $p$, leading to higher corrections at larger momentum. On the other hand, for bulk viscosity the correction factor can be very large and applicability of hydrodynamics is contested. In order to deal with this problem, hybrid models using hydrodynamics coupled with transport models can be employed.

## 2.4 Equation of state

One of the fundamental parameters of hydrodynamic models is the equation of state (EOS), which is a thermodynamic relation between the energy density, pressure, and number density of the fluid. The equation of state can incorporate explicitly the phase transitions in order to study its effects on the QGP and the interaction of the particles with it.[58] In order to calculate the EOS, lattice QCD is commonly used,[75–77] however, due to the sensitivity of the EOS to high momentum modes and the effects of finite lattice spacing, this is usually not an easy task. One common approach is to combine lattice calculations for the deconfined phase with hadron resonance gas (HRG) models to describe the confined phase.[75]

In lattice QCD the equation of state can be parametrized as:[75]

$$\frac{\epsilon - 3P}{T^4} = \frac{d_2}{T^2} + \frac{d_4}{T^4} + \frac{c_1}{T^{n_1}} + \frac{c_2}{T^{n_2}} \,, \qquad (2.72)$$

in which $d_i$, $c_i$, and $n_i$ are parameters, $T$ is the temperature, $P$ is the pressure and $\epsilon$ is the energy density. The equations of state parametrized in this way are labeled as s95p-v1, s95n-v1, and s90f-v1 in which the label states the fraction of ideal entropy at $T = 800$ MeV and the treatment given to the peak of the trace anomaly.

In this work, the EOS labeled s95n-v1 is used, with parameters: $d_2 = 0.2654$ GeV, $d_4 = 6.563 \times 10^{-3}$ GeV$^4$, $c_1 = -4.370 \times 10^{-5}$ GeV$^8$, and $c_2 = 5.774 \times 10^{-6}$ GeV$^9$, with $n_1 = 8$ and $n_2 = 9$.

## 2.5 Initial conditions

In order to solve the differential equations from the hydrodynamic phenomenology, one needs to input some initial state, or boundary conditions. Thus, it is necessary to model the initial energy, baryonic and velocity densities. It is usual, in relativistic heavy ion collisions, to parametrize the initial transverse distributions separated from the longitudinal distributions following the longitudinal boost-invariance of the collision in the mid-rapidity regime.[53] Another important parameter for the initial condition is its initial time $\tau_0$, at which is the thermalization is assumed. If $\tau_0$ is small enough the initial transverse velocity is assumed to be zero,[52] due to the transverse isotropy of the produced particles. This is an important consideration when analyzing transverse distributions; the transverse collective flow seen in data must then be generated by the medium during its expansion.

A common approach to the initial conditions parametrization is to give the energy density in the transverse plane as proportional to the number of collisions that produce wounded nucleons and the number of binary collisions. Wounded nucleons are defined as nucleons that are inelastically excited with nucleon-nucleon collision cross section $\sigma_{NN}$ at least once. Its distribution is obtained via Glauber distributions:[78–81]

$$\frac{d^2 N_{\text{WN}}}{ds^2} = T_A(\boldsymbol{b}_A) \cdot \left(1 - e^{-T_B(\boldsymbol{b}_B)\sigma}\right) + T_B(\boldsymbol{b}_B) \cdot \left(1 - e^{-T_A(\boldsymbol{b}_A)\sigma}\right), \tag{2.73}$$

in which $b$ is the impact parameter, $\boldsymbol{s} = (x, y)$, $\boldsymbol{b}_A = \boldsymbol{s} + b\hat{x}$, $\boldsymbol{b}_B = \boldsymbol{s} - b\hat{x}$, $\sigma$ is the total nucleon-nucleon cross section and $T_X$ is the nuclear thickness function of the nucleus $X$, given by:

$$T_X(x) = \int \rho_X(z, s)\, dz, \tag{2.74}$$

for the nuclear density $\rho_X$. This density is usually parametrized as a Woods-Saxon distribution. The distribution of binary collisions is given by:

$$\frac{d^2 N_{\text{BC}}}{ds^2} = \sigma\, T_A(\boldsymbol{b}_A)\, T_B(\boldsymbol{b}_B). \tag{2.75}$$

One can then write the energy density as:

$$\epsilon(\boldsymbol{s}) = \epsilon_0 \left[ w \frac{d^2 N_{\text{BC}}}{ds^2} + (1 - w) \frac{d^2 N_{\text{WN}}}{ds^2} \right], \tag{2.76}$$

for $\epsilon_0$ the maximum energy density in a central collision, and $w$ being parameters that must be fitted experimentally. The baryon number is considered to be proportional to the energy density: $n(s) \propto \epsilon(s)$.

In order to obtain fluctuating distributions for the initial conditions using the Glauber model, Monte Carlo simulations[46,82,83] are used to generate the wounded nucleon distribution $d^2 N_{WN}/ds^2$ and the binary collisions distribution $d^2 N_{BC}/ds^2$.

Another approach for the initial conditions is based on the color glass condensate (CGC) formalism.[85–91] The CGC is a form of matter that is thought to be created prior to the QGP formation in which the gluons have been saturated due to the high energy of the nucleons. A nucleon is made of three valence quarks bound by gluons. In order to probe the nucleon by means of a given reaction, a characteristic time scale is automatically defined by this reaction which limits its resolution power. Therefore, the probe can only be sensitive to fluctuations with lifetime longer than that of the characteristic time scale. In case of low energy nucleons, only the valence quarks and few fluctuations are really visible. However, the high energy nucleons in relativistic heavy ion collisions are subject to time dilation leading to an increase of the fluctuations lifetime. Because of that, in the CGC framework, more fluctuations are now visible by the probe, which leads to an increased number of gluons. Furthermore, the probability of interaction precisely during the time interval probed in the reaction is diminished, so that the partons can be considered free. The increased gluon density at higher energies can be measured, and it is shown in

$$x := \frac{p_z}{\sqrt{s}} \,, \tag{2.77}$$

in which $p_z$ is the nucleon momentum and $\sqrt{s}$ is the collision energy. This "small-$x$" regime with increased gluon density cannot be understood, at first sight, in terms of PQCD, even if $\alpha_s$ is weak. Furthermore, it would require knowledge about probability of multi-gluon states. However, as it turns out, one can use as scaling the saturation momentum $Q_s$, which is much larger than $\Lambda_{QCD}$, in order to re-enable weak coupling methods. This new scale measures the strength of gluon recombination processes. The CGC formalism is, thus, an effective theory that describe processes in the saturation regime, namely any process with momenta smaller than $Q_s$. It relies on the JIMWLK equation which describes the evolution of the color glass condensate:[85,92–98]

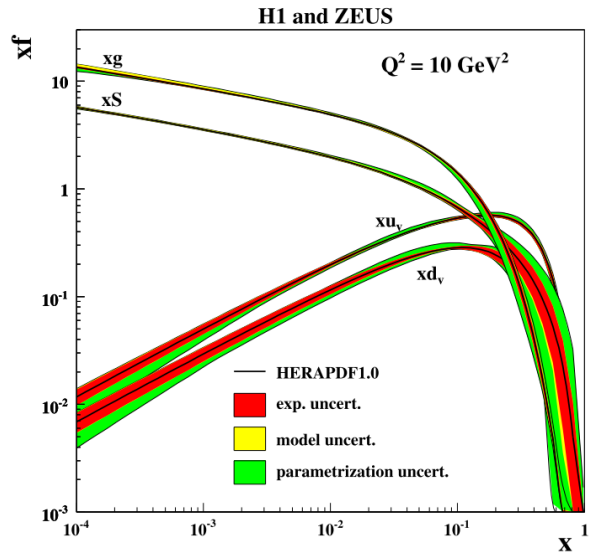

FIGURE 2.1 – Parton distribution in a proton measured in Deep Inelastic Scattering at HERA.[84]

figure 2.1. The quantity $x$ in the abscissas is defined as:

$$\frac{\partial W_\Lambda}{\partial \log \Lambda} = \mathcal{H} W_\Lambda \,, \tag{2.78}$$

in which $\mathcal{H}$ is a quadratic operator in functional derivatives with respect to the static color sources $\rho$. $W[\rho]$ is the probability of a given spatial color distribution $\rho$, and depends on a cutoff scale $\Lambda$ which discriminates fast partons (which can be treated as static sources) from slow partons.

In nuclei collisions, each of the nuclei are described by static color sources $\rho_A$ and $\rho_B$ moving in opposite direction along the light-cone. A Monte-Carlo simulation can be performed in order to obtain fluctuations on these distributions and evaluate different observables.

The MCKLN deals with the high density regime of the gluons in terms of quasi-classical gluon fields and describes the differential cross section of gluon production in nuclei collisions as:[89]

$$E \frac{d\sigma}{d^3 p} = \frac{4\pi N_c}{N_c - 1} \frac{1}{p_T^2} \int \alpha_s(k_T) \, \varphi_A\big(x_1, k_T^2\big) \, \varphi_B\big(x_2, (p_T - k_T)^2\big) \, dk_T^2 \,, \tag{2.79}$$

in which $N_c$ is the number of colors, $\varphi_X\big(x, k_T^2\big)$ is the unintegrated gluon distribution which gives the probability of finding a gluon at a given $x$ and transverse momentum $k_T$ inside the nucleus $X$, and:

$$x_{1,2} = \frac{p_T}{\sqrt{s}} e^{\mp \eta} \,, \tag{2.80}$$

for the pseudo-rapidity $\eta$.

The energy density is obtained from the gluon distribution:[58]

$$\varepsilon = \varepsilon_0 \left( \frac{dN_g}{d\mathbf{r}_T \, dy} \right)^{4/3} \,, \tag{2.81}$$

for $y$ the rapidity.

# Heavy quarks in the QGP

The heavy quarks created in the QGP interact with the medium before generating the particles that are observed in the detector. As explained in chapter 1, the heavy quarks have their masses much greater than the QCD scale parameter and are produced at the very beginning of the system's evolution. Therefore, they are important probes for the study of the quark gluon plasma, retaining information of the whole system's development.

One of the main observables that gives information on this interaction is the nuclear modification factor, the $R_{AA}$. The $R_{AA}$ is a measurement of the suppression factor experienced by particles inside the QGP and it is defined as the ratio between the $p_T$ spectra of the particles of interest (in this case, the heavy mesons $B^0$ and $D^0$ or leptons produced from these mesons via semi-leptonic decays) produced in nuclei collisions with respect to these particles produced in proton-proton collisions. This factor can be written as:

$$R_{AA} = \frac{1}{\langle N_{coll} \rangle} \frac{dN_{AA}/dp_T}{dN_{pp}/dp_T},$$ (3.1)

in which $\langle N_{coll} \rangle$ is the number of binary collisions.

Data from experiments shows that heavy mesons, despite their large mass that causes them not to constitute the bulk part of the medium, experience a high amount of suppression, similar to that exhibited by light particles.[101–104] Although surprising, the nuclear modification factor gives just part of the information. Due to the QGP's fluid nature, also collectivity is observed in the final observables, even for central collisions, contrary to what one could think as the azimuthal anisotropy being due only to the impact parameter of the collision. The plot in figure 3.1 shows experimental data from the ALICE experiment[99,100] of heavy flavor electron from $B^0$ and $D^0$ mesons in PbPb collisions at $\sqrt{s_{NN}} = 2.76$ TeV. On the left side of the figure it is clear that heavy quarks are heavily suppressed when traversing the QGP while on the right side the elliptic flow is evident for semi-central collisions. Both of these data are compared with different theoretical predictions for these observables. One can observe that although these predictions seem to describe the overall behavior of the data, a closer look reveals that describing *both* the $R_{AA}$ and the $v_2$ may not be as easy. For instance, the POWLANG-LQCD predictions for the $R_{AA}$ underestimates the suppression, even thought the POWLANG seem to describe well the result for the elliptic flow. This $R_{AA} \otimes v_2$ puzzle has been around for a fairly long time[105–111] now and much of the heavy flavor analysis in the field is aimed at solving it. Crucial elements influence these results, such as the initial conditions fluctuations for the quark gluon plasma, the heavy quarks energy loss mechanisms inside the medium and the interactions among themselves. Furthermore, different evaluations of the same observables may lead to different biases that could undermine the correct description of these data, as is the case when evaluating some observable that is not sensitive to the initial conditions fluctuations, for instance.

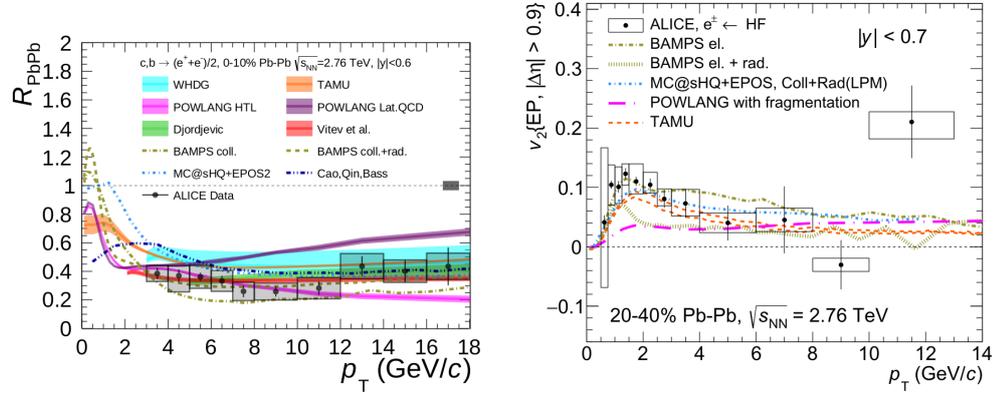

Figure 3.1 – Experimental measurements in PbPb collisions at $\sqrt{s_{NN}} = 2.76$ TeV by the ALICE experiment.[99,100] Left plot shows the nuclear modification factor, defined by equation 3.1 for central collisions, while right plot shows the elliptic flow $v_2$ using event plane method for semi-central collisions. Both results are compared with different theoretical predictions.

In this chapter, the interactions of the heavy quarks inside the QGP is discussed, since it's initial production until the decays to electrons which are detected in the experiments. Also, the flow harmonics analysis is described for different approaches in order to obtain an estimation method that leads to unbiased results.

## 3.1 Heavy quark production

As already exposed in the previous sections, heavy quarks created inside the QGP are not directly accessible and so what is really measured in the experiments are heavy mesons and light particles created from the heavy quarks. Because of that, the study of heavy quarks must account for a series of processes that occur prior to the detection. The final cross section obtained is therefore defined as the convolution of the different stages of the evolution of the system:

$$E\frac{\mathrm{d}^3\sigma}{\mathrm{d}p^3} = E_i\frac{\mathrm{d}^3\sigma_Q}{\mathrm{d}p_i^3} \otimes P(E_i \to E_f) \otimes P(Q \to H_Q) \otimes P(H_Q \to \mathrm{e}^\pm), \qquad (3.2)$$

in which the initial cross section of the heavy quark $Q$ is convoluted with the energy loss, the hadronization, and finally the decay into the semi-leptonic channel. This equation assume the validity of the factorization property of the QCD and allows for a separate study of each aspect of the collision.

Before going into details on all the convoluted functions of this distribution one has to initially discuss how the heavy quarks are created inside the plasma. Since their masses are much greater than the QCD scale $\Lambda_{\mathrm{QCD}}$, their cross section is calculable as a perturbation series in the QCD running coupling $\alpha_s$, evaluated at the mass of the heavy quark $m$. Figures 3.2 and 3.3 show schematically the leading order and some of the next-to-leading order Feynman diagrams of heavy flavor production in nucleus-nucleus collisions. The production processes for heavy flavor include gluon fusion, pair annihilation and further corrections such as pair creation with gluon emission.

Next-to-leading order calculations (NLO)[112] fail at high $p_T$ (i.e. $p_T \gg m$) due to large logarithms of the ratio $p_T/m$ that arise to all orders in the perturbative expansion. Those logarithms for the inclusive transverse momentum distribution can be classified into $\alpha_s^2[\alpha_s\log(p_T/m)]^k$ and $\alpha_s^3[\alpha_s\log(p_T/m)]^k$, respectively leading-logarithmic order (LL) and next-to-leading-logarithmic order (NLL). Two different approaches have been tried in order to deal with this problem[113,114], but led to different problems at different $p_T$ regimes. Later calculations proposed a formalism that merged both approaches in order to obtain all the terms of order $\alpha_s^2$ and $\alpha_s^3$ exactly, including mass effects and also all the logarithmic terms exactly.[115,116]

Schematically, the fixed-order NLO calculation leads to:

$$\frac{\mathrm{d}^2\sigma^{(\mathrm{NLO})}}{\mathrm{d}p_T^2} = A(m)\alpha_s^2 + B(m)\alpha_s^3 + \mathcal{O}\!\left(\alpha_s^4\right), \qquad (3.3)$$

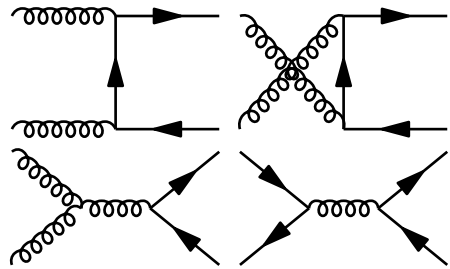

FIGURE 3.2 – Leading order Feynman diagrams of heavy flavor production mechanisms in nuclei collisions.

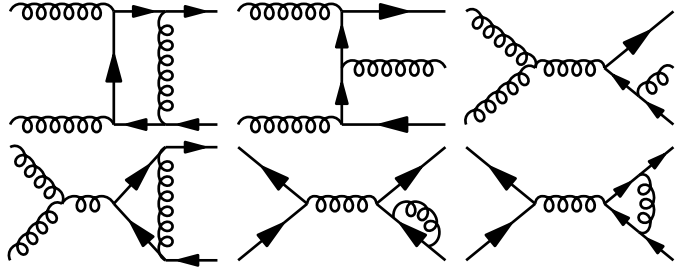

Figure 3.3 – Some of the Next-to-Leading order Feynman diagrams of heavy flavor production mechanisms in nuclei collisions.

while the NLL resummed calculation is given by:

$$\frac{\mathrm{d}^2\sigma^{(\text{NLL})}}{\mathrm{d}p_{\mathrm{T}}^2} = \alpha_s^2 \sum_{i=0}^{\infty} a_i \left(\alpha_s \log \frac{\mu}{m}\right)^i + \alpha_s^3 \sum_{i=0}^{\infty} b_i \left(\alpha_s \log \frac{\mu}{m}\right)^i$$
$$+ \mathcal{O}\left[\alpha_s^4 \left(\alpha_s \log \frac{\mu}{m}\right)^i\right] + \mathcal{O}(\alpha_s^2 \times \text{PST}) , \quad (3.4)$$

in which PST are suppressed terms at high $p_{\mathrm{T}}$ by powers of $m/p_{\mathrm{T}}$ and the dependency on the collision energy, momentum and scale $\mu$ have not been made explicitly.

By combining the results from equations 3.3 and 3.4 and avoiding double counting from both results, the Fixed-Order-Next-to-Leading-Logs (FONLL) approach results in:[115,116]

$$\frac{\mathrm{d}^2\sigma}{\mathrm{d}p_{\mathrm{T}}^2} = A(m)\alpha_s^2 + B(m)\alpha_s^2$$
$$+ \left[\alpha_s^2 \sum_{i=2}^{\infty} a_i \left(\alpha_s \log \frac{\mu}{m}\right)^i + \alpha_s^3 \sum_{i=1}^{\infty} b_i \left(\alpha_s \log \frac{\mu}{m}\right)^i\right] \times G(m, p_{\mathrm{T}})$$
$$+ \mathcal{O}\left[\alpha_s^4 \left(\alpha_s \log \frac{\mu}{m}\right)^i\right] + \mathcal{O}(\alpha_s^4 \times \text{PST}) , \quad (3.5)$$

in which $G(m, p_{\mathrm{T}})$ is an arbitrary function that must approach unity in the limit $m/p_{\mathrm{T}} \to 0$, accounting for the structure of the power-suppressed terms in NLL, and is given by:

$$G(m, p_{\mathrm{T}}) = \frac{p_{\mathrm{T}}^2}{p_{\mathrm{T}}^2 + c^2 m^2} . \quad (3.6)$$

In order to evaluate the cross-section, the fixed-order FO part of the computation needs to have its renormalization scheme changed to the same scheme used by the NLL calculations. Doing that leads to an exact match between the two of them in the massless limit, up to order $\alpha_s^3$. These terms can be written as:

$$\frac{\mathrm{d}^2\sigma^{(\text{FOM0})}}{\mathrm{d}p_{\mathrm{T}}^2} = a_0\alpha_s^2 + \left(a_1 \log \frac{\mu}{m} + b_0\right)\alpha_s^3 + \mathcal{O}(\alpha_s^2 \times \text{PST}) , \quad (3.7)$$

and fonll can finally be written schematically as:

$$\text{FONLL} = \text{FO} + (\text{NLL} - \text{FOM0}) \times G(m, p_T). \tag{3.8}$$

The massless limit fom0 in the above equation becomes meaningless when $p_T$ is the same order as $m$, therefore, the $G(m, p_T)$ function has to be chosen as to suppress the correction $(\text{NLL} - \text{FOM0})$ in that case. It has been shown[115] that the correction becomes relevant when the mass is about one fifth of the transversal mass $m_T = \sqrt{p_T^2 + m^2}$, so that a good choice is to make $c = 5$.

## 3.2 Energy loss

There are two main mechanisms for the energy loss of heavy quarks in the quark gluon plasma. The collisional mechanism is related to quarks losing energy due to (quasi-)elastic scattering, while the radiative mechanism is due to gluon radiation.[108,110,117,118] It has been shown that including both mechanisms in phenomenological studies increase the agreement with data for the intermediate-$p_T$ regime.[119] As a matter of fact, it is observed that elastic scattering is dominant at low-$p_T$ and the radiative scattering, on the other hand, is more important at high-$p_T$.[108] This can be observed in the plots of figure 3.4 that compares, for bottom and charm quarks, the collisional and radiative energy loss mechanisms.

The first elastic scattering calculations are as old as 1982 and various simplifications have been performed such as neglecting the mass dependence in the density and only considering small momentum transfers. Also, calculations were performed for a static medium disregarding the cooling and expansion of the system.[120–122] The multiple scattering of heavy quarks in the qgp can be described as Brownian motion[110,123] which will lead to the Fokker-Planck equation:[123]

$$\frac{\partial D}{\partial t} = \frac{\partial}{\partial p_i}\left[\mathcal{T}_{1i}^{FP}D\right] + \mathcal{T}_2^{FP}\left(\frac{\partial}{\partial p}\right)^2 D, \tag{3.9}$$

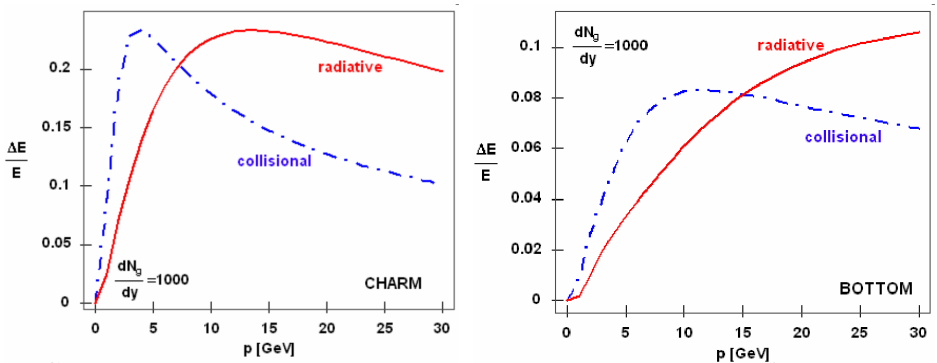

FIGURE 3.4 – Comparison between elastic scattering and gluon radiation induced energy loss for heavy quarks charm (left) and bottom (right).[108]

in which $D$ is the distribution due to the particles motion, $p$ is the initial momentum of the particle and the transport coefficients are given by:

$$\mathcal{T}_{1i}^{FP} = p_i \mathcal{A} \,, \tag{3.10}$$

$$\mathcal{T}_2^{FP} = \mathcal{B}_0 \,. \tag{3.11}$$

for the drag $\mathcal{A}$ and the transverse diffusion $\mathcal{B}_0$.

The Fokker-Planck equation can be also be obtained from a discretization of the Langevin equation.[124] The Langevin dynamics is given by the equation:

$$\frac{\mathrm{d}p_i}{\mathrm{d}t} = \xi_i(t) - \eta_D p_i \,, \tag{3.12}$$

with $\eta_D$ the momentum drag coefficient and the stochastic term $\xi_i(t)$ that delivers random momentum "kicks" uncorrelated in time, given by:

$$\langle \xi_i(t)\xi_j(t') \rangle = \kappa \delta_{ij}\delta(t - t') \,, \tag{3.13}$$

in which $3\kappa$ is the mean-squared momentum transfer per unit time.[124]

For a thermal heavy quark of mass $m \gg T$ and typical momentum $p \approx \sqrt{mT}$, the stochastic solution of equation 3.12 leads to:

$$\langle p^2 \rangle = 3mT = \frac{3\kappa}{2\eta_D} \Rightarrow \eta_D = \frac{\kappa}{2mT} \,, \tag{3.14}$$

and the diffusion constant in space is obtained as:[124]

$$D = \frac{T}{m\eta_D} = \frac{2T^2}{\kappa} \,, \tag{3.15}$$

subject to the determination of $\kappa$. For leading-order calculations it can be shown that the momentum loss per time is very similar to the one obtained with a constant momentum drag coefficient leading to:

$$\frac{\mathrm{d}p}{\mathrm{d}t} \propto p \,. \tag{3.16}$$

Langevin approaches for the computation of collisional energy loss of the heavy quarks have successfully described experimental data in the low transverse momentum regime. In this case, the gluon emission is diminished and thus, radiative energy loss becomes unimportant. Otherwise, it is generally assumed that radiative energy loss will dominate the evolution of the heavy quarks at higher momentum. This is contested by the model adopted by Moore and Teaney[124] who argue that even then, the dominant energy-loss mechanism should be elastic scattering. There are many other different models based on the Langevin approach for the collisional energy loss.[59,125–131]

One can add a new term to the Langevin equation in order to include radiative energy loss of the heavy quarks. The new Langevin equation can then be written as:[110]

$$\frac{\mathrm{d}p_i}{\mathrm{d}t} = \xi_i(t) - \eta_D p_i + (f_g)_i \,, \tag{3.17}$$

in which $f_g = -\mathrm{d}p_g/\mathrm{d}t$ is the recoil force experienced by the heavy quarks from the medium-induced gluon radiation due to gluon momentum $p_g$. This term must be obtained from a gluon distribution function, which leads to the probability of gluon radiation from a heavy quark. One possible model is given by the higher-twist calculations:[110,132–134]

$$\frac{\mathrm{d}N_g}{\mathrm{d}x\,\mathrm{d}k_\perp^2\,\mathrm{d}t} = \frac{2\alpha_s P(x)\hat{q}}{\pi k_\perp^4}\sin^2\left(\frac{t-t_i}{2\tau_f}\right)\left(\frac{k_\perp^2}{k_\perp^2 + x^2 m^2}\right)^4 \,, \tag{3.18}$$

for the gluon splitting function $P(x)$, the gluon formation time $\tau_f$, the heavy quark $m$, the fractional energy taken by the emitted gluon from the heavy quark $x$, and the gluon transverse momentum $k_\perp$.

The last term in the above equation is the reason why the radiative energy loss is suppressed at low momentum and it is known as the "dead cone effect".[135,136] The ratio between the gluon radiation from heavy quarks with respect to that of the light quarks is given by:[136]

$$\frac{\mathrm{d}P_{HQ}}{\mathrm{d}P_0} = \left(1 + \frac{\theta_0^2}{\theta^2}\right)^{-2} = \left(1 + \theta_0^2 x\sqrt{\frac{x}{\hat{q}}}\right)^{-2} \,, \tag{3.19}$$

for $\theta_0 := m/E$ and the characteristic gluon radiation angle $\theta \approx k_\perp/x \sim (\hat{q}x^3)^{1/4}$. The equation 3.19 corresponds to a suppression for the spectrum at small angles $\theta$.

The gluon transport coefficient $\hat{q}$ in equations equations 3.18 and 3.19 can be obtained from the quark diffusion coefficient[110] as:

$$\hat{q} = \frac{2\kappa C_A}{C_F} \,. \tag{3.20}$$

In this case, the introduction of the radiative term in the Langevin equation did not change the number of free parameters.

A different approach for the calculation of heavy quark energy loss inside the QGP uses AdS/CFT correspondence.[126,137–139] One can describe the late-time behavior of an external quark moving with speed $v$ in the $x^1$ direction in an static gauge by following an *ansatz*:[140]

$$x^1(t,r) = vt + \xi(r) + o(t) \,, \tag{3.21}$$

in which $o(t)$ includes other motions of the quark which is neglected. The Lagrangian can then be written as:

$$\mathcal{L} = -\sqrt{1 - \frac{v^2}{h} + \frac{h}{H}\xi'^2} \,, \tag{3.22}$$

and the general coordinate is defined $\pi_\xi := \partial \mathcal{L} / \partial \xi'$. This solves to $\pi_\xi$ being constant and one can write:

$$\xi' = \pi_\xi \frac{H}{h} \sqrt{\frac{h - v^2}{h - \pi_\xi^2 H}}, \qquad (3.23)$$

for $H = 1 + L^4/r^4$ and $h = 1 - r_H^4/r^4$.

The above equation describes a string that trails the external quark. One must now integrate the equation in order to obtain $\xi(r)$. The integration leads to:

$$\xi = -\frac{L^2}{2 r_H} v \left( \arctan \frac{r}{r_H} + \log \sqrt{\frac{r + r_H}{r - r_H}} \right). \qquad (3.24)$$

Plugging this into equation 3.21 and integrating the flow momentum one obtains:

$$\frac{dp_1}{dt} = \sqrt{-g} \frac{-1}{2\pi\alpha'} G_{x^1 v} \, \partial^r X^v = -\frac{r_H^2 / L^2}{2\pi\alpha'} \frac{v}{\sqrt{1 - v^2}}. \qquad (3.25)$$

Finally, by using $L^4 = g_{YM}^2 N \alpha'^2$ and $T = r_H / (\pi L^2)$ the momentum loss can be written as:

$$\frac{dp_1}{dt} = -\frac{\pi \sqrt{g_{YM}^2 N}}{2} T^2 \frac{p_1}{m}, \qquad (3.26)$$

for $g_{YM}^2 = 4\pi g_{string}$. This equation is one of the energy loss models that inspires the parametrizations used in this work.

In this work, the main focus has been pushed away from the exact energy loss mechanism that governs the evolution of the heavy quarks. Instead, a general parametrization of the energy loss models is employed in which the underlying quantities obtained from various different models are represented by few parameters in the simulations. This way the analysis could focus on the main aspects of different models without a strong commitment with a single choice.

## 3.3 Hadronization and decay

Once the heavy quarks have been produced in the quark gluon plasma and subsequently interacted with it during its hydrodynamic expansion, the bulk medium undergoes a phase transition from the QGP to hadronic matter. The heavy quarks are not in full equilibrium with the medium so that in order to evaluate the heavy quarks spectra one must resort to microscopic hadronization mechanisms.[59] Two different mechanisms are in play during heavy quark hadronization: independent fragmentation of partons and coalescence of quarks.

It is expected that coalescence are dominating in the low-$p_T$ regime due to the abundance of partons in the phase-space.[141–145] In this scenario, three quarks or a pair of quark and anti-quark that happens in a highly dense phase space can recombine into a baryon or a meson. The wave function of the hadron determines the amplitude of this process.

The fragmentation picture, on the other hand, is the dominant mechanism at high-$p_T$. In this scenario a single parton have a given probability, the fragmentation function, of hadronize into a hadron which carries a fraction $z$ of the momentum of the parent parton.

In this work, only the fragmentation of the heavy quarks have been implemented and coalescence has been neglected, as the main interest is to study the effects of the hydrodynamical evolution on the high-$p_T$ heavy quarks. In order to obtain the fragmentation functions different perturbative calculations in QCD have been employed,[146–149] and non-perturbative effects have been treated as corrections to the models. The Lund fragmentation model[150–154] takes a different approach in which the production of particles will obey a saturation hypothesis and takes semi-classical considerations to obtain the hadronization in terms of a stochastical process. In this scenario an original pair of quark and anti-quark is assumed to be created at a single space-time point and start to go apart, leading to a stretching of the string field which eventually breaks up into new pairs.

Another common approach is to parametrize the fragmentation function disregarding the dynamics of the processes. These models can work as translations from the partonic stage to the hadronic stage and although the physical meaning of the parameters is doubtful, these kind of models do a good job in order to perform experiment and planning analysis. Some of the most common models within this approach that are used in relativistic heavy ion collisions include the fragmentation functions due to Kartvelishvili[155] for mesons $\bar{q}c$, given by:

$$f_C^c(z) = \frac{\Gamma(2 + \gamma - \alpha_c - \alpha_q)}{\Gamma(1 - \alpha_c)\Gamma(1 + \gamma - \alpha_q)} z^{-\alpha_c}(1 - z)^{\gamma - \alpha_q}, \tag{3.27}$$

with $\gamma = 3/2$, $\alpha_c \in [-2, -4]$ and $\alpha_q = 1/2$. Also, the Peterson fragmentation function[156] obtained from $e^-e^+$ annihilation:

$$D(z) \propto \frac{1}{z\left(1 - \frac{1}{z} - \frac{\varepsilon}{1-z}\right)^2}, \tag{3.28}$$

in which $\varepsilon \approx m_q^2/m_H^2$ is the ratio between the quark mass and the hadron mass. Finally, based on the Peterson function, the Collins fragmentation function[157] for heavy quarks can be derived as:

$$D(z) \propto \left[\frac{1-z}{z} + \frac{(2-z)\varepsilon}{1-z}\right] \frac{1 + z^2}{\left(1 - \frac{1}{z} - \frac{\varepsilon}{1-z}\right)^2}. \tag{3.29}$$

In the above equations, $z$ is defined as the ratio between the hadron and the originating quark energy and longitudinal moment:

$$z = \frac{E^{\text{meson}} + p_\ell^{\text{meson}}}{E^{\text{quark}} + p_\ell^{\text{quark}}}, \tag{3.30}$$

however, this variable is not experimentally accessible and thus should be converted to:[158]

$$z = x := \frac{\sqrt{x_E^2 - x_{\min}^2}}{\sqrt{1 - x_{\min}^2}}, \tag{3.31}$$

in which $x_{\min} = \frac{2m_H}{\sqrt{s}}$, with $\sqrt{s}$ being the center of mass energy, and:[159]

$$x_E := \frac{2E_H}{\sqrt{s}}. \tag{3.32}$$

With this conversion, it is possible to compare the fragmentation functions of equations 3.27 to 3.29 in figure 3.5.

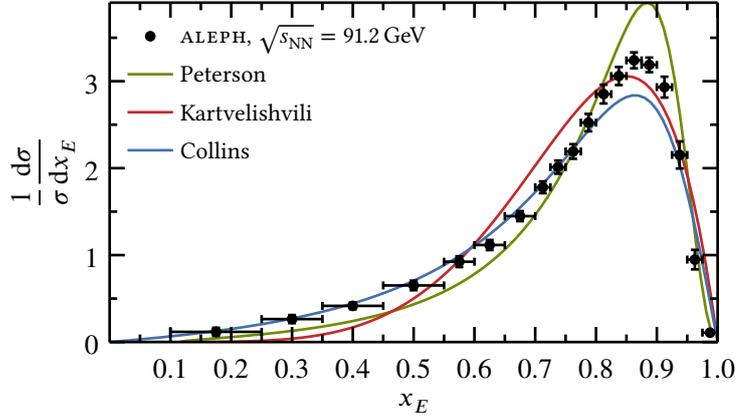

FIGURE 3.5 – Fragmentation functions from equations 3.27 to 3.29 compared with experimental data from ALEPH collaboration.[159]

In this work the Peterson fragmentation function is parametrized and fit to FONLL predictions for heavy mesons $B^0$ and $D^0$ in order to include the energy loss parametrization between the heavy quark production and hadronization. The heavy mesons decay into the semi-leptonic channels ($X \to e^{\pm}$) and the electron spectra are the final result from the system's evolution that is analyzed.

Before going into details on how to implement a computer simulation for the evolution of heavy quarks in the quark gluon plasma, let us review the azimuthal anisotropy analysis procedures, which consists of one of the most important tools to the study of the QGP.

## 3.4 Azimuthal anisotropy

One of the most important observables in relativistic heavy ion collisions is the azimuthal anisotropy of particles. The initial anisotropic spatial distribution of particles that form the QGP, coupled with the nearly perfect liquid behavior of the plasma, creates pressure gradients that transfer this anisotropy to the momentum space in the final state. As a matter of fact, this observable is the main evidence of this behavior of the plasma.[160–162] It is also closely related to initial state fluctuations in such a way that, to clearly describe the azimuthal anisotropy of the collision, one has to take into account that events fluctuate. These fluctuations generate *all* Fourier flow harmonics in each event, and disregarding this fact by evaluating average events leads to the higher order harmonics being washed out in the final analysis.

It has already been implied that the azimuthal distribution of particles are usually expanded into Fourier series:

$$E\frac{\mathrm{d}^3 N}{\mathrm{d}p^3} = \frac{1}{2\pi}\frac{\mathrm{d}^2 N}{p_\mathrm{T}\,\mathrm{d}p_\mathrm{T}\,\mathrm{d}y}\left(1 + \sum_{n=1}^{\infty} 2v_n \cos[n(\varphi - \Phi_\mathrm{R})]\right), \qquad (3.33)$$

in which $\varphi$ is the azimuthal angle and $\Phi_\mathrm{R}$ is defined as the reaction plane angle, related to the direction of the impact parameter. The schemes in figure 3.6 show the contribution of each order of the flow harmonic. For instance, the second harmonic is usually called the elliptic flow, while the third is the triangular flow.

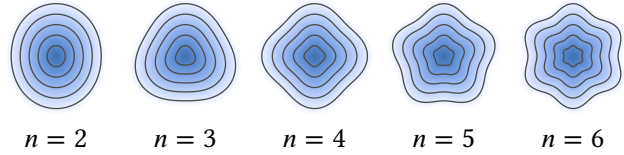

$$n = 2 \qquad n = 3 \qquad n = 4 \qquad n = 5 \qquad n = 6$$

FIGURE 3.6 – Illustration of the different contributions of each of the first Fourier harmonics to the expansion of equation 3.33.

Azimuthal analysis consists in evaluating the values of the $v_n$ harmonics. At first sight, the initial geometry in the collisions would lead to all the symmetry planes to coincide with the reaction plane defined by $\Phi_\mathrm{R}$, however, fluctuations in the initial spatial distribution of particles easily breaks this assumption. The scheme in figure 3.7, although exaggerated, shows that depending on the participating nucleons distribution, the angle $\Phi_n$ can have a different direction that the expected reaction plane. These are called the participant planes, and are differently associated for each of the Fourier $n$-harmonics.

Typically, one can try to estimate the participant plane $\Phi_n$. One of the methods that aims at accomplish that is the event-plane method. The event-plane angle $\psi_n$ is defined as the angle of symmetry of the particle azimuthal distribution in the transverse plane, explicitly, it can be written as:[163–167]

$$\psi_n = \frac{1}{n}\arctan\frac{Q_{n,y}}{Q_{n,x}}, \qquad (3.34)$$

in which the flow vector in the harmonic $n$ is defined as:

$$\boldsymbol{Q}_n = \begin{pmatrix} |\boldsymbol{Q}_n|\cos(n\psi_n) \\ |\boldsymbol{Q}_n|\sin(n\psi_n) \end{pmatrix} = \frac{1}{N}\sum_{j=1}^{N}\begin{pmatrix} \cos(n\varphi_j) \\ \sin(n\varphi_j) \end{pmatrix}, \qquad (3.35)$$

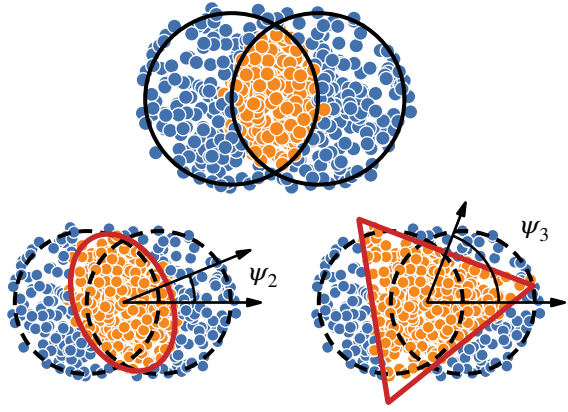

Figure 3.7 – Scheme of the distribution of nucleons in a nuclei collision. The scheme exaggerates the fluctuations of the nucleon distributions which can lead to participant planes for all the Fourier harmonics to differ from the impact parameter direction.

for the sum running through the $N$ particles with azimuthal angles $\varphi_j$ each. Note that this is essentially a one-particle probability distribution analysis. Because of that, the distribution may differ from the participant plane angle, being highly dependent on the number of particles used to generate the distribution. Indeed, the obtained value for the harmonic $v_n$ using event plane method must be corrected by the event-plane resolution,[163,168] leading to:

$$v_n = \frac{\langle \cos[n(\varphi - \psi_n)] \rangle}{\langle \cos[n(\psi_n - \Phi_R)] \rangle} \,, \tag{3.36}$$

in which the denominator represents the resolution correction obtained, for instance, via sub-event method.[163] This correction is not trivial and can heavily affect the estimation of the harmonics, for instance, the elliptic flow using event plane method, and resolution correction using sub-event method, depends on flow fluctuations as:

$$v_2\{\text{EP}\} = \langle v_2^\alpha \rangle^{1/\alpha} \,, \tag{3.37}$$

in which $\alpha \approx 1$ for high resolution while it can change to $\alpha \approx 2$ if the resolution is low.[49]

Similarly to event plane method, the scalar product method weights the event averages using $|\boldsymbol{Q}_n|$:[49,169]

$$v_n = \frac{\langle |\boldsymbol{Q}_n| \cos[n(\varphi - \Psi_n)] \rangle}{\langle \cos[n(\psi_n - \Phi_R)] \rangle} \,, \tag{3.38}$$

where the denominator represents the resolution correction. This procedure leads to an unambiguous estimation of the flow harmonics:

$$v_n\{\text{SP}\} = \sqrt{\langle v_n^2 \rangle} \,. \tag{3.39}$$

Another method which is equivalent to the scalar product, and thus gives unambiguous estimation of the second moment of the distribution, is the 2-particle correlation.[170,171] Furthermore, by increasing the number of correlated particles, one can obtain unambiguous measurements of the even moments of the distribution $\langle v_n^{2k} \rangle$. Let us redefine the flow vector from equation 3.35 as a complex number:

$$Q_n \rightarrow Q_n = e^{in\varphi}, \tag{3.40}$$

in which $\varphi$ is the azimuthal angle of a given particle. The azimuthal correlation between two particles is defined as:[172]

$$\langle Q_{n,1} Q_{n,2}^{\dagger} \rangle = \langle\!\langle e^{in(\varphi_1 - \varphi_2)} \rangle\!\rangle = \langle v_n^2 \rangle, \tag{3.41}$$

in which the first average is taken over all particle combinations in a single event while the outer average is over all the considered events. For a higher number of correlated particles the exponential will contain the azimuthal angle for each of the particles. In order to evaluate the multi-particle correlations numerically and experimentally, one needs to combine all the possibilities among the particles of interest, which can lead to an incredible amount of iterations.

In order to avoid this problem one can calculate multi-particle cumulants in terms of moments of the $Q$-vectors. There exists a couple of methods to this goal. The direct cumulants (or $Q$-cumulants) method[160,173,174] leads to non-biased cumulants due to interference between harmonics. In the $Q$-cumulants one define the average $m$-particle azimuthal correlations for a single event:

$$\langle 2 \rangle := \left\langle e^{in(\varphi_1 - \varphi_2)} \right\rangle := \frac{1}{P_{M,2}} \sum_{i,j}' e^{in(\varphi_i - \varphi_j)}, \tag{3.42}$$

$$\langle 4 \rangle := \left\langle e^{in(\varphi_1 + \varphi_2 - \varphi_3 - \varphi_4)} \right\rangle := \frac{1}{P_{M,4}} \sum_{i,j,k,l}' e^{in(\varphi_i + \varphi_j - \varphi_k - \varphi_l)}, \tag{3.43}$$

and similarly

$$\langle 6 \rangle := \left\langle e^{in(\varphi_1 + \varphi_2 + \varphi_3 - \varphi_4 - \varphi_5 - \varphi_6)} \right\rangle, \tag{3.44}$$

$$\langle 8 \rangle := \left\langle e^{in(\varphi_1 + \cdots + \varphi_4 - \varphi_5 - \cdots - \varphi_8)} \right\rangle, \tag{3.45}$$

in which $P_{n,m} = n!/(n-m)!$, and the primed sum symbol indicates that the indexes must all be taken different.

The $m$-particle correlator is then defined as the average over all the events of the azimuthal correlations:

$$\langle\!\langle m \rangle\!\rangle := \frac{\sum_{\text{events}} W_{\langle m \rangle} \langle m \rangle}{\sum_{\text{events}} W_{\langle m \rangle}}, \tag{3.46}$$

in which the weights $W_{\langle m \rangle}$ are used to minimize the bias due to multiplicity variations in the set of events used to estimate the $m$-particle correlations. Finally, the unbiased estimators of the reference $m$-particle cumulants are written as:

$$c_n\{2\} = \langle\langle 2 \rangle\rangle \, , \tag{3.47}$$

$$c_n\{4\} = \langle\langle 4 \rangle\rangle - 2\langle\langle 2 \rangle\rangle^2 \, , \tag{3.48}$$

$$c_n\{6\} = \langle\langle 6 \rangle\rangle - 9\langle\langle 4 \rangle\rangle\langle\langle 2 \rangle\rangle + 12\langle\langle 2 \rangle\rangle^3 \, , \tag{3.49}$$

$$c_n\{8\} = \langle\langle 8 \rangle\rangle - 16\langle\langle 6 \rangle\rangle\langle\langle 2 \rangle\rangle - 18\langle\langle 4 \rangle\rangle^2 + 144\langle\langle 4 \rangle\rangle\langle\langle 2 \rangle\rangle^2 - 144\langle\langle 2 \rangle\rangle^4 \, . \tag{3.50}$$

One can define the differential cumulants by changing one of the correlating particles with the particle of interest:[175]

$$d_n\{2\} = \langle\langle 2' \rangle\rangle \, , \tag{3.51}$$

$$d_n\{4\} = \langle\langle 4' \rangle\rangle - 2\langle\langle 2 \rangle\rangle\langle\langle 2' \rangle\rangle \, , \tag{3.52}$$

$$d_n\{6\} = \langle\langle 6' \rangle\rangle - 6\langle\langle 4' \rangle\rangle\langle\langle 2 \rangle\rangle - 3\langle\langle 4 \rangle\rangle\langle\langle 2' \rangle\rangle + 12\langle\langle 2 \rangle\rangle^2\langle\langle 2' \rangle\rangle \, , \tag{3.53}$$

$$\begin{aligned} d_n\{8\} = {} & \langle\langle 8' \rangle\rangle - 12\langle\langle 6' \rangle\rangle\langle\langle 2 \rangle\rangle - 4\langle\langle 6 \rangle\rangle\langle\langle 2' \rangle\rangle - 18\langle\langle 4 \rangle\rangle\langle\langle 4' \rangle\rangle + 72\langle\langle 4' \rangle\rangle\langle\langle 2 \rangle\rangle^2 \\ & + 72\langle\langle 4 \rangle\rangle\langle\langle 2 \rangle\rangle\langle\langle 2' \rangle\rangle - 144\langle\langle 2 \rangle\rangle^3\langle\langle 2' \rangle\rangle \, , \end{aligned} \tag{3.54}$$

in which the primes indicate the substitution. As an example, the correlation between two particles leads to:

$$\langle\langle 2 \rangle\rangle \approx \langle v_n^2 \rangle \Rightarrow \langle\langle 2' \rangle\rangle \approx \langle v_n \bar{v}_n \cos[n(\psi_n - \bar{\psi}_n)] \rangle \, . \tag{3.55}$$

The bar sign in the above equation indicates the evaluation for the particles of interest while the absence of the bar indicates the evaluation for the reference particles.

The differential $v_n$ are finally evaluated as:

$$v_n\{2\}(p_T) = \frac{d_n\{2\}(p_T)}{\left(c_n\{2\}\right)^{1/2}} \, , \tag{3.56}$$

$$v_n\{4\}(p_T) = \frac{-d_n\{4\}(p_T)}{\left(-c_n\{4\}\right)^{3/4}} \, , \tag{3.57}$$

$$v_n\{6\}(p_T) = \frac{d_n\{6\}(p_T)}{\left[4(c_n\{6\})^5\right]^{1/6}} \, , \tag{3.58}$$

$$v_n\{8\}(p_T) = \frac{-d_n\{8\}(p_T)}{\left[33(-c_n\{8\})^7\right]^{1/8}} \, . \tag{3.59}$$

# Heavy quarks evolution in an event-by-event expanding QGP

Given all the theoretical framework presented in the previous chapters, and in order to study the evolution of heavy quarks inside the quark gluon plasma, a simulation framework, from now on referred to as DABMod (D and B mesons Modular framework),[176–180] has been developed using C++ programming language with the aid of ROOT[181,182] and PYTHIA8[183] libraries. It consists of a Monte-Carlo simulation of heavy quarks traversing an expanding hot quark gluon plasma in which they interact by means of an energy loss model. The final spectra obtained from the core of the simulation is then used to evaluate different observables on an event-by-event approach, such as the nuclear modification factor and the azimuthal anisotropy as well as observables derived from these.

The simulation has been built with a modular paradigm as its main feature, taking for granted the QCD factorization. The purpose of this approach is to enable the easy selection of different models for each stage of the system's evolution, such as the energy loss, medium backgrounds, or hadronization processes so one can investigate separately the effects of each one of those stages on the final results.

Event-by-event analysis is implemented in the code in a way that each execution of the program is meant to deal with only one choice of event, completely independent of other possible choices. A High-Performance Computing (HPC) cluster is then used to evolve all the desired events, using straightforward parallelization of the program execution, which leads to intermediate results. These results must undergo a reduction analysis which combines all the events in order to evaluate the final observables. In this work, the SAMPA cluster, from the High Energy Physics Instrumentation Center (HEPIC) at the University of São Paulo, has been used to perform all the computations.

In the simulation, bottom and charm quarks are sampled within the transverse plane at mid-rapidity of the QGP medium at an initial time $\tau_0$ with their initial momentum given by pQCD calculation. The hydrodynamical evolution of the background medium is performed independently of the evolution of the heavy quarks so no effect from the probes on the medium is considered. Each sampled heavy quark travels along the transverse plane with a velocity $v$ and a constant direction $\varphi_{\text{quark}}$ and loses energy by means of some parametrization of $dE/dx$. Once the heavy quarks reach a certain position $(x, y)$ where the temperature of the medium is below a chosen decoupling temperature $T_d$, they undergo a hadronization process leading to heavy mesons $B^0$ and $D^0$, which in turn decay into electrons and positrons. Various quantities are stored during the execution of the program, such as the heavy quarks positions and $p_T$ spectra in order to obtain the desired observables meant for study.

In order to setup the execution, some parameters are fixed using available data from the Particle Data Group[184]. Those parameters are presented in the table 4.1.

Table 4.1 – Constant values used on all executions of the simulation.

| Name | Symbol | Value | |
|------|--------|-------|------|
| Bottom mass | $m_b$ | 4.18 | GeV |
| $B^0$ mass | $m_B$ | 5.27 | GeV |
| Charm mass | $m_c$ | 1.275 | GeV |
| $D^0$ mass | $m_D$ | 1.86 | GeV |
| Electron mass | $m_e$ | 5.1099 | keV |

This chapter presents the details on how each stage of the system's evolution is evaluated.

## 4.1 Initial conditions

The initial conditions' purpose in the simulation is two-folded: they serve as a starting point for the hydrodynamic evolution of the plasma and also for setting up the spatial distribution of the sampled heavy quarks that traverse the medium. They are to be considered the actual input of the simulation so that very different approaches can be used. In this work the results are obtained from the MCKLN[89–91] initial conditions, based on the CGC formalism. In order to perform the hydrodynamical expansion of the system the energy density distributions are smoothed by a smoothing parameter $\lambda = 0.3$ fm which scales the microscopic spatial gradients of the fluctuations in the medium.[185] This parameter serves as a lower bound for scales at which the hydrodynamic model chosen for the simulation provides accurate results.[186]

In order to run the simulation, events for both LHC energies, namely $\sqrt{s_{NN}} = 2.76$ TeV and $\sqrt{s_{NN}} = 5.02$ TeV, have been selected for centralities in the range of 0–50%. The centrality selection was done using the average number of participants for each event $\langle N_{\text{part}} \rangle$ and the events have been binned in 1%-wide centrality classes. A small centrality bin is required in order to perform the event-by-event analysis with the results from the simulation. The plots in figure 4.1 show the events generated from the MCKLN for $\sqrt{s_{NN}} = 2.76$ TeV and $\sqrt{s_{NN}} = 5.02$ TeV, grouped in 10%-wide bins, in function of multiplicity.

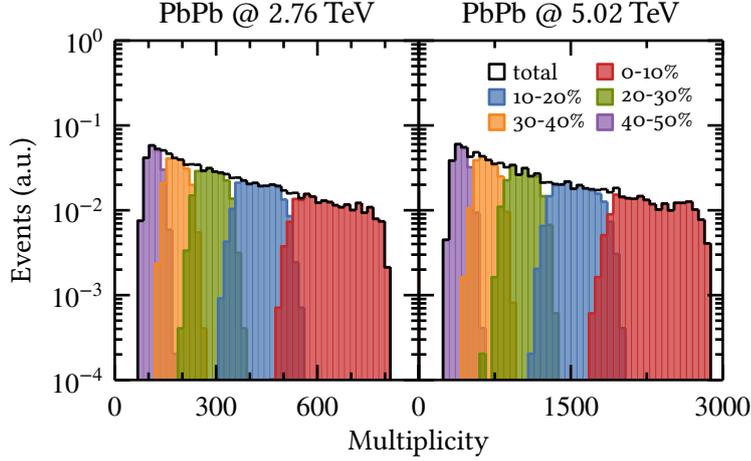

FIGURE 4.1 – Centrality classification of MCKLN events for $\sqrt{s_{NN}} = 2.76$ TeV (left) and $\sqrt{s_{NN}} = 5.02$ TeV collision energies in function of multiplicity. Event count is represented in arbitrary units.

Once the events are sorted, they are selected for the hydrodynamical evolution of the system. An event-by-event analysis is needed in order to correctly evaluate the desired observables, however, too few heavy quarks are actually produced in a single event and in order to obtain reasonable statistics for the results, heavy quarks must be oversampled. This oversample can be justified if one considers all the possibilities for the fluctuations in a given centrality class; clearly there should be very similar events if a high enough number of events is considered. These events can then be classified by geometrical similarity and events in the same class can be treated as the same. Thus, event-by-event in this context should not be regarded as single real one but rather a grouping of possible distributions. The results obtained from the simulation are then probabilities and the oversampling works as a tool in order to evaluate these probabilities. For the results presented in this work $1 \times 10^5$ heavy quarks have been sampled for the $R_{AA}$ evaluations and $1 \times 10^7$ heavy quarks for the $v_n$ evaluations. Bottom and charm quarks are sampled and simulated separately and all the evaluations are performed for both heavy quarks independently.

The plots in the figures 4.2 and 4.3 show examples of initial conditions energy distributions for PbPb collision energies of $\sqrt{s_{NN}} = 2.76$ TeV and $\sqrt{s_{NN}} = 5.02$ TeV, respectively. They present the sample events classified in different centrality classes where it's possible to note that when centrality is increased the total area corresponding to the participant distribution is reduced. This is made clearer by the bottom panel showing the average for all events in a given centrality class. Also notable are the differences in the fluctuations as the high density lumps can vary from very thin

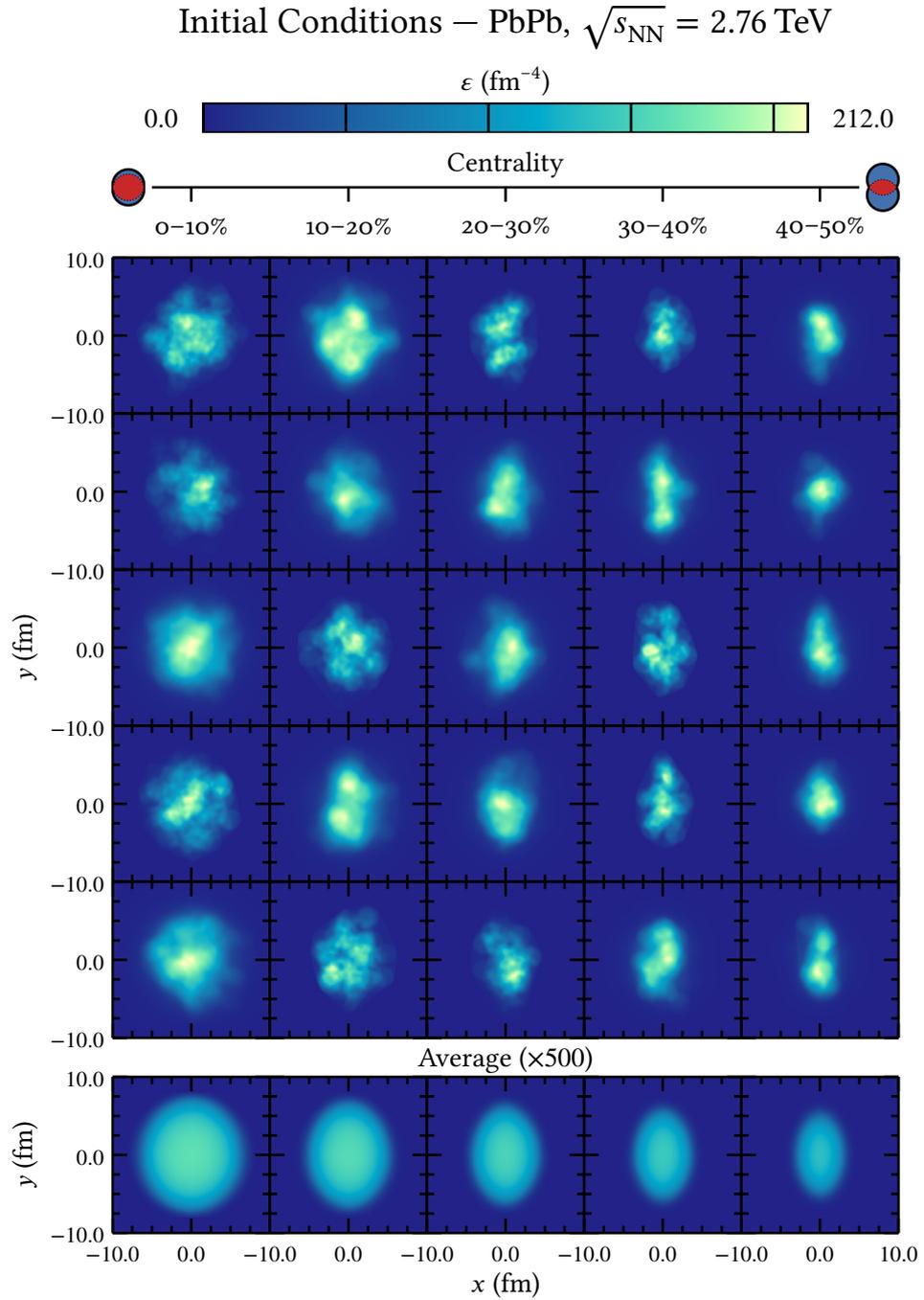

Figure 4.2 – Sample events showing the initial conditions for PbPb collisions at $\sqrt{s_{NN}} = 2.76$ TeV and for different centrality classes (columns) from most central (left) to semi-central collisions (right). Each line is a single example. The bottom panels show the average of all the events used in the simulation for a given centrality class where the energy density scale has been multiplied by a factor of 500 in order to make the colors visible in the same scale as the others.

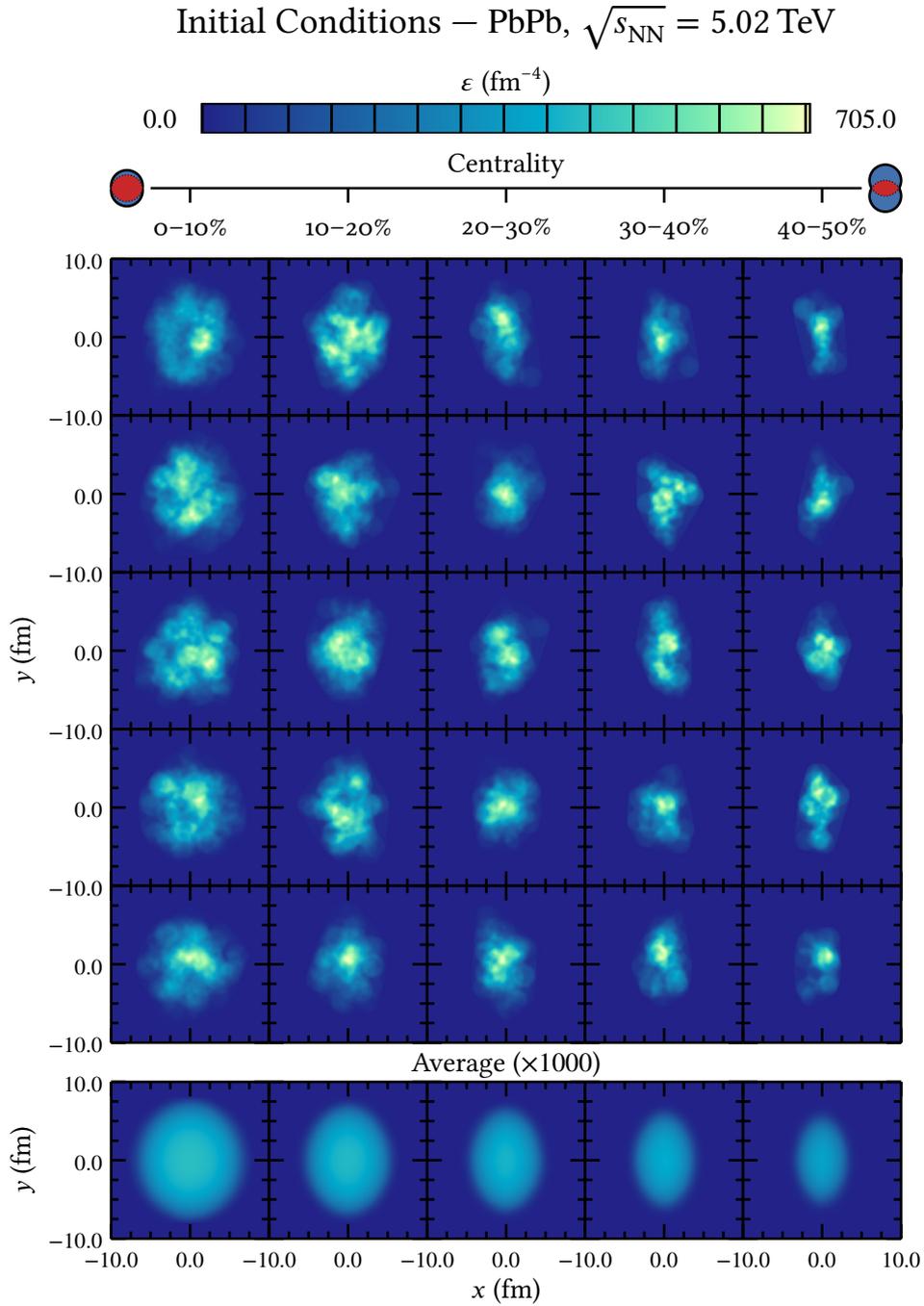

FIGURE 4.3 – Sample events showing the initial conditions for PbPb collisions at $\sqrt{s_{NN}} = 5.02$ TeV and for different centrality classes (columns) from most central (left) to semi-central collisions (right). Each line is a single example. The bottom panels show the average of all the events used in the simulation for a given centrality class where the energy density scale has been multiplied by a factor of 1000 in order to make the colors visible in the same scale as the others.

to more smeared distributions.

Similar simulations for the evolution of heavy quarks in the QGP have selected the initial spatial distribution of heavy quarks using the distribution of binary collisions in a Monte-Carlo Glauber model.[187,188] This quantity is related to the energy density of the distribution and as a first simplification, in this framework, the spatial distribution of heavy quarks is obtained from the energy density given by the MCKLN approach.

Once the heavy quarks are sampled within the medium at the initial stage of the simulation they evolve interacting with a hydrodynamical background. Let us describe the medium evolution prior to go into details of the quark-medium interaction.

## 4.2 Hydrodynamics

Each heavy quark sampled from the initial conditions travels along the transverse plane of an evolving plasma that is obtained from a relativistic hydrodynamic evolution of the initial conditions created by the MCKLN approach. Although heavy quarks may affect the hydrodynamical evolution by medium recoil,[189] this work disregards this effect and the medium acts as background for the heavy quark probes. The background samples are then obtained prior to the heavy quark evolution and are stored so that multiple executions of the program can be made.

In order to evaluate the profiles that will be used in the evolution of the heavy quarks inside the medium, the event-by-event relativistic viscous hydrodynamical model developed at the University of São Paulo (Viscous Ultrarelativistic Smoothed Particle Hydrodynamics — v-USPhydro)[63,109,185,190] is used to evolve the MCKLN events and create the local temperature and medium velocities that are used in the energy loss calculations. This code is implemented using a mesh-free Lagrangian method implementation of the Smoothed Particle Hydrodynamics (SPH)[66−72] to solve the equations of a 2D + 1 relativistic hydrodynamics, which presents an overall fast computational time compared to different approaches using grid-based computations and has been successfully tested against analytical viscous hydrodynamics.[191] It is assumed a boost-invariant longitudinal expansion[53] in which, for central rapidity, the system is invariant under Lorentz boosts. The evolution is setup to start at an initial time $\tau_0 = 0.6$ fm, with shear viscosity given by $\eta/s = 0.11$. The simulation used the lattice-based equation of state (EOS) s95n-v1[75] of equation 2.72 and the freeze-out temperature $T_{FO}$ for the Cooper-Frye[74] was set to $T_{FO} = 120$ MeV. Furthermore, the length scale $h$ for the SPH particles was set to $h = 0.3$ fm following the smoothing parameter defined for the MCKLN initial conditions. No analysis from the perspective of changing these parameters have been done in this work as the main goal is to evaluate the energy loss effects on the final results.

The hydrodynamics simulation provides transverse plane profiles over time for the energy density, the local temperature and the flow velocity in both $\hat{x}$ and $\hat{y}$ directions. Those quantities are used in DABMod in order to evaluate the energy loss experienced by the heavy quarks during the system's evolution. Furthermore, from the Cooper-Frye freeze-out, charged soft particle spectra are obtained which are used to obtain the integrated flow harmonic coefficients. Those are later correlated with the heavy flavor sector obtained from DABMod.

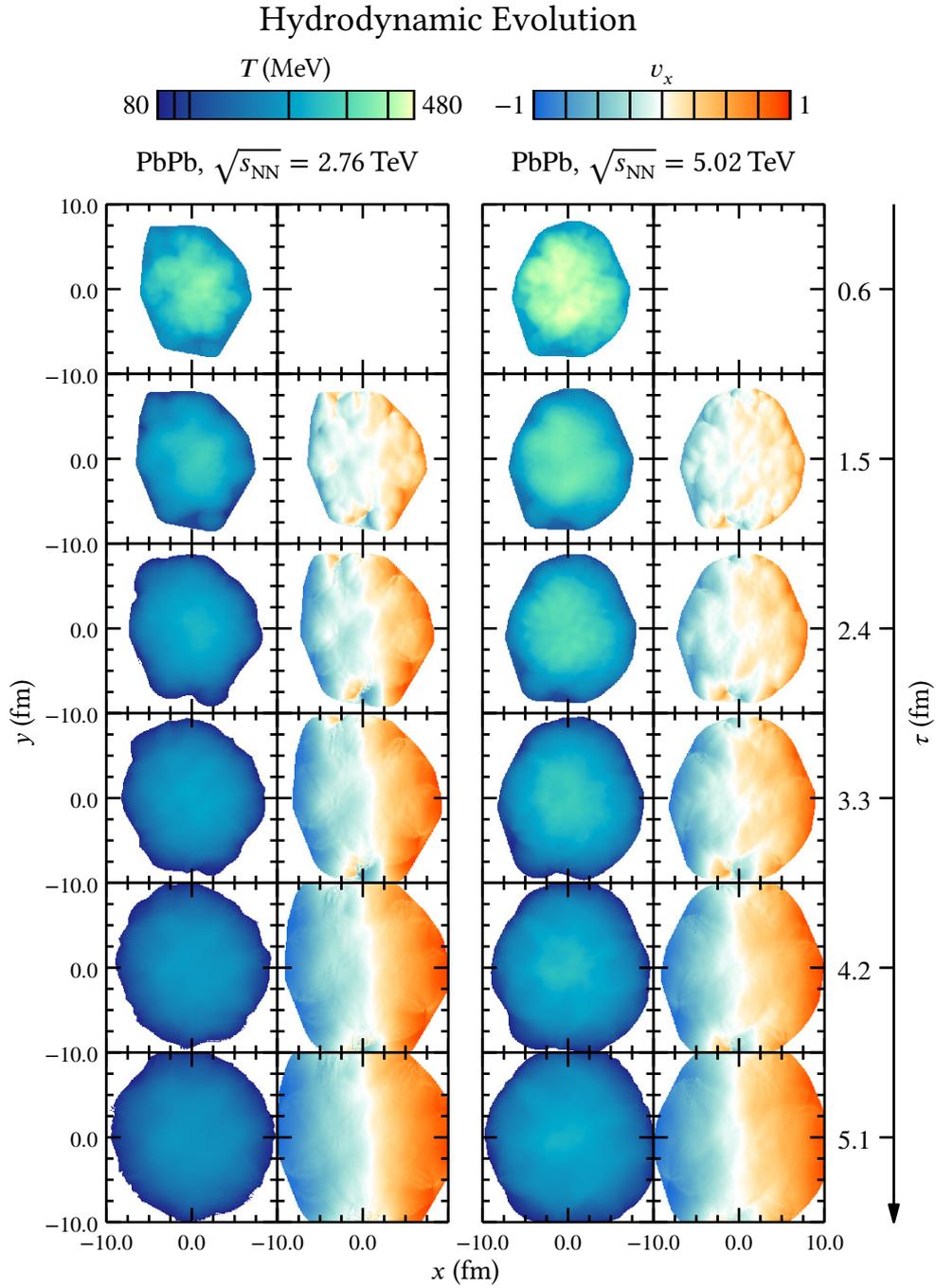

FIGURE 4.4 – Beginning of the hydrodynamical evolution for $\tau \leq 5.1\,\mathrm{fm}$ for PbPb collisions at $\sqrt{s_{\mathrm{NN}}} = 2.76\,\mathrm{TeV}$ (left) and $\sqrt{s_{\mathrm{NN}}} = 5.02\,\mathrm{TeV}$ (right). For each collision example the local temperature $T$ (left) and the velocity in the $\hat{x}$ direction $v_x$ (right) in the transverse azimuthal plane are shown. It is possible to observe the cooling of the medium and the expansion away from the center.

In figure 4.4, two sample events for PbPb collisions at $\sqrt{s_{\mathrm{NN}}} = 2.76\,\mathrm{TeV}$ and $\sqrt{s_{\mathrm{NN}}} = 5.02\,\mathrm{TeV}$ have their temperature and $v_x$ velocity shown for the beginning of the system's evolution. The color scales are the same for the same type of information and a couple of features can be directly observed from the figure. It is clear the expansion of the QGP medium over time which is shown from the temperature plots. This also reflects on the velocity plots which show that although the simulation begins with a static medium, $v_x$ magnitude quickly increases symmetrically around the center. Also, over time the medium tends to become more uniform and the fluctuations become less pronounced.

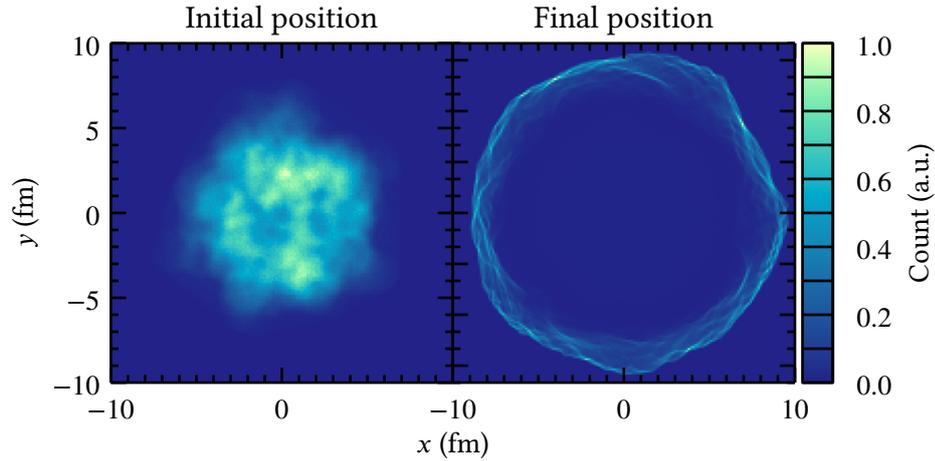

Figure 4.5 – Example of initial distribution of heavy quarks in one event (left) that leads to the final distribution (right) where heavy quarks have mostly been pushed away from the center of the azimuthal plane.

Heavy quarks sampled from the initial conditions will follow the medium's flow during the system's evolution due to the interaction with the QGP. The figure 4.5 shows on the left the initial position of the heavy quarks, prior to the evolution, distributed according to the energy density. After the system evolves, the final position of the same heavy quarks, before hadronization, is far from the center of the azimuthal plane, as it can be seen in the right plot. Also, due to spatial inhomogeneity of the medium, the shape of the distribution reflects these fluctuations. The following sections will describe in details how the interaction of the heavy quarks with the QGP is implemented in the simulation.

## 4.3 Heavy quark production

The initial momentum distribution is selected according to PQCD calculation using First-Order Next-to-Leading-Logs (FONLL).[115,116] The momentum probability in the reference proton-proton collision is obtained from the cross section, proportional to the invariant yield:

$$E \frac{d^3\sigma}{dp^3} \propto E \frac{d^3N}{dp^3} = E \frac{d^3N}{dp_x\,dp_y\,dp_z}\,, \qquad (4.1)$$

in which $E$ is the energy, $p$ the momentum, and $N$ the number of partons. By defining the beam axis in the direction $\hat{x}$ and performing a coordinate change of $dp_x = E\,dy$, for the rapidity $y$ and $dp_y\,dp_z = p_T\,dp_T\,d\varphi$, with $p_T$ the transverse momentum and $\varphi$ the azimuthal angle, one can rewrite the invariant yield as:

$$E \frac{d^3N}{dp^3} = \frac{d^3N}{p_T\,dp_T\,d\varphi\,dy}\,. \qquad (4.2)$$

Furthermore, it is assumed isotropic production in the azimuthal direction so the above equation becomes:

$$E \frac{d^3N}{dp^3} = \frac{1}{2\pi} \frac{d^2N}{p_T\,dp_T\,dy}\,. \qquad (4.3)$$

Finally, the evolution of the heavy quarks in the simulation is performed in the mid-rapidity regime:

$$E \frac{d^3N}{dp^3} = \frac{1}{2\pi} \frac{dN}{p_T\,dp_T}\bigg|_{y=0}\,. \qquad (4.4)$$

In the simulation, heavy quarks are first set to have initial $p_T$ following an uniform distribution between a defined range $[p_T^{min}, p_T^{max}]$, then the distribution histograms are filled using $dN/dp_T$ from equation 4.4 as a weighting factor. This is done in order to obtain reasonable statistics for the high $p_T$ regime, where the probability given by equation 4.4 is very small, while at the same time, providing the correct $p_T$ distribution of the heavy quarks. The quark initial direction $\varphi$ is taken into account and is uniformly random among all possible azimuthal angles.

The idea behind the proton-proton reference yield is to provide means for evaluating the nuclear modification factor. The $R_{AA}$, previously defined in equation 3.1, is a ratio between the differential yields in nucleus-nucleus collisions with respect to proton-proton collisions, scaled by the nuclear overlap function $\langle T_{AA} \rangle = \langle N_{coll} \rangle / \sigma_{inelastic}^{pp}$. Generally speaking, what it means is that if nature were to behave in such a way that the nuclei collisions consisted of a scaling of multiple proton-proton collisions, with no nuclear effects, the expected value for the $R_{AA}$ would equals unity. As it turns out, this is not actually the case. However, by exploiting this definition in the simulation, one can take the overlap function into account by using the FONLL spectra for the heavy quarks sampling using the energy density distribution from the initial conditions. The $R_{AA}$ is then evaluated by considering the ratio of spectra after interacting with the medium (in this case, by means of the energy loss) with respect to the ones without medium interaction. This approach has also been used in other similar computations:[124]

$$R_{AA}(p_T, \varphi) = \frac{\left(\mathrm{d}N(p_T, \varphi)/\mathrm{d}p_T\right)^{\text{interacting}}}{\left(\mathrm{d}N(p_T)/\mathrm{d}p_T\right)^{\text{non-interacting}}} . \tag{4.5}$$

The $R_{AA}$ is evaluated in the simulation for both heavy quarks independently and the FONLL distribution itself is used to compute the total contribution by correctly weighting the $R_{AA}$ for each heavy quark.

In order to include the energy loss in the computations one cannot use the FONLL predictions directly for the heavy mesons and electrons spectra, however, these predictions are still used as references for the choice of parameters of the next steps of the simulation. Let us address the hadronization and decays before going into details on how the energy loss parametrizations are done.

## 4.4 Hadronization and decay

The hadronization of the heavy quarks is assumed to occur when the local temperature of the medium falls below a certain value $T_d$. When this happens, the Peterson fragmentation function[156] is employed and only fragmentation is performed. The low $p_T$ quark coalescence[142,144,192] is not implemented in the simulation, which can lead to under predictions for $p_T \lesssim 10$ GeV.[193,194] This is not, however, much of concern for the high $p_T$ results that are presented in this work.

From the previously defined Peterson fragmentation function:

$$f(z) = \frac{1}{z\left(1 - \frac{1}{z} - \frac{\varepsilon}{1-z}\right)^2} , \tag{4.6}$$

$z$ is defined as the fraction of the original heavy quark that is carried to the originated heavy meson:

$$z = \frac{E^{\text{meson}} + p_\ell^{\text{meson}}}{E^{\text{quark}} + p_\ell^{\text{quark}}} . \tag{4.7}$$

In the above equation, the index $\ell$ indicates the direction of the originating heavy quark. Although the heavy meson may have a slightly different propagation direction, both are assumed to be collinear and the heavy meson $p_T$ can be obtained from manipulating the equations:

$$p_T^{\text{meson}} \approx \frac{z^2\left(E^{\text{quark}} + p_T^{\text{quark}}\right)^2 - m^2}{2z\left(E^{\text{quark}} + p_T^{\text{quark}}\right)} , \tag{4.8}$$

in which $m$ is the heavy meson mass. The simulation draws a random value for $z$ for each heavy quark that undergoes fragmentation in order to perform the Monte Carlo computation of $p_\mathrm{T}^\mathrm{meson}$ yields.

Following what has been discussed in the previous section, in order to evaluate the $R_\mathrm{AA}$ at the meson level, two distinct spectra must be evaluated, one which includes the medium interaction and other that the interaction is not present. The latter is obtained by directly performing the fragmentation on the heavy quarks sampled in the simulation regardless of the local temperature of the medium. In this case, for consistency with the FONLL predictions, the $\varepsilon$ parameter of the Peterson fragmentation function must be fit. This fitting is performed prior to the full execution of the simulation for both bottom and charm mesons and is verified to maintain the same value for different beam energies.

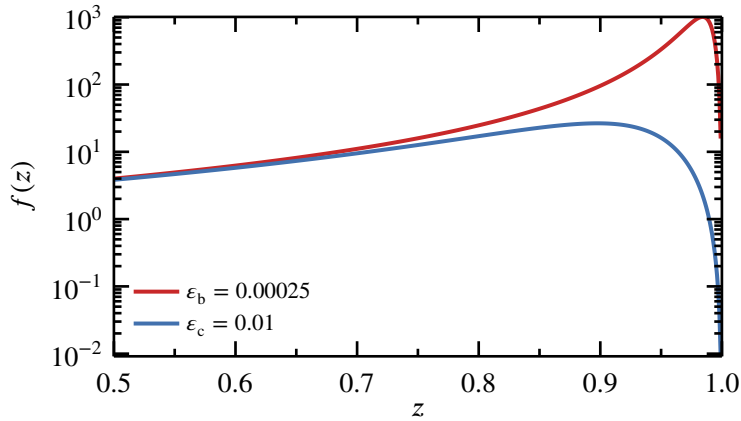

FIGURE 4.6 – Peterson fragmentation function for bottom and charm quarks showing the $\varepsilon$ values that were fitted using FONLL spectra as reference.

The values of $\varepsilon$ fitted for bottom and charm quarks in the simulation are shown in Figure 4.6. The fitting has been performed by finding the parameter that minimized $Q^2$ function, given by:

$$Q^2(p_\mathrm{T}; \varepsilon) = \left[ \frac{1}{\delta\left(\mathrm{d}N(p_\mathrm{T})/\mathrm{d}p_\mathrm{T}\right)} \left( \frac{\mathrm{d}N^\mathrm{sim}}{\mathrm{d}p_\mathrm{T}}(p_\mathrm{T}; \varepsilon) - \frac{\mathrm{d}N}{\mathrm{d}p_\mathrm{T}}(p_\mathrm{T}) \right) \right]^2 , \qquad (4.9)$$

in which $\mathrm{d}N/\mathrm{d}p_\mathrm{T}$ indicates the FONLL prediction for the heavy meson, $\mathrm{d}N^\mathrm{sim}/\mathrm{d}p_\mathrm{T}$ the simulation computing for the same meson and the term $\delta\left(\mathrm{d}N(p_\mathrm{T})/\mathrm{d}p_\mathrm{T}\right)$ is the error associated with the FONLL spectrum.

Once the spectra for the heavy mesons have been obtained for both the interacting and non-interacting heavy quarks, the mesons are made to decay into the semi-leptonic channels using PYTHIA8[183]. In order to do that, an event is setup within PYTHIA8's framework containing a single particle, the heavy meson, with all the cinematic properties obtained from the fragmentation. All decay channels, except for the relevant ones, are turned off so that every single heavy meson will decay into electrons and positrons. From the list of resulting particles of the decays, $e^{\pm}$ are selected to construct the final spectra of the simulation. In order to account for the enforced decay channel, the correct branching ratios from the Particle Data Group[184], shown in table 4.2 are applied.

TABLE 4.2 – Branching ratios for the semi-leptonic decay channels used within the simulation.

| Decay | Ratio |
|---|---|
| $B^0 \to e^{\pm} + \cdots$ | 0.1086 |
| $D^0 \to e^{\pm} + \cdots$ | 0.103 |

The electron spectra from non-interacting heavy quarks are also compared to

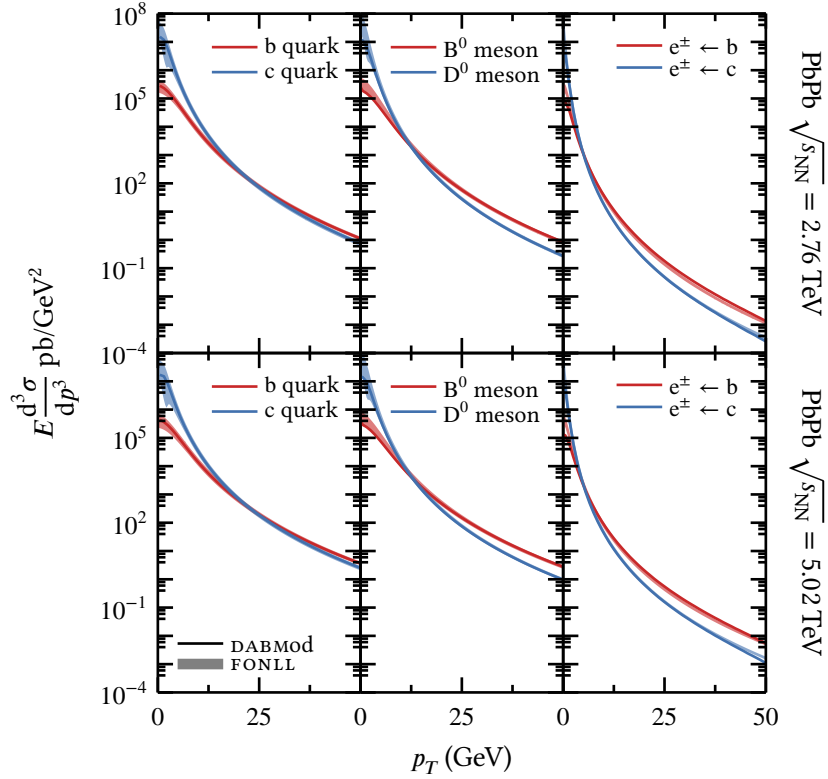

FIGURE 4.7 – Comparison between the simulation spectra obtained from Peterson fragmentation function and PYTHIA8's decays with respect to FONLL predictions for PbPb collisions at $\sqrt{s_{NN}} = 2.76$ TeV (upper panels) and $\sqrt{s_{NN}} = 5.02$ TeV (lower panels). The left panels show the heavy quark spectra, which should match by construction. The middle panels show the spectra after the fragmentation function has been applied. The right panels show the comparison for the electron spectra after applying the correct branching ratio.

FONLL predictions in order to determine that the correct setup has been made in the simulation. The comparison between the results from the simulation and the FONLL cross-sections are shown in figure 4.7 for all the stages of the system's evolution and for both studied beam energies. From the plots it is possible to note that all the calculations fall within the uncertainties bands from the pQCD predictions. From this result it is possible now to include the energy loss parametrizations in order to evaluate the $R_{AA}$ and $v_n$'s for the heavy ion collisions.

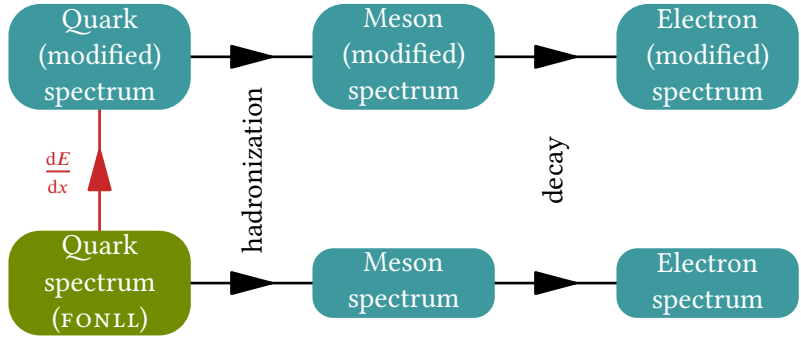

FIGURE 4.8 – Program flow scheme showing the start point with quark spectrum obtained from FONLL and forking into two different branches. The upper branch includes the energy loss and the lower branch does not include any interaction with the medium. In order to evaluate $R_{AA}$ the ratio between the upper and lower branches is computed.

An schematic depiction of the program flow is shown in figure 4.8 showing the two branches that are created within the simulation by allowing one of them to interact with the medium using some energy loss parametrization. The $R_{AA}$ evaluation is performed for three distinct levels, quarks, mesons and electrons, by including the energy loss parametrization in the code. The following section will describe in details how this is implemented.

## 4.5 Energy loss parametrization

The interaction of the heavy quarks inside the medium, once they are sampled from the initial conditions, is implemented by means of an energy loss parametrization. This is done inspired by different works on the study of jets:[195–201]

$$-\frac{dE}{dx} = -\frac{dP}{d\tau}(\boldsymbol{x}_0, \varphi, \tau) = \kappa(T)P^\alpha(\tau)\tau^z T^c \zeta, \tag{4.10}$$

where $\kappa(T)$ and $T[\boldsymbol{x}(\tau), \tau]$ describes the local temperature along the jet path at a time $\tau$, $P^\alpha(\tau)$ describes the dependence on the jet energy and $\zeta$ is the term describing the energy loss fluctuations along the path-length. In this work a simplification of this equation is employed for the heavy quark interactions:

$$-\frac{dE}{dx}(T, v; \lambda) = f(T, v; \lambda)\Gamma_{\text{flow}}\zeta, \tag{4.11}$$

where $T$ is the medium temperature, $v$ the heavy quark velocity and $\lambda$ is some parameter that must be fitted for each model to be studied. By adopting a shape for the $f(T, v; \lambda)$ function, various parameters from equation 4.10 are automatically selected. The factor $\Gamma_{\text{flow}} = \gamma\left[1 - v_{\text{flow}}\cos(\varphi_{\text{quark}} - \varphi_{\text{flow}})\right]$, with $\gamma = \left(1 - v_{\text{flow}}^2\right)^{-1/2}$, takes into account the boost from the local rest frame of the fluid[202] by relating the quark direction in the azimuthal plane $\varphi_{\text{quark}}$ with the underlying flow $\varphi_{\text{flow}}$ in such a way that heavy quarks propagating with the same direction as the flow will lose less energy than heavy quarks in other directions. Furthermore, heavy quarks with the opposite direction will be pushed in the direction of the medium.

In the framework of the simulation, energy loss is implemented via integration of the $\mathrm{d}E/\mathrm{d}x$ equation by choosing an arbitrarily small displacement interval $\Delta x$ where the variation of energy is evaluated:

$$\Delta E = \left.\frac{\mathrm{d}E}{\mathrm{d}x}\right|_E \times \Delta x\,. \tag{4.12}$$

The average velocity of the heavy quark within this displacement is evaluated in order to obtain the time interval $\Delta t$. As it happens, if the velocity is too small, the time interval will be bigger than the resolution of the hydrodynamic evolution. In this case, a new $\Delta x$ is chosen from the resolution $\Delta t_0$. The new position and four-momentum of the heavy quark is then evaluated and the steps are repeated until the medium temperature is below $T_{\text{d}}$ so that the fragmentation is set to occur.

In this work, different energy models have been selected for study. The first one, inspired by conformal AdS/CFT calculations that evaluate the drag force on an external quark moving in a thermal plasma of $\mathcal{N} = 4$ super-Yang-Mills theory. This leads to to a dependency in the momentum of the heavy quark and the squared temperature of the medium:[140]

$$f(v, T; \lambda) = \lambda v T^2\,. \tag{4.13}$$

The second choice is made based on an study made on the temperature dependence of the heavy quark drag coefficient. This work pointed out that a non-decreasing dependence near the phase transition is favoured in order to obtain a simultaneous description of both $R_{\text{AA}}$ and $v_2$:[105]

$$f(v, T; \lambda) = \lambda\,. \tag{4.14}$$

Inspired by the two energy loss models presented in equations 4.13 and 4.14, it is worth looking individually at the $T$ and $v$ dependence, so two more models are tested in the simulation:

$$f(v, T; \lambda) = \lambda T^2\,, \tag{4.15}$$

for the $T^2$ dependence and another for the $v$ dependence as:

$$f(v, T; \lambda) = \lambda v\,. \tag{4.16}$$

Finally, the drag force evaluated from lattice QCD is used to determine the last energy loss model that is explored:[203]

$$f(v, T; \lambda) = \lambda T^2 \frac{F_{\text{drag}}}{\sqrt{\lambda}} .$$ (4.17)

It is worth noting that the modular approach adopted by the program allows for easily changing between these models. Also, using the same techniques, more robust energy loss models can be directly input in the simulation for further exploring the effects of the energy loss in the commonly studied observables.

In all the models presented, the free parameter $\lambda$ must be chosen. This is done by fitting the $R_{\text{AA}}$ results that are obtained from the simulation with available experimental data from ALICE and CMS experiments for each heavy quark. Data for PbPb collisions at $\sqrt{s_{\text{NN}}} = 2.76$ TeV is used for the $D^0$ meson $R_{\text{AA}}$ and sets the parameter $\lambda$ for the charm quark simulations. In order to obtain results for bottom quark, electron $R_{\text{AA}}$ data must be used instead as no data for $B^0$ mesons is available. In this case, by using the already defined $\lambda_{\text{c}}$ for charm quark, the same parameter for the bottom quark is varied so the total contribution to the electron $R_{\text{AA}}$ matches the data.

For PbPb collisions at $\sqrt{s_{\text{NN}}} = 5.02$ TeV there is recent data for $B^0$ mesons, however, the results are calculated for the whole centrality range of 0–100%. This work presents a comparison of $B^0$ meson data and the results obtained from the simulation at a different centrality range but future improvement on the data is required in order to make valuable predictions from this result.

In order to find the correct value for the $\lambda$ parameter, the differential $R_{\text{AA}}$ is evaluated for a finite number of choices of the parameter and then the fit with the data is performed by a least squares algorithm in which bi-linear interpolation is performed in order to obtain the values outside the $\lambda \times p_{\text{T}}$ grid. If different simulation parameters, such as the $T_{\text{d}}$, lead to no change in the evaluated value for $\lambda$, the grid is refined in order to make it possible to separate them.

Once $\lambda_{\text{c}}$ and $\lambda_{\text{b}}$ are fitted for the simulation, energy loss fluctuations may be included. In this study, the fluctuations $\zeta$ in the equation 4.11 is implemented as a random constant for each heavy quark traversing the QGP. Different probability distributions for $\zeta$ can be implemented and in this work three implementations are used. The first one is the Gaussian distribution:

$$f_{\text{gauss}}(\zeta) = \frac{1}{\sigma\sqrt{2\pi}} \exp\left[-\frac{1}{2}\left(\frac{\zeta-1}{\sigma}\right)^2\right],$$ (4.18)

with standard deviation values of $\sigma_1 = 0.1$ and $\sigma_2 = 0.3$. Also, inspired by the work with jet quenching[195], uniform and linear probabilities function have been used, defined by:

$$f_{\text{uniform}}(\zeta) = \frac{1}{2}, \qquad\qquad \text{for } 0 \leq \zeta \leq 2,$$ (4.19)

$$f_{\text{linear}}(\zeta) = \frac{2}{3} - \frac{2}{9}\zeta, \qquad\qquad \text{for } 0 \leq \zeta \leq 3.$$ (4.20)

As $\zeta$ is a multiplicative factor, when $\zeta = 1$ the energy loss experienced by the heavy quark is not altered, while $\zeta < 1$ implies a decrease in the experienced energy loss and $\zeta > 1$ increases its value. The linear and uniform distributions provide very extreme cases while the small values for the standard deviation for the Gaussian distribution, provides much smaller fluctuations. From these choices it is possible to study both the magnitude and the shape of the fluctuations and their effects on the observables.

The energy loss is setup in order to evaluate the $R_{AA}$ for all the events in all centrality classes, which in turn, is used to perform event-by-event analysis and leads to $v_n$ estimatives for each event. The next section describes the final analysis that is performed from the data obtained from the simulation.

## 4.6 Event-by-event analysis

The spectra obtained from the simulation in each hydro event is used to perform event-by-event analysis. In this case, first a constraint is chosen to select the events. Usually, this is the centrality of the event, however, events may be selected by other parameters such as the integrated $v_n$ in the soft sector, event multiplicity, average number of participants or other physical property in order to perform event-shape engineering analysis.

Before averaging over all the events, the $R_{AA}(p_T, \varphi)$ spectra are used to evaluate the differential heavy flavor azimuthal anisotropy from the Fourier harmonics $v_n^{heavy}(p_T)$:

$$\frac{R_{AA}(p_T, \varphi)}{R_{AA}(p_T)} = 1 + 2 \sum_{n=1}^{\infty} v_n^{heavy}(p_T) \cos\left[n\left(\varphi - \Psi_n^{heavy}(p_T)\right)\right],\qquad(4.21)$$

in which:

$$v_n^{heavy}(p_T) = \frac{\frac{1}{2\pi} \int_0^{2\pi} \mathrm{d}\varphi \, \cos\left[n\left(\varphi - \psi_n^{heavy}(p_T)\right)\right] R_{AA}(p_T, \varphi)}{R_{AA}(p_T)},\qquad(4.22)$$

and the event plane angles are defined as:

$$\psi_n^{heavy}(p_T) = \frac{1}{n} \arctan\left(\frac{\int_0^{2\pi} \mathrm{d}\varphi \, \sin(n\varphi) \, R_{AA}(p_T, \varphi)}{\int_0^{2\pi} \mathrm{d}\varphi \, \cos(n\varphi) \, R_{AA}(p_T, \varphi)}\right).\qquad(4.23)$$

The final results for a given event selection are then computed. The nuclear modification factor $\langle R_{AA}(p_T, \varphi)\rangle_{events}$ is the average over all the events considered and the differential azimuthal anisotropy $v_n(p_T, \varphi)$ is computed from the correlation between $v_n^{heavy}(p_T, \varphi)$ in the heavy flavor sector and the integrated $v_n^{soft}$ of all charged particles in the soft sector using the $Q$-vector cumulants method,[160,165,174,204-206] described in the section 3.4.

In order to obtain the $p_T$ integrated azimuthal anisotropy over an interval $\Delta p_T$, the same procedure is employed after the integration has been performed for the $R_{AA}(p_T, \varphi)$ spectra.

Events with larger multiplicity have smaller statistical uncertainty. In order to refine the results, this fact is exploited by weighting the events in the cumulants evaluation using the multiplicity:[207,208]

$$W_{\langle m \rangle} = \prod_{n=0}^{n=m-1} (M - n),$$ (4.24)

where $M$ is the event multiplicity and $\langle m \rangle$ is a shorthand notation for the single-event average $m$-particle azimuthal correlations. The above weighting bias can have a non-negligible impact on the final results for the correlations if a wide centrality bin is used. Due to this fact, events are further classified into 1%-wide centrality bins and then reaveraged into wider bins after the multiplicity weighting is done. This has been proven to remove the effect due to the bias.[208]

Finally, by using the event-by-event results, the correlations between the soft and heavy sectors of the event are explored using event-shape techniques[176,178,204,209] by classifying the events in a given centrality class using the integrated $v_n$ in the soft sector.

## 4.7 Error estimates

The errors associated with the calculations are obtained via jack-knife resampling[210] where the averages over the $n$ events are reevaluated with one of the events removed. This procedure is repeated until all the events have been removed once from the sampling in order to obtain the variance:

$$\mathrm{Var}(\mathcal{O})_{\text{jackknife}} = \frac{n-1}{n} \sum_{i=1}^{n} \left( \bar{\mathcal{O}}_i - \bar{\mathcal{O}} \right)^2,$$ (4.25)

where $\bar{\mathcal{O}}$ is the estimator for the observable $\mathcal{O}$ with all the events and $\bar{\mathcal{O}}_i$ is the estimator from the subsample without event $i$. The errors from the initial hypotheses of the simulation, such as FONLL errors, have not been considered during the computations, thus all the errors presented in this work are directly related to the Monte Carlo.

The next chapter presents the analysis and results that has been obtained from the described simulation.

# Results

Let us now discuss the results obtained from DABMod simulation from what has been described in chapter 4. One of the main constraints of the program, in order to obtain the final results, is the correct determination of the energy loss proportion factor $\lambda$ present in equations 4.13 to 4.17. This factor directly affects how much energy the heavy quarks will deposit in the medium while the general form of the energy loss models is related to how this interaction depends on the plasma properties. The straightforward observable related to the energy loss is the nuclear modification factor, already described, therefore, in order to find a reasonable value for this parameter, experimental data for $R_{AA}$ is used. However, other parameters of the simulation may affect the energy loss as well, making it necessary to select values for $\lambda$ every time one of these parameters change. In this work, the decoupling temperature $T_d$ is going to be explored within a wide range, from $T_d = 120\,\text{MeV}$ to $T_d = 160\,\text{MeV}$. The choice of this range was motivated by the width of the cross-over transitions from lattice QCD[211] defined using the chiral condensate, as the hadronization processes can become very cumbersome and lots of technicalities regarding these processes are not addressed here. The main quenching mechanism for the high energy heavy quarks is the path-length differences experienced by them during the interaction with the QGP. It follows naturally then that by changing the hadronization temperature, one should be able to notice differences in the energy loss related observables; the nuclear modification factor and the azimuthal anisotropy.

The results presented in this chapter will start by evaluating the effects of the hadronization temperature in each energy loss model. Following, a scan of $\lambda$ parameters for the different models is done assuming the wide range for the temperatures. The nuclear modification factor is studied and compared with experimental data in order to filter out models that falls too far from the expected behavior. After this selection is done, the remaining models are further explored by analyzing the azimuthal anisotropy from the simulation. A special care is taken at this stage where different ways of analyzing the results are proposed and compared. Finally, the energy loss fluctuations are included for one of the models in order to study this effect on the results.

## 5.1 Energy loss factor scan

In order to obtain the differential nuclear modification factor that is used for comparison with data, events in the 0–10% centrality class are selected and the simulation is performed for PbPb collisions at $\sqrt{s_{NN}} = 2.76$ TeV and $\sqrt{s_{NN}} = 5.02$ TeV. The program has been executed for about one thousand events in this centrality class for each beam energy and the number of heavy quarks (bottom and charm separately) sampled per event could be set to a lower value of 1 million, due to the $R_{\mathrm{AA}}$ being very robust with regards to the statistics of the simulation. The chosen values were enough so the errors associated with the Monte Carlo were within the line widths in the plots. The resolution of the $\lambda$ partition has been selected so that the closest values to be compared could be identified by a difference of at least two digits, as the errors associated with experimental data used for the fitting are much larger than the resolution of the simulation and a proper evaluation of $\lambda$ errors would be meaningless at this point.

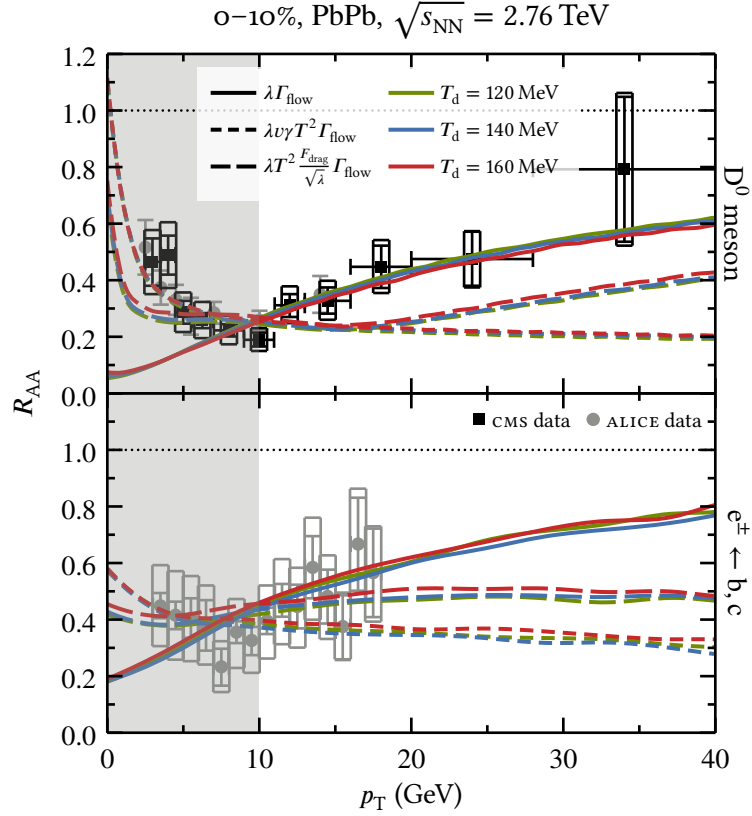

FIGURE 5.1 – Fit of the $\lambda$ parameter for three energy loss models and different temperatures $T_d$ for PbPb collisions at $\sqrt{s_{NN}} = 2.76$ TeV. The upper panel presents the fit for the charm quark using $D^0$ meson data from the ALICE[104] and CMS[212] experiments while the bottom panel shows the fit meson and the bottom panel for the bottom quark using electron data from the ALICE[213] experiment. The shaded area corresponds to the $p_T$ region where other effects may be important.

Before starting the scan of $\lambda$ values for each temperature $T_{\mathrm{d}}$, one needs to make sure the general shape of the spectra is not affected by the choice of the latter. The plots in figure 5.1 present the $\lambda$ fit for three energy loss models, from equations 4.13, 4.14 and 4.17, by selecting different temperatures $T_{\mathrm{d}}$ for each. The upper plot shows the D$^0$ meson $R_{\mathrm{AA}}$ compared with experimental data from both ALICE[104] and CMS[212] while the lower plot shows the combined contribution from bottom and charm quarks into heavy flavor electrons compared to ALICE data[213]. In order to select the $\lambda$ parameter, the experimental point at $p_{\mathrm{T}} \sim 10$ GeV has been used as reference, the same procedure as has been used by other different works in order to compare simulation computations with data.[109,195] In the figure, the gray-dashed area at the low $p_{\mathrm{T}}$ regime may have a non-negligible effect from heavy quark coalescence, which is not taken into account in this work as it has been exposed earlier. As it can be verified in the plots, different energy loss models have very different behavior with regards to the $R_{\mathrm{AA}}$. However, when changing the hadronization temperature, the overall shape of the curves is retained, even for a large range of variation. This is consistent with idea that this temperature only sets a stopping point for the heavy quarks traversing the medium. The same behavior is verified for both heavy quarks and is carried upon the total contribution for the heavy flavor electrons.

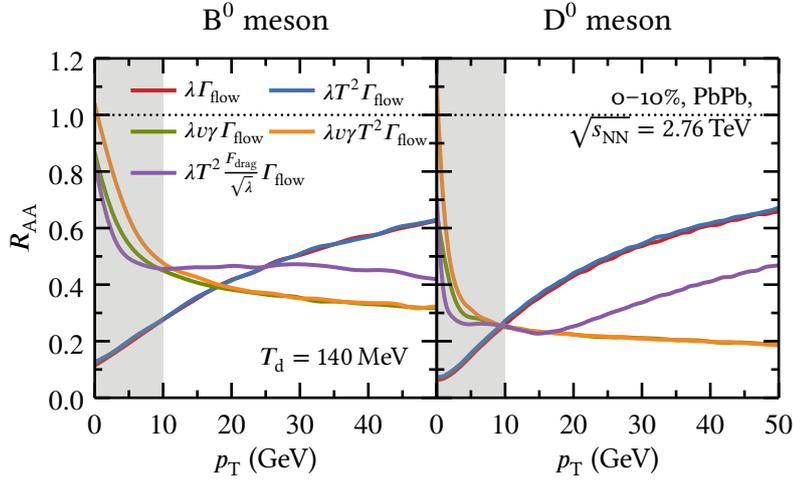

FIGURE 5.2 – Nuclear modification factor for B$^0$ meson (left) and D$^0$ meson (right) for 0–10% central PbPb collisions at $\sqrt{s_{\mathrm{NN}}} = 2.76$ TeV and decoupling temperature $T_{\mathrm{d}} = 140$ MeV. Different energy loss models, as described by equations 4.13 to 4.17, are compared. The shaded area corresponds to the $p_{\mathrm{T}}$ region where other effects may be important.

From what have been concluded from figure 5.1, one can explore other energy loss models for a single temperature $T_d$ in order to test the shapes of the $R_{AA}$ curves. The plots presented in figure 5.2 show on the left, the $R_{AA}$ results for B$^0$ mesons for all the energy loss models described in equations 4.13 to 4.17. The same is shown for D$^0$ mesons in the right panel of the figure. All of these results are evaluated for $T_d = 140$ MeV. Some of the characteristics of each model are already clear by looking at these plots. One can note that both equations that does not involve the velocity of the heavy meson, 4.14 and 4.15, behave very similar and the temperature dependence with the term $T^2$ does not seem to play an important role in these models. The same conclusion can be drawn from the other two models that have a dependence on the $v\gamma$ factor, namely the ADS/CFT model from equation 4.13 and the equation 4.16. It seems that for the $R_{AA}$ evaluation, the fluctuations of temperature in the medium are not determinant for the interaction of the heavy quarks. This fact can give a clue as to why, for so long, average events have been used for studying energy loss models, often comparing the predictions with $R_{AA}$ data, other than, of course, the statistics limitations in the past.

On the other hand, the $v\gamma$ dependence, which translates into a momentum dependence if one considers the rest mass $m_0$ as being absorbed by the constant $\lambda$, does have a clear effect on the $R_{AA}$ shapes. Without it, the shape tends to increase with the momentum, while by adding this term, the energy loss experienced by heavy quarks with high transversal momentum is increased, leading in turn to a slight decrease in the nuclear modification factor over $p_T$. The lattice-based model from equation 4.17 has a more subtle dependence on $v$ so that the curve lies in between the other two extremes. Also noticeable in the figure, by comparing the B$^0$ and D$^0$ mesons for this particular model, is the strong dependence on the parton masses. In table 5.1 presents the values found for $\lambda_b$ and $\lambda_c$ that were used in the figure 5.2. The values from the table makes even more clear that the energy loss factor depends on the mass of the heavy quarks with $\lambda_b$ being greater the $\lambda_c$ for all the studied cases.

Comparing the results from figure 5.2 with figure 5.1, it is clear that only the energy loss models based on equations 4.14 and 4.15 are able to describe the experimental data for the $D^0$ meson $R_{AA}$ for $p_T \gtrsim 10$ GeV and that the other models present problems, either by leading to a different slope on the curves or they could be fitted only for a much higher cut in the $p_T$ range. Therefore, further analysis on the $\lambda$ determination will be only performed for the two mentioned energy loss models.

TABLE 5.1 – Values of the energy loss factor $\lambda$ fitted for different models for B$^0$ and D$^0$ mesons using $R_{AA}$ data for $T_d = 140$ MeV.

| Energy loss | $\lambda_b$ | | $\lambda_c$ | | Unit |
|---|---|---|---|---|---|
| $\lambda \Gamma_{\text{flow}}$ | 0.86 | | 0.75 | | GeV fm$^{-1}$ |
| $\lambda T^2 \Gamma_{\text{flow}}$ | 23.0 | | 19.5 | | GeV$^{-1}$ fm$^{-1}$ |
| $\lambda v\gamma \Gamma_{\text{flow}}$ | 0.214 | | 0.102 | | GeV fm$^{-1}$ |
| $\lambda v\gamma T^2 \Gamma_{\text{flow}}$ | 5.8 | | 2.6 | | GeV$^{-1}$ fm$^{-1}$ |
| $\lambda T^2 \frac{F_{\text{drag}}}{\sqrt{\lambda}} \Gamma_{\text{flow}}$ | 4.6 | $\times 10^{-5}$ | 3.0 | $\times 10^{-5}$ | |

Furthermore, looking closely at figure 5.2 and comparing both $B^0$ and $D^0$ mesons $R_{AA}$ for those models, one can note that there's actually little difference between them for the whole range of $p_T$ shown in the plots. This does not seem to be the case when J/ψ $R_{AA}$, which can be regarded as an indirect measurement of B mesons, is compared to D mesons. In fact, cms measurements on the J/ψ $R_{AA}$ seems to indicate that charm quarks are more quenched than bottom quarks.[214] Figure 5.3, reproducing those studies, show a clear distinction between both mesons for for all centralities, represented by the number of binary collisions $\langle N_{part} \rangle$. However, the same studied showed that meson $R_{AA}$ also depends heavily on the considered rapidity range.

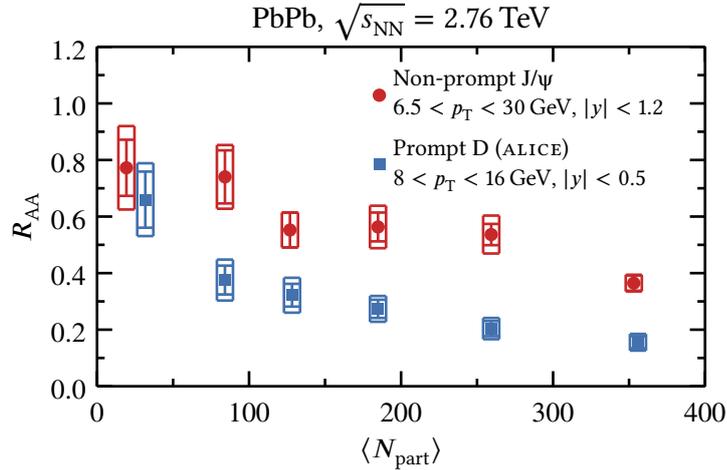

FIGURE 5.3 – Non-prompt J/ψ and prompt D meson $R_{AA}$ versus centrality for events flatly distributed across centrality. Data from cms[214] and alice.[215,216]

Taking this into account, a more recent data measurement from the alice experiment[217] with a smaller rapidity range of $|y| < 0.5$ was considered for a new $\lambda$ selection for the constant energy loss model. By using this new data it is expected that the computations comparison be more consistent as the simulation is evaluated for the mid-rapidity range. The results for this simulation are shown in figure 5.4 where $D^0$ and $B^0$ mesons are compared with an average of $D^0$, $D^+$, and $D^{*+}$. In this new evaluation one note that the $D^0$ meson are consistent with the data, furthermore, there is a clear distinction between both heavy mesons as should be expected from what has been discussed and in agreement with the suggestion that $B^0$ meson should give a larger $R_{AA}$ than $D^0$ meson.

The values for the $\lambda$ parameter found for this new run were $\lambda_b = 0.72$ GeV fm$^{-1}$ for the bottom quark and $\lambda_c = 0.99$ GeV fm$^{-1}$ for the charm quark. Although these values cannot be compared with those from the table 5.1 due to the difference in the temperature $T_d$, the relation between both heavy quarks has been changed due to a much lower $D^0$ meson $R_{AA}$ which, in turn, increases its $\lambda_c$ parameter.

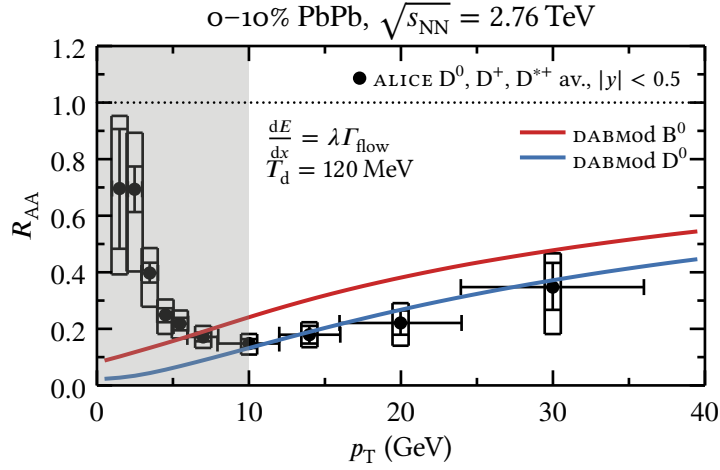

Figure 5.4 – Nuclear modification factor for $B^0$ meson and $D^0$ meson compared with Alice data [217] for central PbPb collisions at $\sqrt{s_{NN}} = 2.76$ TeV and energy loss from equation 4.14. No data for $B^0$ meson $R_{AA}$ is available. The shaded area corresponds to the $p_T$ region where other effects may be important.

From what has been discussed so far, it is clear that comparison with experimental data should be addressed with special care as different experimental setup may result in fairly different outcomes. However, although the comparison with experimental data in this work are not final, they present an important and needed tool in order to check for the simulation framework consistency.

Following the same procedure as with PbPb collisions at $\sqrt{s_{NN}} = 2.76$ TeV, the simulation has been performed for the two chosen energy loss models from equations 4.14 and 4.15 for the higher beam energy of $\sqrt{s_{NN}} = 5.02$ TeV. Both heavy quarks have been studied and the $\lambda$ parameter has been determined, however, differently from what has been previously done for $\sqrt{s_{NN}} = 2.76$ TeV collisions, the $\lambda_b$ for the bottom quark has been determined from $B^+$ meson $R_{AA}$ recent measurement from the CMS experiment. [218] The drawback is that the centrality range used in the experimental analysis was 0–100% while only events for 0–50% have been simulated in this work. Also, a wide rapidity range of $|y| < 2.4$ has been used and the results still have poor statistics, with only five data points distributed onto a wide range of $p_T$. However, this was the first direct measurement of open B meson $R_{AA}$ performed in the LHC, which will lead the path for new developments in this regard. The results presented here is therefore preliminary.

The plots in figure 5.5 show the differential $R_{AA}$ for $B^0$ meson (left) and $D^0$ meson (right) for PbPb collisions at $\sqrt{s_{NN}} = 5.02$ TeV evaluated from DABMod simulation compared with the experimental data [218,219] for two different temperatures $T_d$ and only for the energy loss model from equation 4.14, which has lead to reasonable results for the lower beam energy. The results are consistent with the measurements from the CMS experiment for both heavy mesons, while the general shape of the $R_{AA}$ curves is very similar to the behavior previously obtained. Also, the variation in the temperature $T_d$ does not change the $R_{AA}$ behavior, confirming what has been previously found in the simulation. The $\lambda$ parameters found for these runs are summarized in the table 5.2. From the values obtained from the table it is now possible to verify that with the increase of the temperature $T_d$, the $\lambda$ values are also increased. This is primarily due to the diminished path-length of the heavy quarks inside the medium so that the energy loss rate must compensate for it in order to obtain the same $R_{AA}$.

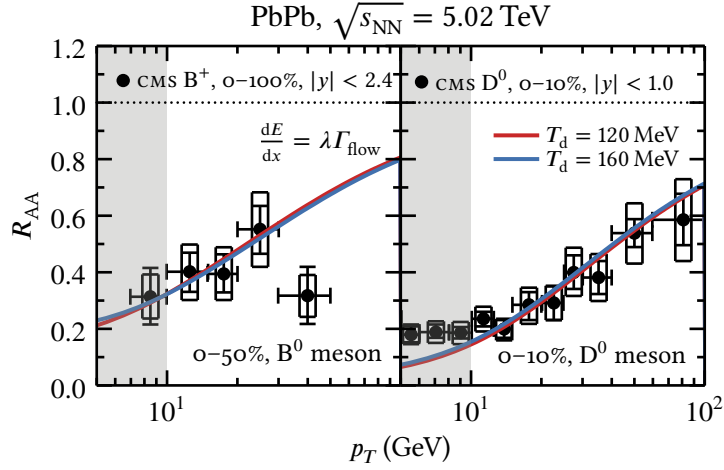

FIGURE 5.5 – Nuclear modification factor for $B^0$ meson (left) and $D^0$ meson (right) for PbPb collisions at $\sqrt{s_{NN}} = 5.02$ TeV compared with experimental data from the CMS collaboration.[218,219] Evaluation has been done using energy loss model from equation 4.14. As data for B meson does not match the centrality range of the simulations, the comparison presented here is preliminary. The shaded area corresponds to the $p_T$ region where other effects may be important.

TABLE 5.2 – Values of the energy loss factor $\lambda$ fitted for PbPb collisions at $\sqrt{s_{NN}} = 5.02$ TeV and for $T_d = 120$ MeV and $T_d = 160$ MeV.

| $T_d$ (MeV) | $\lambda_b$ | $\lambda_c$ | Unit |
|---|---|---|---|
| 120 | 0.6 | 0.8 | GeV fm$^{-1}$ |
| 160 | 0.9 | 1.1 | GeV fm$^{-1}$ |

As the $R_{AA}$ serves as the basis for the $v_n$ evaluations, having correctly determined the values for the $\lambda$ parameter and checking the consistency of the results with the experimental data is the primary baseline for the remaining analysis. The results presented so far have successfully described the available measurements within error bars and the event-by-event analysis of the heavy flavor azimuthal anisotropy is then possible.

## 5.2   Evaluation of the azimuthal anisotropy

The differential $R_{\mathrm{AA}}(p_{\mathrm{T}}, \varphi)$ obtained from DABMod simulation is used, as described in section 4.6, in order to obtain the integrated heavy flavor $v_n^{\mathrm{heavy}}$ for each hydrodynamic event. This quantity is completely independent of the soft sector particles and is obtained simply via Fourier expansion of the azimuthal distribution of the $R_{\mathrm{AA}}$. Although the Fourier harmonics from the azimuthal distribution of particles are usually associated with the spatial distribution of the colliding nuclei, event-by-event fluctuations have a big impact on how the participant nucleons interact during the collisions which, in turn, affect the measured distribution.[220] In order to include the important role played by the initial state fluctuations and their dynamical evolution within the system, one needs to evaluate the correlations among the participant nucleons in an event-by-event basis, as detailed in section 3.4. This work study the correlations between the heavy flavor sector and the soft sector, based on the approach that has been used for the hard-soft sector of charged particles.[109] The differential $v_n(p_{\mathrm{T}})$ is evaluated using the $Q$-vector cumulants method[160,165,174,204–206] and a new way of looking at the correlations is explored by changing the simulation parameters. The study is focused on the second and third order flow harmonics, $v_2$ and $v_3$ respectively.

Figure 5.6 shows the distribution of events' integrated $v_n^{\mathrm{heavy}}$ from the heavy flavor sector with respect to the $v_n^{\mathrm{soft}}$ from charged particles for $n = 2$ (left) and $n = 3$ (right) for semi-central PbPb collisions at $\sqrt{s_{\mathrm{NN}}} = 5.02$ TeV. Each point in the scatter plots correspond to the evaluated values for a single event. The distributions show that for both elliptic and triangular flow, there is a linear correlation between the heavy and the soft sectors on an event-by-event basis. The same correlation has been observed between low-$p_{\mathrm{T}}$ and high-$p_{\mathrm{T}}$ soft particles,[109] leading to the conclusion that the initial state geometrical fluctuations, which is responsible for the elliptic flow for low-$p_{\mathrm{T}}$ particles is also responsible for fluctuations in the path-length for the high-$p_{\mathrm{T}}$ jets. From this perspective, the same path-length anisotropy is observed in the heavy flavor sector.

Using event-shape engineering technique[209] on the distribution of figure 5.6, the hydrodynamic events are characterized by the flow harmonics in the soft sector $v_n^{\mathrm{soft}}$. The corresponding elliptic and triangular flow for the $B^0$ and $D^0$ mesons can be computed, which gives the probability that a soft event will correspond to a particular value of $v_n^{\mathrm{heavy}}$. This is shown in figure 5.6 as the black lines over the scatter distribution of events. This observable encodes the event-by-event fluctuations of the heavy flavor azimuthal anisotropy and the correlation with the charged particles. One can then study the slope of these correlations with respect to different collision conditions leading to $v_n$ fluctuations. If no fluctuation were to be observed, a straight flat line would be observed in the plots.

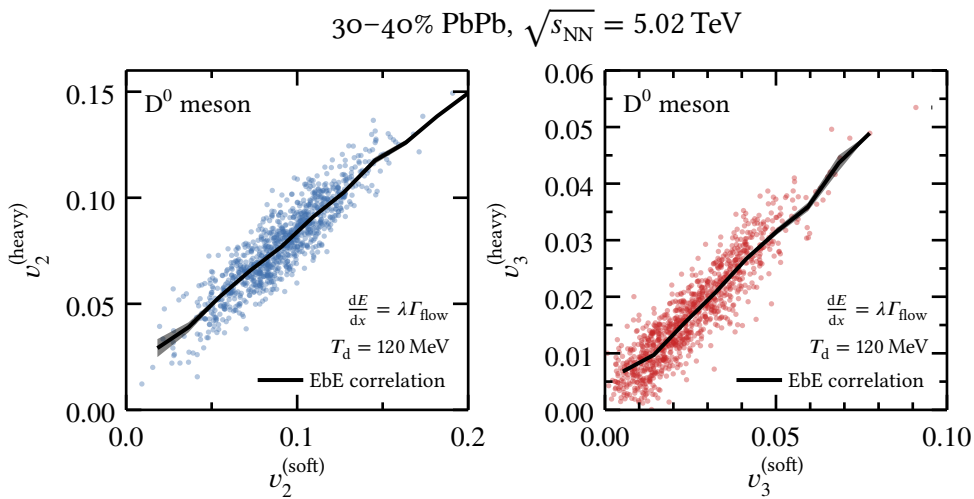

FIGURE 5.6 – Event-by-event distribution of $D^0$ meson $v_2$ (left) and $v_3$ (right) of the heavy sector with respect to the soft sector for 30–40% PbPb at $\sqrt{s_{NN}} = 5.02$ TeV. The lines show the correlation obtained from event-shape engineering techniques by classifying the events using the soft $v_n$.

Experimentally, this measurement can be performed by binning the values of $v_2^{\text{soft}}$ and evaluating the $v_2^{\text{heavy}}$ for the set of events within a given bin. A similar approach has already been adopted by the ATLAS collaboration for high-$p_T$ identified hadrons evaluation of symmetric cumulants and correlations.[204]

One can also look at the alignment of the evaluated event plane angles $\psi_n^{\text{heavy}}$ with respect to the soft sector $\psi_n^{\text{soft}}$. As the cumulants evaluation from equation 3.56 contains a cosine term of the difference between the event plane, one can visualize its correlation from the mean of this term. The figure 5.7 shows the $\psi_n$ correlations for $n = 2$ and $n = 3$ for PbPb collisions at both studied energies. The same constant energy loss model from equation 4.14 has been used and the plots show the results for both extremes of temperature $T_d$ from 120 MeV to 160 MeV.

As the blue lines indicate, the heavy flavor $\psi_2^{\text{heavy}}$ for both energies and temperatures is highly correlated with the soft sector, with $\left\langle \cos\left[2(\psi_2^{\text{soft}} - \psi_2^{\text{heavy}})\right]\right\rangle > 0.95$ for most of the centrality range considered. However, the same does not happen for the triangular flow. Although the event plane angles for $n = 3$ are still highly correlated when $T_d = 120$ MeV, for both collision energies, if this temperature is increased it is clear the $\sqrt{s_{NN}} = 2.76$ TeV collisions result in a decorrelation of the angles with increasing centrality. This decorrelation will reflect into a suppression of the corresponding $v_3(p_T)$ for the cumulants evaluation. The lower energy collisions have an average lower temperature within the system throughout the whole evolution of the QGP, which will fall below $T_d$ sooner, in comparison to events for $\sqrt{s_{NN}} = 5.02$ TeV. Because of this, the path-lengths of the heavy quarks in this type of collision is diminished and the azimuthal anisotropy is suppressed. The effect is more pronounced for non central collisions as the overlap region between the nuclei, being smaller, leads to even lower average temperatures.

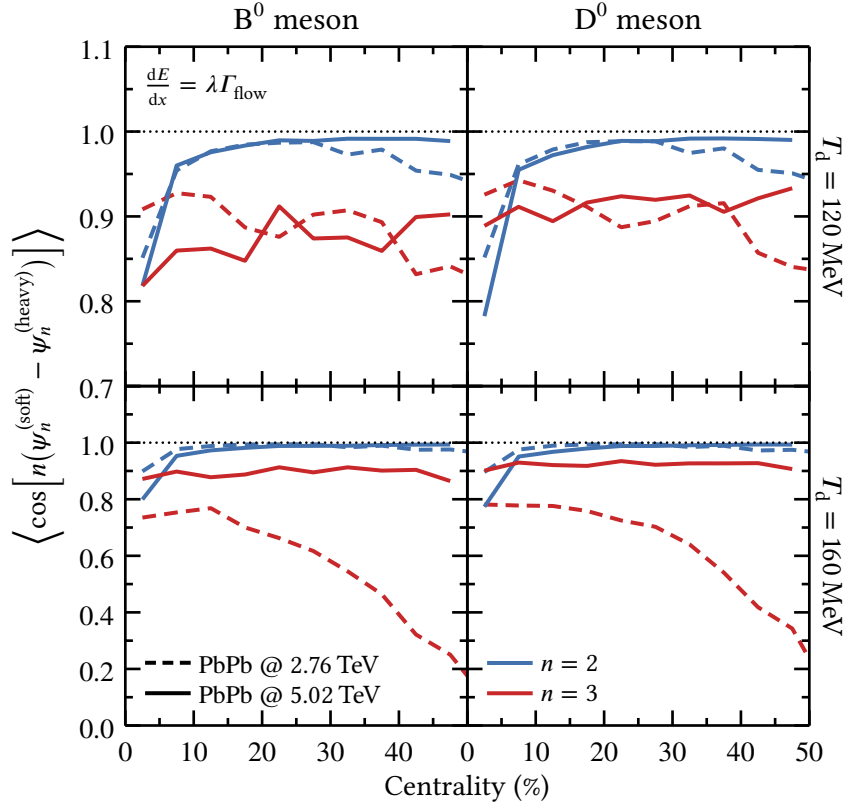

FIGURE 5.7 – Correlation of the event plane angles $\psi_n$ between the heavy and soft sector for $B^0$ meson (left) and $D^0$ meson (right) versus event centrality. Energy collisions are compared for temperatures $T_d = 120$ MeV (top) and $T_d = 160$ MeV (bottom) and for $n = 2$ and $n = 3$.

The path-length difference becomes clearer when the one looks at the final position of the heavy quarks evolved during an event in the simulation, as shown in figure 5.8. The figure shows the spatial distribution of the heavy quarks prior to the hadronization, after they have interacted with the medium in a PbPb collision at $\sqrt{s_{NN}} = 2.76$ TeV comparing both temperatures of $T_d = 120$ MeV (left) and $T_d = 160$ MeV (right). Both distributions originated from the same initial density and, as it can be seen in the figure, led to very different final states. While heavy quarks have been completely pushed away from the collision axis in the case of the lower temperature $T_d$, some of them are still distributed along a wider radius range when this temperature is increased. Moreover, the radius of the circular area in the right plot is evidently smaller than the one from the left and the average displacement of the heavy quarks from their original point is shorter.

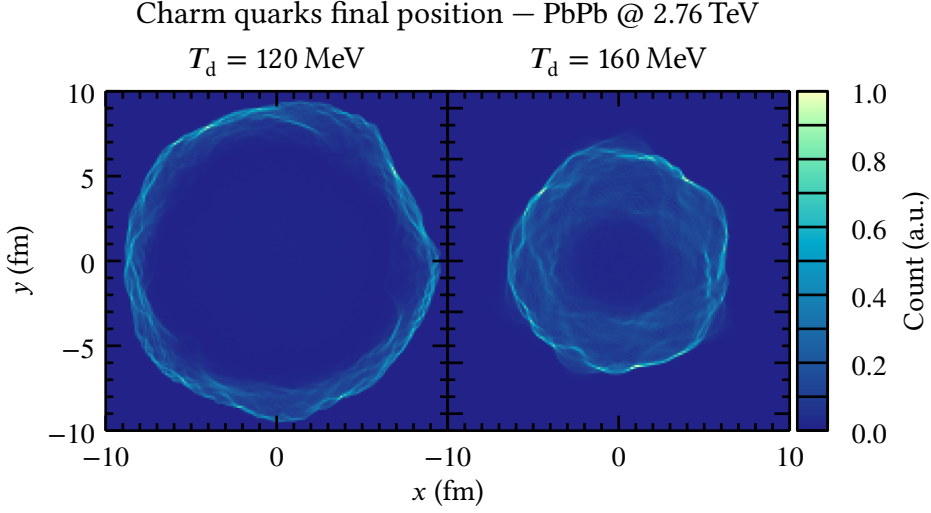

FIGURE 5.8 – Final position of charm quarks after evolution of an event for PbPb collision at $\sqrt{s_{\mathrm{NN}}} = 2.76$ TeV for temperatures $T_{\mathrm{d}} = 120$ MeV (left) and $T_{\mathrm{d}} = 160$ MeV (right).

Before going into more details on the study of the correlations, let us first recall the energy loss models defined by equations 4.13 to 4.17, for which the nuclear modification factor has been evaluated in section 5.1. The same PbPb collision at $\sqrt{s_{\mathrm{NN}}} = 2.76$ TeV has been simulated but for 20–40% in order to evaluate the $v_2\{2\}$ for all the energy loss models described, using the same temperature $T_{\mathrm{d}} = 140$ MeV that has been used for the $R_{\mathrm{AA}}$ evaluation. The results for these events are summarized in figure 5.9 in which the elliptic flow is shown for $\mathrm{B}^0$ mesons (left) and $\mathrm{D}^0$ mesons (right). As with the $R_{\mathrm{AA}}$ results, the low-$p_{\mathrm{T}}$ regime has been shaded in the plots as a reminder that coalescence effects are expected to play a role in this region.

The comparison between figures 5.9 and 5.2 shows that even though some energy loss parametrizations lead to the same result for the nuclear modification factor, the elliptic flow, on the other hand, can clearly distinguish between them. This is the case for the models described by equations 4.14 and 4.15, as well as 4.16 and 4.13. This fact indicates that the fluctuations of temperature due to the initial state are no longer washed out when evaluating the azimuthal anisotropy and that this observable, in contrast with the $R_{\mathrm{AA}}$, is very sensitive to these event-by-event fluctuations. Furthermore, from the two energy loss models selected in section 5.1, the scenario in which there is no dependence on the temperature leads to the larger values for the elliptic flow. This result is consistent to what has been previously found using smooth initial conditions[105] that suggests that the drag of the medium cannot decrease as $T \to T_{\mathrm{c}}$ as the interaction would be too weak when the bulk medium has developed its elliptic flow, even though the $R_{\mathrm{AA}}$ could be made very small with a strong suppression happening at the beginning of the system's evolution. This time hierarchy between the $R_{AA}$ and the elliptic flow in the collision is also extended to the triangular flow as the correlation study in the next sections is going to expose.

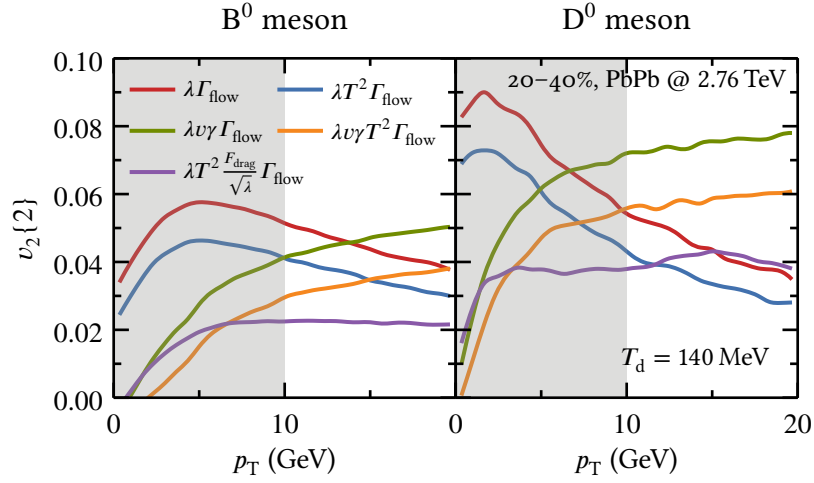

FIGURE 5.9 – Elliptic flow $v_2\{2\}$ for B$^0$ meson (left) and D$^0$ meson (right) for 20–40% PbPb collisions at $\sqrt{s_{NN}} = 2.76$ TeV and decoupling temperature $T_d = 140$ MeV. Different energy loss models, as described by equations 4.13 to 4.17, are compared. The shaded area corresponds to the $p_T$ region where other effects may be important.

## 5.3 Heavy-soft correlations

Using the event-shape engineering heavy-soft correlations defined in section 5.2 it is possible to explore different parameters of the simulation and obtain information on how these parameters affect the azimuthal anisotropy for heavy mesons. This work focus on the study of the two energy loss parametrizations that are consistent with experimental data for the nuclear modification factor and have been previously selected in section 5.1. In order to obtain reasonable statistics for the evaluations, mainly for $v_3$, 10 million heavy quarks have been sampled from the initial conditions in the simulation for both bottom and charm quarks. The $\lambda$ parameter have been fixed as described previously from the 0–10% central events and are maintained for other centrality bins. The integrated $p_T$ range for all the simulations are set to $8$ GeV $\leq p_T \leq 13$ GeV although a proper experimental evaluation is needed in order to determine which $p_T$ range would lead to statistically viable experimental results. Also, the results are presented for D$^0$ meson only due to the inconsistencies previously found for the B$^0$ meson when comparing the nuclear modification factor with data.

The first result is shown in figure 5.10 in which the correlations for the two collision energies at 30–40% centrality are represented using different colors while the decoupling temperature is shown with different line styles. In the left side of the figure the constant energy loss model from equation 4.14 is presented for both $v_2\{2\}$ at the top and $v_3\{2\}$ at the bottom. The same is shown on the left for the $T^2$ dependent energy loss from equation 4.15. The figure shows that the decoupling temperature $T_d$ affects differently the elliptic and triangular flow. The difference observed when its value is increased from from 120 MeV to 160 MeV is much smaller for $v_2$ for both energy loss models. This indicates that the triangular flow takes longer to build up than the elliptic flow, as the decoupling temperature is mostly related with the path-length experienced by the heavy quarks, which adds the $v_3$ later in the hierarchical sequence described in the last section. Furthermore, the difference is less evident for the $\sqrt{s_{NN}} = 5.02$ TeV collision as a higher average temperature in the medium takes longer to cool below the decoupling temperature. This can be verified by examining figure 4.4 that shows the evolution of the system's temperature profile for both collision energies. Moreover, by recalling the figure 5.7, it was expected that the triangular flow should be heavily affected by the increase in the decoupling temperature.

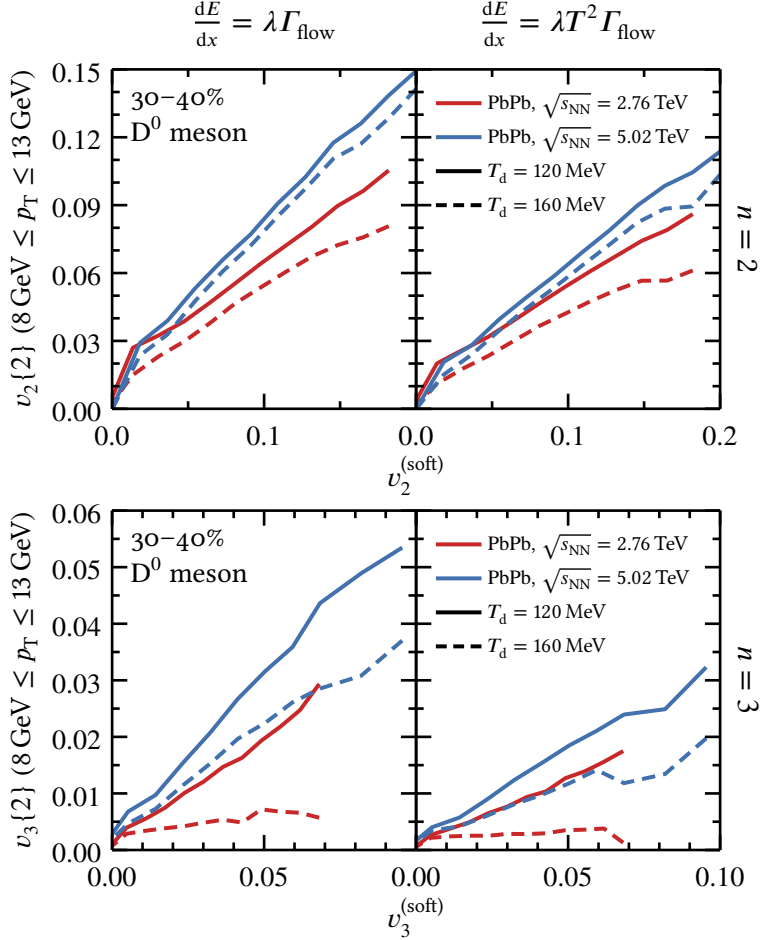

Figure 5.10 – Correlations between the $D^0$ meson $v_2$ (top) and $v_3$ (bottom) with respect to the soft sector for 30–40% PbPb collisions comparing two energy loss models based on equations 4.14 (left) and 4.15 (right).

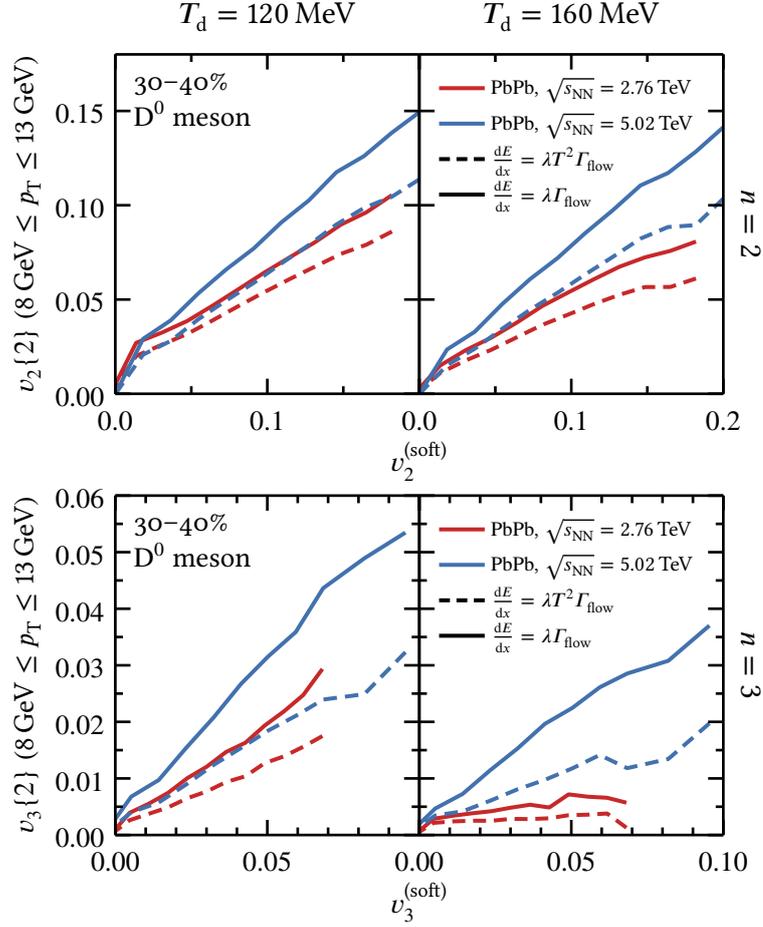

FIGURE 5.11 – Correlations between the $D^0$ meson $v_2$ (top) and $v_3$ (bottom) with respect to the soft sector for 30–40% PbPb collisions comparing the hadronization temperature $T_d = 120$ MeV (left) and $T_d = 160$ MeV (right).

The energy loss dependence of these observables is made clearer in figure 5.11 which reorder the way of showing the same results by grouping the decoupling temperature into the same plots with $T_d = 120$ MeV on the left and $T_d = 160$ MeV on the right while showing different energy loss models in each panel. The elliptic flow results confirm what has been previously described in figure 5.9 that the constant energy loss model from equation 4.14 leads to a higher value of $v_2$ than the $T^2$ dependent model from equation 4.15, regardless of the temperature and the collision energy. Furthermore, this effect is also observed, even more evidently, for $v_3$. The conclusion that can be drawn from these results are consistent with what has been shown so far that the temperature dependent energy loss will lead to a weak interaction at later stages of the evolution where the triangular flow should build up.

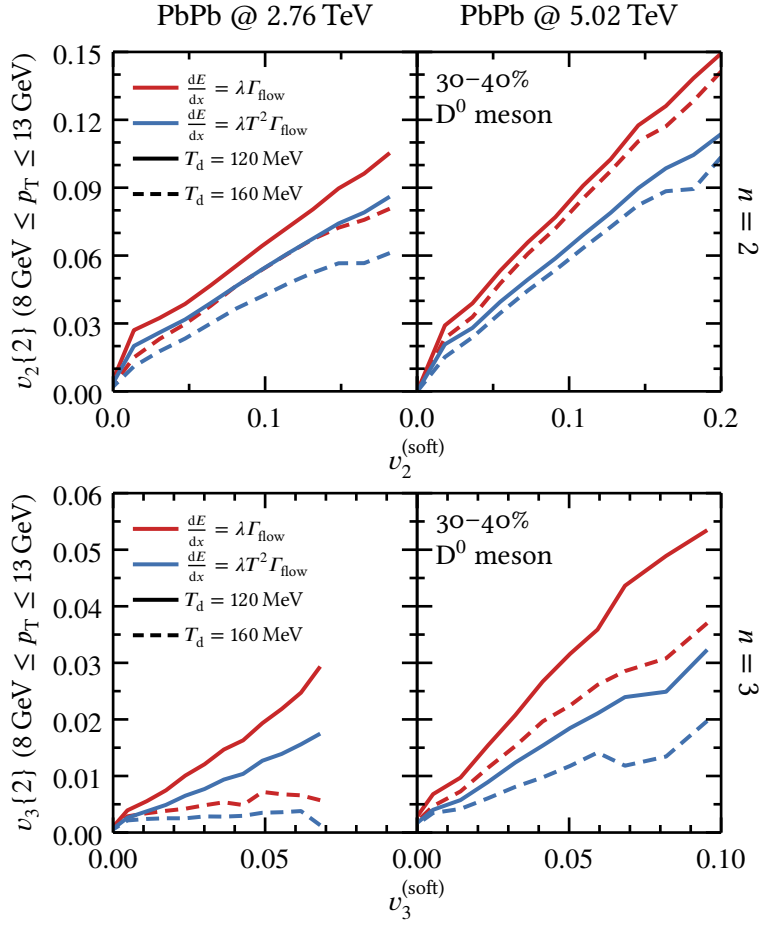

FIGURE 5.12 – Correlations between the D$^0$ meson $v_2$ (top) and $v_3$ (bottom) with respect to the soft sector for 30–40% PbPb collisions comparing the two collision energies $\sqrt{s_{NN}} = 2.76$ TeV (left) and $\sqrt{s_{NN}} = 5.02$ TeV (right).

Figure 5.12 consists of another permutation of the same results where the collision energies have been separated and the comparison within each panel of the figure is done for the energy loss models within the range of decoupling temperature. The results shown in this manner emphasizes the separation of the energy loss models. It is observed that for the $\sqrt{s_{NN}} = 5.02$ TeV collisions, both models are clearly differentiated, even when considering the wide range of temperature $T_d$. In the case of the $\sqrt{s_{NN}} = 2.76$ TeV, the elliptic flow is still separate, although the $v_3$ shows a very large overlap between both models, mostly due to the shorter path-length of the heavy quarks that has been discussed.

Another way to look at the decoupling temperature dependence on the azimuthal anisotropy is the evaluation of the convergence of the multi-particle cumulants as the number of correlated particles increase. The higher order of cumulants are usually expected to suppress the so-called "non-flow" contributions to the fluctuations, although they cannot be completely removed by constructing cumulants.[221] Figure 5.13 shows the multi-particle cumulants for $m = 2$, 4, 6, and 8 for 30–40% PbPb collisions at $\sqrt{s_{NN}} = 2.76$ TeV of $D^0$ mesons (top) and $B^0$ mesons (bottom) where the ratio $v_n\{m\}/v_n\{2\}$ is shown at the bottom panels for each plot. The comparison between both decoupling temperatures shows that for a lower temperature the ratios are closely converging while a higher temperature results in the ratios being farther apart. This convergence is another indication of the event-by-event fluctuations building up the elliptic flow over time during the evolution of the system.

One of the main questions in the field regarding heavy quarks concerns the extent with which they couple with the medium and, consequently, flow with the expanding QGP formed in the heavy ion collisions. In this regards, the correlations calculated with the simulation can be used in order to obtain information about this coupling. The linear correlation between heavy quarks and soft particles can provide a novel signature of collectivity in the heavy sector. These results can be compared with the cumulants evaluation of $v_2$ and $v_3$. The figure 5.14 shows the $D^0$ meson (left) and electron from heavy flavor (right) elliptic flow from 2-particle cumulants $v_2\{2\}$ for PbPb collisions at $\sqrt{s_{NN}} = 2.76$ TeV compared to experimental data from the ALICE experiment [100,215] where the range of decoupling temperature is represented as a band. The cumulants show a separation between the models similar to that observed from the correlations in figure 5.12, although the centrality range of both results is slightly different. The comparison with the experimental data shows that the results from the simulation are below the expected measurements, however, these measurements at this collision energy are concentrated in the lower $p_T$ regime where other effects not considered in the simulation may have an important role. Furthermore, the errors of the measured data points are significantly large. Despite of that, the computations are consistent with the general behavior of the data for both results.

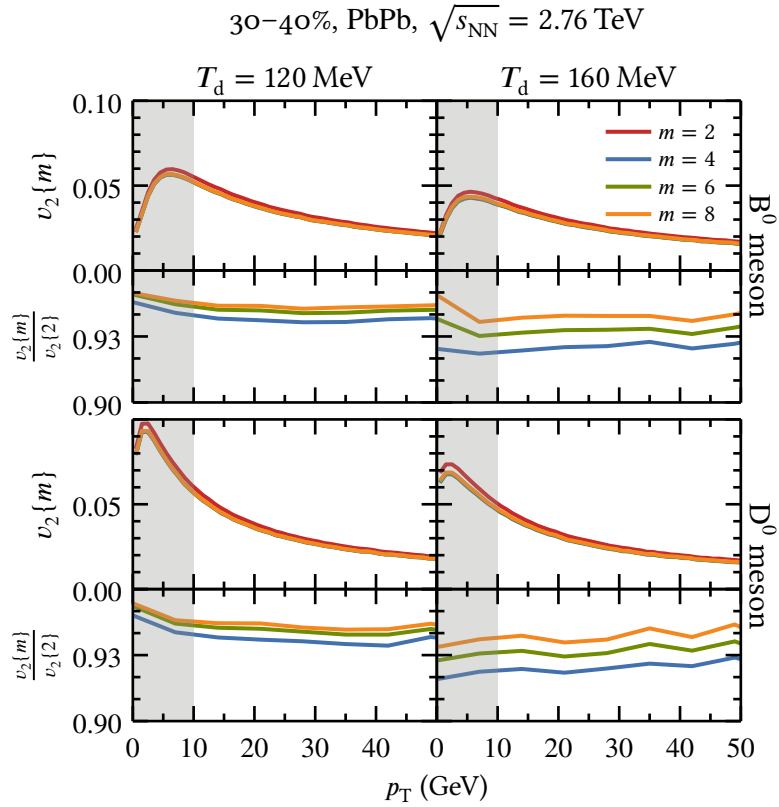

FIGURE 5.13 – Multi-particle cumulants of $v_2$ for $B^0$ meson (top) and $D^0$ meson (bottom) for 30–40% PbPb collisions at $\sqrt{s_{NN}} = 2.76$ TeV. Two different temperatures $T_d$ are shown with $T_d = 120$ MeV on the left and $T_d = 160$ MeV on the right. The shaded area corresponds to the $p_T$ region where other effects may be important.

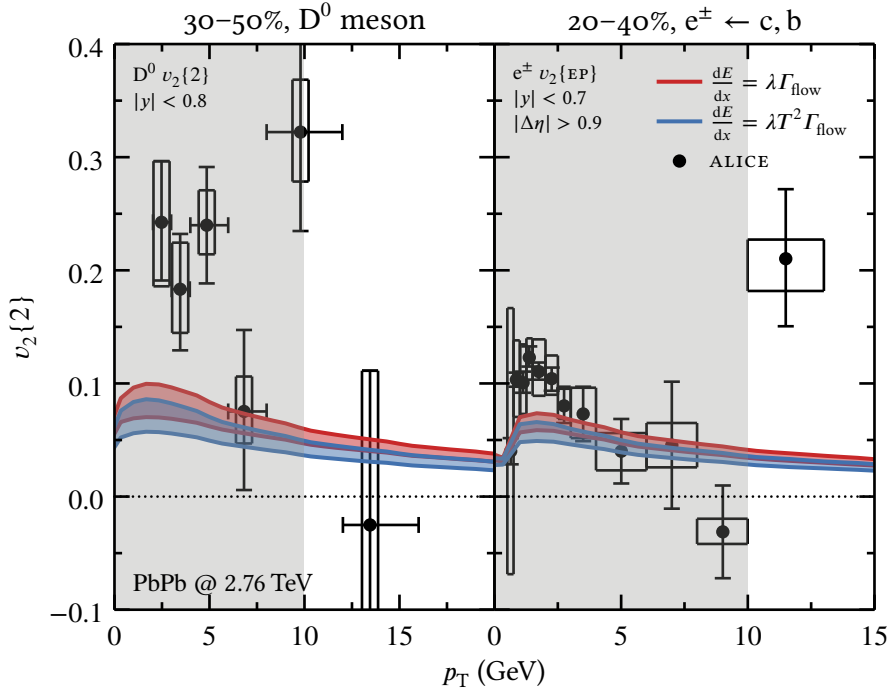

Figure 5.14 – Differential elliptic flow $v_2\{2\}$ for $D^0$ meson (left) and electron from heavy flavor (right) compared to experimental data from the ALICE experiment [100,215] for PbPb collisions at $\sqrt{s_{NN}} = 2.76$ TeV. The band for each energy loss model represent the temperature range of 120 MeV $\leq T_d \leq$ 160 MeV. The shaded area corresponds to the $p_T$ region where other effects may be important.

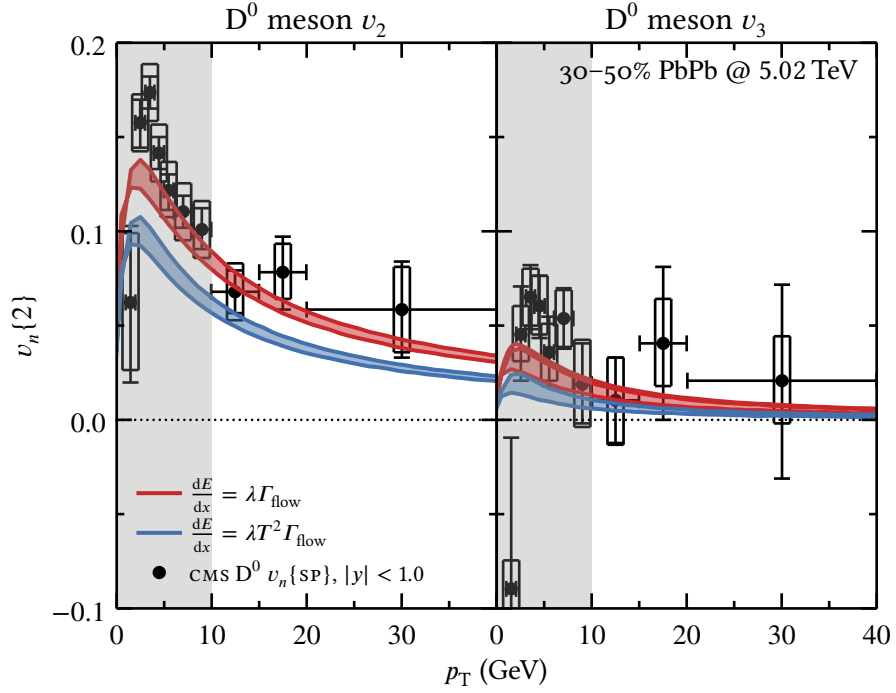

Figure 5.15 – Differential elliptic flow $v_2\{2\}$ (left) and triangular flow $v_3\{2\}$ (right) for D⁰ meson compared to experimental data from the CMS experiment[222] for PbPb collisions at $\sqrt{s_{NN}} = 5.02\,\mathrm{TeV}$. The band for each energy loss model represent the temperature range of $120\,\mathrm{MeV} \leq T_d \leq 160\,\mathrm{MeV}$. The shaded area corresponds to the $p_T$ region where other effects may be important.

Figure 5.15 compares both energy loss models shown in different bands with experimental data from the CMS experiment for the D⁰ meson $v_2$ (left) and $v_3$ (right) of 30–50% PbPb collisions at $\sqrt{s_{NN}}$ PbPb collisions. The elliptic flow using the constant energy loss model from equation 4.14 although systematically below the data points are within the error bars for nearly all the $p_T$ range considered. The triangular flow $v_3$ is also consistent with the experimental data within error bars. Furthermore, the separation between the models seen in the plots is also similar to the observed separation in figure 5.12.

The comparison between figures 5.12, 5.14 and 5.15 show that the correlations predictions are consistent with other methods for evaluating the heavy-flavor azimuthal anisotropy. The main difficulty when evaluating particle correlations for heavy flavor is the statistics involved in the measurements and further development in the experimental field must be done in order to determine useful parameters for this event-shape analysis. However, this kind of analysis has already been performed for high-$p_T$ soft particles[204,223] and the increasing collision energies in the experiments should allow for new results to come up.

The plot in figure 5.16 shows the differential elliptic flow for D⁰ meson in semi-central PbPb collisions at $\sqrt{s_{\text{NN}}} = 5.02$ TeV in which an event shape engineering technique has been employed to separate events with low $q_2$ from the ones with high $q_2$. The reduced flow vector is defined as:[160,223,225,226]

$$q_n^2 = \frac{|\boldsymbol{Q}_n|^2}{M} = \langle 2 \rangle (M - 1) + 1 \,, \tag{5.1}$$

in which $M$ is the total number of particles used to evaluate the flow vector $\boldsymbol{Q}_n$.

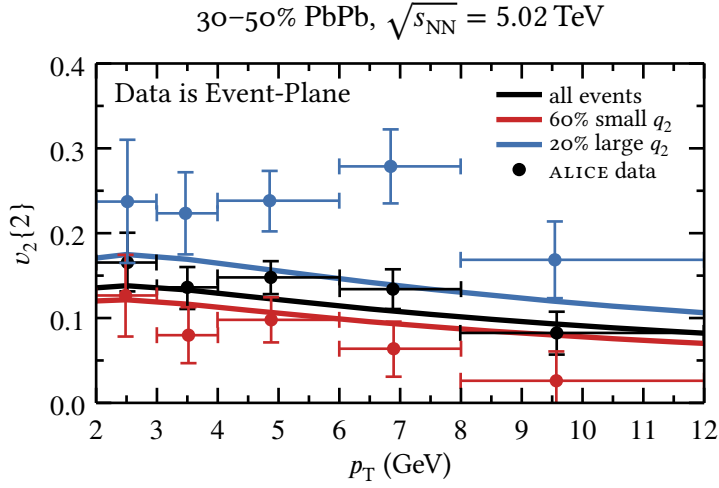

FIGURE 5.16 – Differential elliptic flow $v_2\{2\}$ for $D^0$ meson in 30–50 PbPb collisions at $\sqrt{s_{\text{NN}}} = 5.02$ TeV using event shape technique to separate events by the reduced flow vector $q_2$ as defined by equation 5.1. Results from DABMOD simulation are compared to experimental data.[224]

In the figure, for comparison, it is shown also the differential $v_2\{2\}$ for all the events in the centrality range. Also, the results from DABMOD simulation are compared with experimental data from the ALICE experiment[223] showing the same behavior, even though the event plane method has been used to evaluate the data. Also, the comparison with experimental data should also take into account the $p_{\text{T}}$ interval cuts used in the detectors.

## 5.4 Energy loss fluctuations

The energy loss models described by equations 4.13 to 4.17, do not include fluctuations on the path-length[119,199,227–229] that could arise from the spatial geometry of the medium in the elastic energy loss and gluon number fluctuations in the radiative energy loss. In this work, these fluctuations are parametrized in the energy loss by the $\zeta$ term in equation 4.11 and the probability density $f(\zeta)$ of this term is selected from the equations 4.18 to 4.20,[195] for skewed and Gaussian fluctuations about the mean energy loss. These probabilities are shown in figure 5.17. The parameter is selected once for each heavy quark traversing the medium.

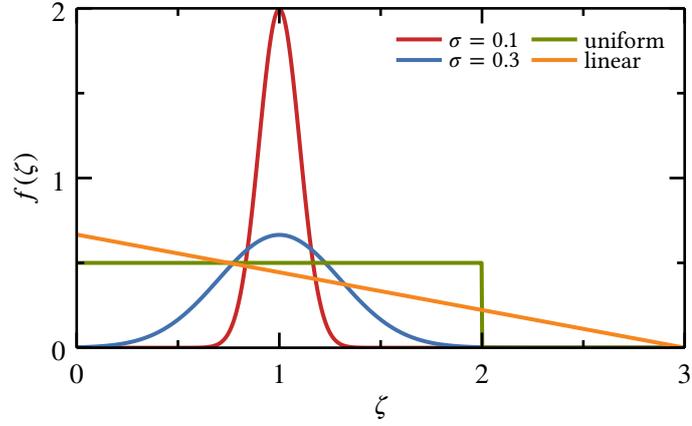

Figure 5.17 – Energy loss fluctuation probabilities from equations 4.18 to 4.20 with the Gaussian standard deviation for $\sigma = 0.1$ and $\sigma = 0.3$ which are used in the simulations.

The simulation is then executed from the beginning and a new $\lambda$ parameter is obtained from the $R_{AA}$ results as the fluctuations directly affect how the heavy quarks interact with the medium. The plot in figure 5.18 shows the results for the $D^0$ meson nuclear modification factor from 0−10% PbPb collisions at $\sqrt{s_{NN}} = 2.76$ TeV for all the fluctuations considered applied to the energy loss model from equation 4.14 and decoupling temperature $T_d = 120$ MeV. The simulation results are compared with the experimental data from the ALICE experiment[217] and also with the previous run without energy loss fluctuations. Instead of using a single value of $p_T$ as constraint for all the simulations as the approach adopted by similar works,[195] the $\lambda$ parameter has been fit using the procedure described in section 4.5.

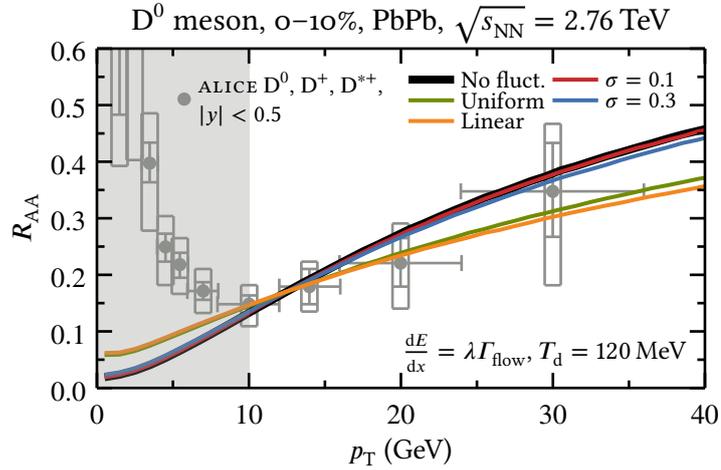

Figure 5.18 – Differential nuclear modification factor for $D^0$ meson compared with ALICE data[217] for central PbPb collisions at $\sqrt{s_{NN}} = 2.76$ TeV, energy loss from equation 4.14 and temperature $T_d = 120$ MeV. The energy loss fluctuations from equations 4.18 to 4.20 are compared with the previous result without fluctuations. The shaded area corresponds to the $p_T$ region where other effects may be important.

From the figure it is observed that the fluctuations in the energy loss tend to increase the $R_{AA}$ for low $p_T$ while at the same time increases the quenching for high $p_T$. These observations are similar to what has been previously shown for pions in a recent work[195], where the constraint has been made to match the nuclear modification factor at $R_{AA}(p_T = 10$ GeV). All the results are within the error bars of the experimental data points, however, the errors allow for a wide range of slopes for the $R_{AA}$ curves. Qualitatively, the uniform and linear fluctuations, which are inspired by pQCD calculations of the energy loss,[119,199,227–229] seem to better describe the overall behavior of the data points.

TABLE 5.3 – Values of the energy loss factor $\lambda_c$ for the charm quark fitted for PbPb collisions at $\sqrt{s_{NN}} = 2.76$ TeV and for $T_d = 120$ MeV using different energy loss fluctuation distributions.

| Fluctuations | Value | |
|---|---|---|
| No fluctuation | 0.99 | GeV fm$^{-1}$ |
| Gauss. $\sigma = 0.1$ | 1.01 | GeV fm$^{-1}$ |
| Gauss. $\sigma = 0.3$ | 1.10 | GeV fm$^{-1}$ |
| Uniform | 1.70 | GeV fm$^{-1}$ |
| Linear | 2.02 | GeV fm$^{-1}$ |

In order to quantitatively compare the results obtained from these simulations, table 5.3 shows the values of the parameter $\lambda_c$ for the different fluctuation distributions. The results show that by increasing the variance associated with the fluctuation distributions the $\lambda$ increases. Furthermore, the linearly skewed fluctuation leads to the highest value for $\lambda_c$ with a factor of 2 compared with the simulation without fluctuation, meaning that the overall energy loss is *reduced* by a factor of 2 when the fluctuations are present and the $\lambda_c$ has to compensate for that. The same factor has been predicted for jet quenching for collision at RHIC.[227]

The differential elliptic flow $v_2$ of $D^0$ meson for semi-central PbPb collisions at $\sqrt{s_{NN}} = 2.76$ TeV is shown in figure 5.19 for the constant energy loss model from equation 4.14 and $T_d = 120$ MeV. As with the $R_{AA}$ plot, all the energy loss fluctuations are compared with the results obtained when fluctuations are not present. As expected the increased nuclear modification factor at the low $p_T$ regime leads to a decrease in the elliptic flow while the opposite occurs for the high $p_T$ regime, although the crossing between the two different behaviors does not occur at the same $p_T$. The energy loss fluctuations have a strong impact on the elliptic flow, as can be observed from the plot, specially at low $p_T$ leading the simulation results to further underestimate the measured data. However, as the $p_T$ increases the impact of the fluctuations become less pronounced. As with the nuclear modification factor, this behavior has also been observed for pions in PbPb collisions at the same energy.

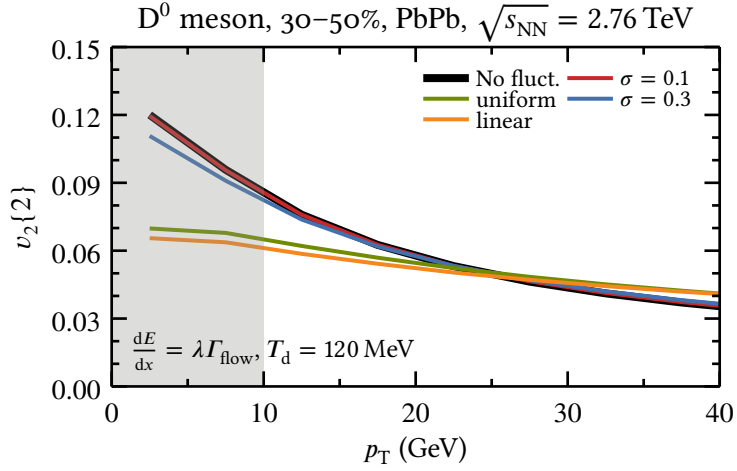

FIGURE 5.19 – Differential elliptic flow $v_2$ for $D^0$ meson for 30–50% PbPb collisions at $\sqrt{s_{NN}} = 2.76$ TeV, energy loss from equation 4.14 and temperature $T_d = 120$ MeV. The energy loss fluctuations from equations 4.18 to 4.20 are compared with the previous result without fluctuations. Experimental data has been suppressed in order to better visualize results by setting the $y$ axis limits, however the comparison with data can be visualized in figure 5.14. The shaded area corresponds to the $p_T$ region where other effects may be important.

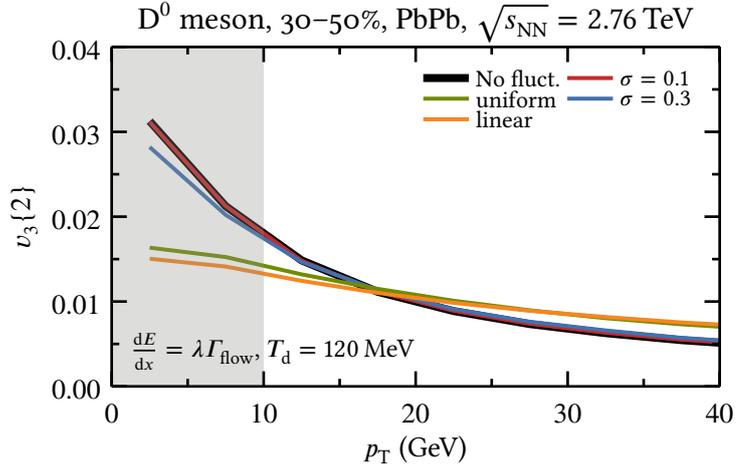

FIGURE 5.20 – Differential triangular flow $v_3$ for $D^0$ meson for 30–50% PbPb collisions at $\sqrt{s_{NN}} = 2.76$ TeV, energy loss from equation 4.14 and temperature $T_d = 120$ MeV. The energy loss fluctuations from equations 4.18 to 4.20 are compared with the previous result without fluctuations. The shaded area corresponds to the $p_T$ region where other effects may be important.

A similar behavior can be observed in figure 5.20 where the differential $D^0$ meson triangular flow $v_3$ for the same centrality is shown to also decrease when fluctuations are included in the simulations. It is also observed that for $p_T \gtrsim 17$ GeV the fluctuations tend to increase $v_3$.

Although no further development as comparing the results obtained from figures 5.18 to 5.20 has been made in this work, it has become clear that even though different energy loss models are able to similarly predict the nuclear modification factor, they can lead to very different results for elliptic and triangular flow. Furthermore, effects as energy loss fluctuations, that may have small impact on the $R_{AA}$ when considering the error bars associated with current measurements, can also greatly affect the observed results for the azimuthal anisotropy. The low $p_T$ regime is specially sensitive to a wide range of effects. Such effects can be implemented in the simulation that has been developed in this work for further studying the complex mechanisms of energy loss and hadronization of heavy quarks in relativistic heavy ion collisions.

# Conclusions

The unique properties of the quark gluon plasma produced in relativistic heavy ion collisions poses a great opportunity for further develop the understanding of the matter that constitutes the Universe. Within this hot and dense medium an enormous amount of particles are created and interact with each other and the study of these particles and interactions is highly involved, requiring strong efforts from both experimentalists and theorists. On the one hand, direct measurements of the matter created in the collision is not possible, making it necessary to infer about the properties of the medium from its sub-products. On the other hand, first principle derivation of the physical processes involved in the collision is very cumbersome, which is made worse when it's not possible to directly test its hypotheses. In this context, the phenomenological approach is highly valuable as one can test the behaviour of the system while setting specific parameters while keeping others fixed in a controlled way and helps closing the bridge between the experiment and the theory by setting different constraints.

Much is yet to be discovered from the heavy ion collisions regarding heavy quarks. One of the main open questions concerns the coupling of the heavy quarks with the medium and the mechanisms of hadronization. The interaction of these heavy quarks within the medium leads to the anisotropic flow observed in the experiments and the extent of these interactions is still not fully understood. Also, as experiments increase the energy of the collisions new results come up and can be used as important probes for the predictions that have been built with current available data. In this regard, with the increasing statistics from the experiments, new ways of looking into data may give different insights on the quantum mechanical fluctuations in initial conditions.

In this work, heavy flavor particles, with special focus in $B^0$ and $D^0$ mesons, have been studied by developing a computer simulation framework. In the simulation, heavy quarks are sampled at the beginning of the system's evolution from energy density profiles given by initial conditions based on the cgc framework. Initial properties of the heavy quarks are selected from pqcd calculations in proton-proton collisions and they are made to traverse the evolving medium background. The evolution of the medium is obtained from a viscous relativistic hydrodynamic code, which provides temperature and velocity profile of the underlying quark gluon plasma. Different parametrization equations for the heavy quark energy loss are employed during the evolution of the system until the decoupling occurs at a chosen temperature $T_d$, which varies in a certain range. The heavy quarks then hadronize via fragmentation leading to $B^0$ and $B^0$ mesons which will then decay into the semi-leptonic channels and the final electron spectra is obtained. The framework allows for each of these steps to be separated in order to obtain spectra for all levels: quark, meson, and electron. The same procedure is executed for about 1000 events in each centrality class in an event-by-event basis so that later analysis can evaluate the particle correlations using the cumulants method, which is a more recent and unambiguous evaluation of the azimuthal anisotropy. From the results of the simulation a new observable that correlates heavy with soft particles in an event shape engineering approach has been explored under changing of the framework parameters.

The simulation predicted a linear correlation between the heavy and the soft sector, even though the mechanisms for the production of flow in both cases differs. From these results it has been concluded that event shape engineering can lead to useful information on the correlation of the particles which is consistent with similar information obtained from the cumulants method and can also provide a novel insight on the mechanisms of heavy flavor coupling with the medium. Further exploring of these results should be feasible during the next runs of the lhc experiments by extending the event-shape engineering methods that have already been applied to high-$p_T$ soft particles. In addition, higher order cumulants such as the triangular coefficient $v_3\{2\}$ have been predicted for the first time for heavy mesons in PbPb collisions from the simulation framework prior to the publication of experimental results. The comparison with these recent data has shown the predictions to be consistent within error expectations. Furthermore, the $v_3$ obtained from the simulation is much more sensitive to the decoupling temperature $T_d$ than the other explored observables, which indicates that $v_3$ can be an important asset in the study of heavy flavor decoupling processes.

The development of this study as well as the results obtained from it have been orally presented at Strangeness in Quark Matter 2016, Physics Meeting 2016, and Quark Matter 2017 conferences. In addition, this work has also been presented in poster form at Quark Matter 2014, Quark Matter 2015, and Hadron Physics 2015. At the time of writing this document, a submission for a Physical Review Letters paper is being reviewed.

The study presented in this work can be easily expanded for soft particles in order to obtain correlations between high-$p_T$ and low-$p_T$ regimes. Also, different energy loss model parametrizations can be explored in order to refine the results obtained here. Further implementation of effects not considered in this work such as heavy quark coalescence is also possible due to the modularization of the framework. Finally, as the $p_T$ spectra obtained from the simulation is highly general, a series of different observables can also be obtained, such as symmetric cumulants. This is definitely not a closed case, and the phenomenological approach presented here opens up for multiple possibilities of studies that can further improve the current understanding on the quark gluon plasma.